\documentclass[]{aa}
\usepackage{graphicx}
\graphicspath{{./figures/}}
\usepackage{natbib}
\usepackage{amssymb,amsmath,amsfonts}
\usepackage{epsfig}
\usepackage{mathptmx}
\usepackage{longtable}
\usepackage{comment}
\usepackage{color}
\usepackage[caption=false]{subfig}
\usepackage{animate}
\usepackage{rotating}
\usepackage{pdflscape}
\usepackage{url}
\usepackage{enumitem}
\usepackage[draft]{hyperref}
\usepackage[normalem]{ulem}
\begin{document} 

\title{The LOFAR Two-metre Sky Survey -- Second Data Release}
\title{The LOFAR Two-metre Sky Survey} 
\subtitle{V. Second data release} 
\authorrunning{Shimwell et~al.}
\titlerunning{The LOFAR Two-metre Sky Survey -- Second Data Release}
\author{T. W. Shimwell$^{1,2}$\thanks{E-mail: shimwell@astron.nl},
M. J. Hardcastle$^{3}$, 
C. Tasse$^{4,5}$,
P. N. Best$^{6}$,
H. J. A. R\"{o}ttgering$^{2}$, 
W. L. Williams$^{2}$,
A. Botteon$^{2}$,
A. Drabent$^{7}$,
A. Mechev$^{2}$,
A. Shulevski$^{2}$
% ...
\\
(Affiliations can be found after the references)}
\institute{}
\date{Accepted December 23, 2021; received October 19, 2021; in original form \today}

\abstract{
\noindent
In this data release from the ongoing LOw-Frequency ARray (LOFAR) Two-metre Sky Survey (LoTSS) we present 120-168\,MHz images covering 27\% of the northern sky. Our coverage is split into two regions centred at approximately 12h45m +44$^\circ$30$\arcmin$ and 1h00m +28$^\circ$00$\arcmin$ and spanning 4178 and 1457 square degrees respectively. The images were derived from 3,451\,hrs (7.6\,PB) of LOFAR High Band Antenna data which were corrected for the direction-independent instrumental properties as well as direction-dependent ionospheric distortions during extensive, but fully automated, data processing. A catalogue of 4,396,228 radio sources is derived from our total intensity (Stokes I) maps, where the majority of these have never been detected at radio wavelengths before. At 6$\arcsec$ resolution, our full bandwidth Stokes I continuum maps with a central frequency of 144\,MHz have: a median rms sensitivity of 83\,$\mu$Jy/beam; a flux density scale accuracy of approximately 10\%; an astrometric accuracy of 0.2$\arcsec$; and we estimate the point-source completeness to be 90\% at a peak brightness of 0.8\,mJy/beam. By creating three 16\,MHz bandwidth images across the band we are able to measure the in-band spectral index of many sources, albeit with an error on the derived spectral index of $>\pm0.2$ which is a consequence of our flux-density scale accuracy and small fractional bandwidth. Our circular polarisation (Stokes V) 20$\arcsec$ resolution 120-168\,MHz continuum images have a median rms sensitivity of 95\,$\mu$Jy/beam, and we estimate a Stokes I to Stokes V leakage of 0.056\%. Our linear polarisation (Stokes Q and Stokes U) image cubes consist of $480 \times 97.6$\,kHz wide planes and have a median rms sensitivity per plane of 10.8\,mJy/beam at 4$\arcmin$ and 2.2\,mJy/beam at $20\arcsec$; we estimate the Stokes I to Stokes Q/U leakage to be approximately 0.2\%. Here we characterise and publicly release our Stokes I, Q, U and V images in addition to the calibrated $uv$-data to facilitate the thorough scientific exploitation of this unique dataset.}

 \maketitle

\newcommand{\textCyril}[1]{\textcolor{blue}{Cyril:} \textcolor{red}{ #1}}
\newcommand{\cCyr}[2]{[\textcolor{blue}{Cyril:}"#1" $\rightarrow$ "\textcolor{red}{ #2}"]}
\def\kMS/{{k\sc{ms}}}
\def\DDF/{{{\sc ddf}acet}}
\def\PSF/{{\sc psf}}
\def\SSD/{{\sc ssd}}
\def\SSDGA/{{\sc ssd-ga}}
\def\LOTTSpipe/{{{LoTSS}-DR1}}
\def\CLEAN/{{\sc clean}}

\section{Introduction}

The LOw Frequency ARray (LOFAR; \citealt{vanHaarlem_2013}) Two-metre Sky Survey (LoTSS; \citealt{Shimwell_2017}) is one of several ongoing very wide area deep radio wavelength sky surveys. Other similar projects with different instruments include the Evolutionary Map of the Universe (EMU; \citealt{Norris_2011,Norris_2021}), the Polarization Sky Survey of the Universe's Magnetism (POSSUM; \citealt{Gaensler_2010}), the APERture Tile In Focus surveys (APERTIF surveys; Hess et al. in prep), the GaLactic and Extragalactic All-Sky MWA-eXtended survey (GLEAM-X; Hurley Walker et al. in prep), the Karl G. Jansky Very Large Array Sky Survey (VLASS; \citealt{Lacy_2020}) and the  Global Magneto-Ionic Medium Survey (GMIMS; \citealt{Wolleben_2019}, \citealt{Wolleben_2021}). LoTSS also forms part of a broader LOFAR Surveys Key Science Project (LSKSP; \citealt{Rottgering_2011}) that is striving to map the low-frequency ($<200$\,MHz) northern sky with a series of surveys spanning a range of depths, frequencies, and areas.  The 120-168\,MHz LoTSS survey is the highest frequency very wide-area LOFAR surveys project and is complemented by the ongoing very wide area  42-66\,MHz LOFAR Low Band Antenna Sky Survey (LoLSS; \citealt{deGasperin_2021}) and the even lower frequency 14-30\,MHz LOFAR Decametre Sky Survey (LoDSS) which has recently started. Furthermore,  narrower area, but far deeper surveys of several fields with exceptionally high quality auxiliary data are also being carried out, namely the LoTSS and LoLSS Deep Fields (\citealt{Tasse_2021}, \citealt{Sabater_2021}, \citealt{Kondapally_2021}, \citealt{Duncan_2021}, Best et al. in prep and \citealt{Williams_2021}) 
as well as moderate depth observations (or otherwise tailored data processing) towards targets of particular scientific interest (the H-ATLAS North Galactic Pole, North Ecliptic Pole, Virgo cluster, Coma cluster, Corona Borealis supercluster, Abell 2255, Abell 399-401, GJ 1151, GJ 412 and others).

The capabilities of LOFAR, and the amount of observing time secured to date, have enabled LoTSS to achieve a unique combination of sensitivity ($\sim$100$\mu$Jy/beam) coupled with high resolution ($\sim$6$\arcsec$) and an accurate recovery of very extended (up to degree scales) objects -- all at a low radio frequency of 144\,MHz.
The emission mechanism for radio sources is generally synchrotron and the sources 
typically increase in integrated flux density ($S_{I}$) with decreasing frequency ($\nu$), with the emission often characterised by $S_\nu \propto \nu^{\alpha}$ where the conventional spectral index ($\alpha$) is $-0.7$ (e.g. \citealt{Condon_2002}).
With its properties, LoTSS is therefore able to detect, and precisely characterise, an exceptionally high density of radio sources. The source density far exceeds ($>$ 8 times) that of pioneering very wide-area higher-frequency surveys such as the NRAO VLA Sky Survey (NVSS; \citealt{Condon_1998}), Faint Images of the Radio Sky at Twenty-Centimeters (FIRST; \citealt{Becker_1995}), Sydney University Molonglo Sky Survey (SUMSS; \citealt{Bock_1999} and \citealt{Mauch_2003}), WEsterbork Northern Sky Survey (WENSS; Rengelink et al. 1997) and Westerbork In the Southern Hemisphere (WISH; \citealt{DeBreuck_2002}) as well as that of current state-of-the-art low-frequency surveys such as  the TIFR  GMRT  Sky  Survey alternative data release (TGSS-ADR1; \citealt{Intema_2017}), GaLactic and  Extragalactic  All-sky MWA (GLEAM; \citealt{Wayth_2015} and \citealt{HurleyWalker_2017}), LOFAR Multifrequency Snapshot Sky Survey (MSSS; \citealt{Heald_2015}) and the Very Large Array Low-frequency Sky Survey Redux (VLSSr; \citealt{Lane_2014}). Thus, LoTSS, and other forthcoming radio surveys with significantly improved sensitivities, resolutions, or other unique properties such as fractional bandwidth, frequency- or time-resolution, are dramatically enriching our view of the radio Universe. Specifically, the suite of ongoing LOFAR surveys will enable us to probe the 14-168\,MHz northern sky over very wide areas with a sensitivity of $\sim0.1\times \left(\frac{\nu}{144\,\textrm{MHz}}\right)^{-2.5}$\,mJy/beam and a resolution of $\sim 6 \times \left(\frac{144\,\textrm{MHz}}{\nu}\right)$\,arcsec whilst narrow areas will be mapped with a factor of up to ten improved sensitivity.

To realise LoTSS, extensive development has been required to build
strategies that correct the severe ionospheric distortions which vary
rapidly with both time and direction on the sky. If uncorrected, these effects
prohibit high fidelity imaging at low frequencies (see e.g. \citealt{Lonsdale_2005} and \citealt{Intema_2009}). 
Furthermore, each individual LoTSS pointing (of which there are 3168 across the Northern sky) corresponds to a very large dataset (8.8\,TB) and 
thus such strategies
must be able to run routinely and efficiently in order to produce the
desired maps within a reasonable time period.

A further challenge,
common to all radio surveys, is that even once high fidelity maps are produced,
to increase the scientific value
of the radio catalogues we need procedures that 
carefully associate the detected sources and cross-match them with
other auxiliary catalogues to deduce information that is vital to
understand the nature of the detected radio sources. Over time our
methods have improved. For example, in the preliminary LoTSS data
release (LoTSS-PDR; \citealt{Shimwell_2017}) we presented a catalogue
of 44,500 radio sources but at that time we were unable to routinely
correct for ionospheric errors over very wide areas of the sky and
were thus limited in resolution, sensitivity and fidelity. This was
followed by the first LoTSS data release (LoTSS-DR1; \citealt{Shimwell_2019}) that mapped the same area but utilised an automated and robust direction dependent calibration pipeline to produce a much larger radio catalogue of 325,694 components. In LoTSS-DR1 we also performed significant post processing of the radio catalogues to enhance their scientific potential. The 325,694  components were carefully grouped into 318,520 distinct radio sources and  73\% of these were matched to optical or infrared host galaxies (\citealt{Williams_2019}) and, where possible, photometric redshifts were estimated (\citealt{Duncan_2019}). 

Our aims within the LSKSP are not only to provide
publicly available radio images and catalogues of the sky but also to increase our understanding of the detected sources through a coordinated scientific exploitation of the images and auxiliary data.  
To date, with this approach, the LOFAR surveys have facilitated numerous scientific studies\footnote{https://lofar-surveys.org/publications.html} in core areas of radio astronomy such as the physics of active galactic nuclei, particle acceleration in galaxy clusters, large scale structure and star formation. 
Furthermore, the breadth of scientific studies continues to expand to include topics ranging from cosmological studies (\citealt{Siewert_2020}) through to pulsars (\citealt{Tan_2018}), supernovae remnants (\citealt{Arias_2019}) and even exoplanets (\citealt{Vedantham_2020}). Meanwhile, valuable synergies are being established such as those with the LOFAR Magnetism Key Science Project\footnote{https://lofar-mksp.org/}, APERTIF imaging surveys, Extended Baryon Oscillation Spectroscopic Survey (eBOSS; \citealt{Dawson_2016}), extended ROentgen Survey with an Imaging Telescope Array (eROSITA; \citealt{Predehl_2021}) and the William  Herschel  Telescope  Enhanced Area  Velocity  Explorer survey of LOFAR selected sources (WEAVE-LOFAR; \citealt{Smith_2016}) which are each enabling new scientific studies (e.g. \citealt{Ghirardini_2021}, \citealt{Morganti_2021}, \citealt{OSullivan_2020}, \citealt{Wolf_2021}). Finally, the LOFAR surveys are also having a large technical impact with studies of calibration and imaging techniques (e.g. \citealt{deGasperin_2019}, \citealt{Tasse_2021} and \citealt{vanWeeren_2021}, \citealt{Morabito_prep} and \citealt{Sweijen_2022}), efficient distributed processing
(\citealt{Drabent_2019} and \citealt{Mechev_2019}), photometric
redshift estimators (\citealt{Duncan_2019}) and automated source classification
(e.g. \citealt{Mostert_2021} and \citealt{Mingo_2019}). Excitingly, despite all
of these advances, the LOFAR surveys data still retain vast, and largely untapped, potential. For example, 96\% of existing LoTSS observations have been conducted with the full international LOFAR telescope which now includes 14 stations outside of the Netherlands and are archived at high (1\,s) time and frequency (12.1875\,kHz) resolution. Presently, due to resource limitations and ongoing technical developments, during regular LoTSS processing we significantly average the data and remove the international stations. We thus do not yet fully realise the higher sensitivity sub-arcsecond wide-field imaging (e.g. \citealt{Morabito_prep}, \citealt{Sweijen_2022}), source variability (e.g. \citealt{Vedantham_2020,callingham2021}) and spectral line (see e.g. \citealt{Emig_2020}, \citealt{Salas_2019}) capabilities of the data.

In this publication we present our second LoTSS data release (LoTSS-DR2) and a characterisation of the associated images. This builds significantly upon our previous work by making use of our enhanced direction dependent calibration and imaging processing pipeline (see \citealt{Tasse_2021}) as well as improved processing efficiency and automation (see e.g. \citealt{Drabent_2019} and \citealt{Mechev_2019}).
These improvements enable us to present images spanning 5,634 square degrees (27\%) of the Northern sky, and a catalogue containing 4,396,228 radio sources -- the largest catalogue of radio sources released to date. In addition to radio continuum catalogues and images at multiple resolutions, we also release  polarisation images and calibrated $uv$-datasets. All data products associated with this release have the Digital Object Identifier (DOI) 10.25606/SURF.LoTSS-DR2 and are available via the collaboration's webpage\footnote{\url{https://www.lofar-surveys.org/}}, the ASTRON Virtual Observatory\footnote{\url{https://vo.astron.nl}} and the SURF Data Repository\footnote{\url{https://repository.surfsara.nl/}}.

In Sect. \ref{sec:observations} we describe the observations and data processing before presenting an assessment of the image quality in Sect. \ref{sec:image_quality}. In Sect. \ref{sec:data_release} we outline the products that have been publicly released. In Sect. \ref{sec:future_prospects} we highlight future prospects before summarising in Sect. \ref{sec:summary}. 

\begin{figure*}   \centering
   \includegraphics[width=\linewidth]{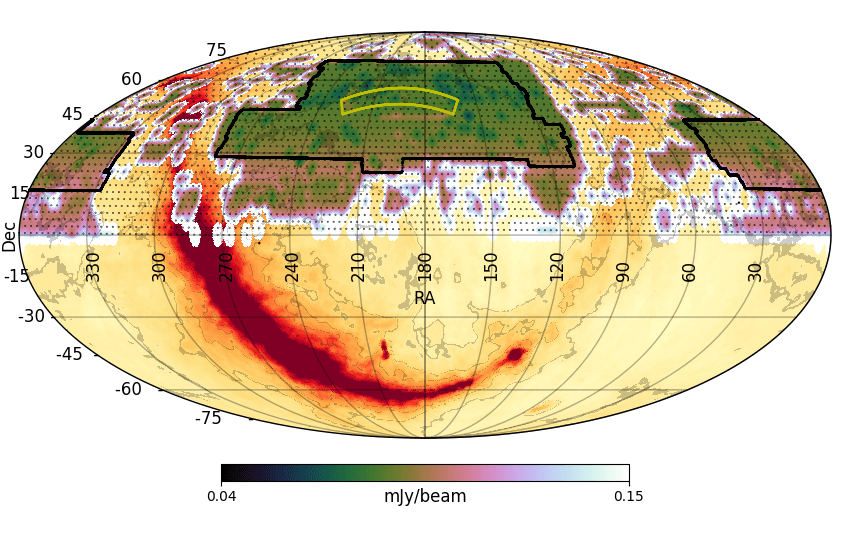}
   \caption{The status of the LoTSS observations as of April 2021 and the approximate  current  sensitivity coverage (accounting for station projection and typical sensitivities achieved to date) overlaid on  the  \cite{Haslam_1982}  408\,MHz  all-sky image (corresponding to yellow to deep red colours, with associated contours). The  yellow and  black  outlines  show  the  LoTSS-DR1 and LoTSS-DR2 areas respectively and the small black dots show the 3168 LoTSS grid positions. LoTSS-DR1 included 63 pointings, LoTSS-DR2 includes 841 pointings; we have fully observed 1623 pointings including those of DR2, a further 154 pointings are partly observed and observations still need to be conducted for 1391 pointings to complete the survey.}
   \label{fig:DR2-region}
\end{figure*}

\section{Observations and data processing}
\label{sec:observations}

As shown in Fig. \ref{fig:DR2-region}, LoTSS-DR2 consists of 841 pointings and it covers a total of 5634 square degrees which corresponds approximately to our contiguous coverage at the time of beginning the LoTSS-DR2 processing run. The data release is formed by two contiguous regions that are centred at approximately 12h45m00s +44$^\circ$30$\arcmin$00$\arcsec$ (RA-13 region) and 1h00m00s +28$^\circ$00$\arcmin$00$\arcsec$ (RA-1 region) and span 4178 and 1457 square degrees, or 626 and 215 pointings, respectively. The data were taken between 2014-05-23 to 2020-02-05 as part of the LoTSS projects LC2\_038, LC3\_008, LC4\_034, LT5\_007, LC6\_015, LC7\_024, LC8\_022, LC9\_030, LT10\_010 and the co-observing projects LC8\_014, LC8\_030, DDT9\_001, LC9\_011, LC9\_012, LC9\_019, LC9\_020, COM10\_001, LC10\_001, LC10\_010, LC10\_014, LT10\_012, LC11\_013, LC11\_016, LC11\_019, LC11\_020, LC12\_014. All the data that were processed as part of this data release are stored in the LOFAR Long Term Archive (LTA{\footnote{\url{https://lta.lofar.eu/}}}) with approximately 62\% in Forschungszentrum J\"{u}lich\footnote{\url{http://www.fz-juelich.de}}, 32\% in SURF\footnote{\url{https://www.surf.nl/}} and the remaining 6\% in Pozna\'{n}\footnote{\url{http://www.man.poznan.pl/online/pl/}}. The vast majority of pointings were observed for a total of 8\,hrs with 48\,MHz (120-168\,MHz) of bandwidth which allows for two pointings to be observed simultaneously with current LOFAR capabilities. 
 However, primarily due to the co-observing program\footnote{The \url{https://www.lofar-surveys.org/co-observing.html}} through which we exploit the multi-beam capability of LOFAR and accumulate LoTSS data simultaneously with observations conducted for other projects, for 18 of the pointings in LoTSS-DR2 we have used data that has the same frequency coverage but a total integration time of $\sim$16\,hrs. 
The overall observing time utilised for this data release is 3451\,hrs and the volume of archived data that was processed is 7.6\,PB. Thus the average data size for an 8\,hr pointing (two observed simultaneously) is 8.8\,TB but there is significant variation because data  
that have been recorded since 2018-09-11 are typically five times smaller than those before this date due to Dysco compression (\citealt{Offringa_2016}) being utilised by the radio observatory prior to ingesting data into the LTA in more recent observations.

To process the data they are first `staged' in the LTA; staging is the
procedure of copying data from tape to disk and is necessary to make
the large archived datasets available for transfer to a compute
cluster. The data are then processed with a direction independent (DI)
calibration pipeline that is executed on compute facilities at
Forschungszentrum J\"{u}lich and SURF (see \citealt{Mechev_2017}
and \citealt{Drabent_2019}). These compute clusters are connected to
the local LTA sites with sufficiently fast connections to mitigate the
difficulties that would be experienced if we were to download these
large datasets to external facilities. Unfortunately data transfer issues are not yet fully mitigated as we currently do not process data on a compute cluster local to the Pozna\'{n} archive and instead we copy these data (6\% of LoTSS-DR2) to Forschungszentrum J\"{u}lich or SURF for processing. 

The DI calibration
pipeline\footnote{\url{https://github.com/lofar-astron/prefactor}}
used for this data processing follows the same procedure as that used
in LoTSS-PDR and LoTSS-DR1. This method is described in \cite{vanWeeren_2016}
and \cite{Williams_2016} and makes use of several software packages
including the Default Pre-Processing Pipeline (DP3;
\citealt{van_Diepen_2018}),  LOFAR SolutionTool (LoSoTo;
\citealt{deGasperin_2019}) and AOFlagger (\citealt{Offringa_2012}).
The pipeline corrects for direction independent errors such as the
clock offsets between different stations, ionospheric Faraday
rotation, the offset between XX and YY phases and amplitude
calibration solutions (see \citealt{deGasperin_2019} for a detailed
description of these effects). The \cite{Scaife_2012} flux density scale is
used for the amplitude calibration and we use TGSS-ADR1 sky models\footnote{The TGSS-ADR1 catalogues have gaps in the region around 8h45m +31$^\circ$30$\arcmin$ and here we use the \cite{Scheers_2011} LOFAR Global Sky Model instead} of our
target fields for an initial phase calibration, although both the
amplitude and phase calibration are refined during subsequent
processing. For regular LoTSS processing we have set up the pipeline to reduce the data volume, typically by a factor of 64 by averaging both in time and frequency. This is because the archived LoTSS data typically have a frequency resolution of 16 channels per 0.195\,MHz subband and a time resolution of 1\,s to facilitate future studies with the international LOFAR stations as well as spectral and time dependent studies,  but such high time and frequency resolution data is not required for 6$\arcsec$ imaging. During the DI calibration the data are therefore averaged to a frequency resolution of 2 channels per 0.195\,MHz subband and a time resolution of 8\,s.

Once the DI calibration pipeline is complete, the smaller, more averaged, output datasets can be downloaded to other compute clusters for further processing with a more computationally expensive direction dependent (DD) calibration and imaging pipeline\footnote{\url{https://github.com/mhardcastle/ddf-pipeline}}. The DD routine is an improvement upon that used in LoTSS-DR1 and again makes use of kMS (\citealt{Tasse_2014} and \citealt{Smirnov_2015}) for direction dependent calibration, and of DDFacet (\citealt{Tasse_2018}) to apply the direction dependent solutions during imaging. Compared to LoTSS-DR1, the most significant changes are the fidelity of faint diffuse emission and the increased dynamic range (see Sect. \ref{sec:emission_recovery} and \ref{sec:dynamic_range} respectively). The LoTSS-DR2 DD pipeline and its performance are described in detail in \cite{Tasse_2021}; however, for completeness we briefly summarise the procedure below.

We begin the processing with just a quarter of the DI calibrated channels (spaced across the frequency coverage) by creating a wide-field
($8.3^\circ \times 8.3^\circ$) image. Using the resulting sky model we
revise the direction independent calibration and tessellate the field
into 45 different directions. The recalibrated data are imaged to
update the sky model, and with the new model, calibration solutions are derived towards each of the 45 directions simultaneously. Then, we image the wide-field again but this time applying the phase corrections from the direction dependent calibration solutions which allows us to produce a further improved sky model.
Here we perform an initial refinement of the
flux density scale through the bootstrap procedure described by
\cite{Hardcastle_2016}, which was also used in the LoTSS-DR1
processing.  The flux density scale is further refined during mosaicing but this initial refinement helps ensure emission is described by a power-law which aids the deconvolution.
Direction dependent calibration solutions are again
derived from the up-to-date sky model and this time both the amplitude
and phase are applied in the subsequent imaging step. Using these
solutions, together with the updated sky model, we predict the
apparent direction-independent view of the sky and perform a further
direction-independent calibration step using that model and a further
imaging step. All the data are then included for the first time and
direction-independent followed by direction-dependent calibration solutions
are derived using the latest sky model. The data are then imaged
again, and further direction-dependent calibration solutions are
derived from the resulting sky model before the final imaging steps
are conducted with the latest calibration solutions. 

The final imaging
steps result in: (i) full-bandwidth high (6$\arcsec$) and low (20$\arcsec$)
resolution Stokes I images; (ii) three 16\,MHz bandwidth high (6$\arcsec$) resolution Stokes I images with central frequencies of 128, 144 and 160\,MHz; (iii)  Stokes Q and U low (20$\arcsec$) and very
low (4$\arcmin$) resolution undeconvolved image cubes with a frequency
resolution of 97.6\,kHz; (iv) and a Stokes V full-bandwidth low
(20$\arcsec$) resolution undeconvolved image. Here only Stokes I products are deconvolved due to the deconvolution capabilities of DDFacet at the time of processing.
Once the data are processed, the final products are archived and an
automated quality assessment of the image is conducted to assess the
astrometry, flux density scale accuracy and noise level.

Some notable aspects of the DD pipeline processing include the improvement of 
the astrometric accuracy of the final high resolution Stokes I images
by performing a facet-based astrometric alignment (as in LoTSS-DR1) with sources in the 
the Pan-STARRS optical catalogue (\citealt{Flewelling_2020}) and applying
appropriate shifts when imaging (see \citealt{Shimwell_2019}).
To deconvolve thoroughly, throughout the processing we refine the masks used for deconvolution,
we also continuously propagate previously derived deconvolution components to subsequent
imaging steps to avoid having to fully deconvolve at each imaging
iteration, and we regularise the calibration solutions to effectively
reduce the number of free parameters that are applied when imaging.
Moreover, as characterised in Sect. 3.3 of \cite{Shimwell_2019} and detailed in \cite{Tasse_2018}, by using a facet-dependent point spread function we account for time-averaging and bandwidth-smearing effects (e.g. \citealt{Bridle_1999}) for deconvolved sources, this would otherwise be significant (a $\sim30\%$ reduction in peak brightness at a distance of 2.5$^\circ$ from the pointing centre) when imaging at 6$\arcsec$ with 2 channels per 0.195\,MHz subband and a time resolution of 8\,s. Finally, we note that the restoring beam used in DDFacet for each image product type is kept constant over the data release region and that all image products are made with a $uv$-minimum of 100\,m with the $uv$-maximum varied to provide images at different resolutions - the highest resolution 6$\arcsec$ images use baselines up to 120km (i.e. all LOFAR stations within the Netherlands).

The DD calibration has been primarily conducted on the LOFAR-UK compute
facilities\footnote{\url{https://lofar-uk.org/lucf.html}} hosted at
the University of Hertfordshire, but a small fraction of processing was also carried out
on the Italian LOFAR computing facilities\footnote{url{http://www.lofar.inaf.it/index.php/en/analisi-dati-en/computationa-data-analysis}} and compute clusters at Leiden University and the University of Hamburg. The DI and DD processing, as well as
the observational status and quality indicators are all kept track of
in central MySQL databases which are updated during the data processing. This allows us to easily coordinate automated processing across many different compute clusters with minimal user interaction.

The mosaicing and cataloguing follow the same procedure as used
for LoTSS-DR1 which is described in \cite{Shimwell_2019}. This implies a mosaic is produced for each pointing by reprojecting all neighbouring pointing images onto the same frame as the central pointing and averaging together the images using weights equal to the station beam attenuation combined with the image noises. Poorly calibrated facets, which are generally caused by severe ionospheric or dynamic range effects, are identified in each image as those with larger than 0.5$\arcsec$ astrometric errors (derived from cross matching with Pan-STARRS) and these regions are blanked in the individual pointing images prior to mosaicing. On average this results in 15$\pm$22\% of the pixels within 30\% of the primary beam power level being excluded for a given pointing. Unlike in LoTSS-DR1, we  further refine the flux density scale of the images during the mosaicing procedure by applying the method that is described in Sect. \ref{sec:flux_scale}. Sources are detected on the mosaiced images using  \textsc{PyBDSF} (\citealt{Mohan_2015}) with wavelet decomposition and a 5$\sigma_{LN}$ peak detection and 4$\sigma_{LN}$ threshold to define the boundaries of source islands, where $\sigma_{LN}$ is the local background noise. During source detection, \textsc{PyBDSF} characterises emission with Gaussian components which are automatically combined into distinct sources to create the source catalogue. This automated association of Gaussian components into final sources is limited because of various reasons such as the complexity and the extent of the source structures, the angular separation between components of the emission related to the same source, and the entanglement of emission from distinct objects. As described in Sect. \ref{sec:value_added_cats}, our attempts to refine the \textsc{PyBDSF} catalogues through source association/deblending, and cross-identification with optical/infrared (e.g. \citealt{Williams_2019} and \citealt{Kondapally_2021}) are ongoing. 

\begin{figure*}   \centering
   \includegraphics[width=0.9\linewidth]{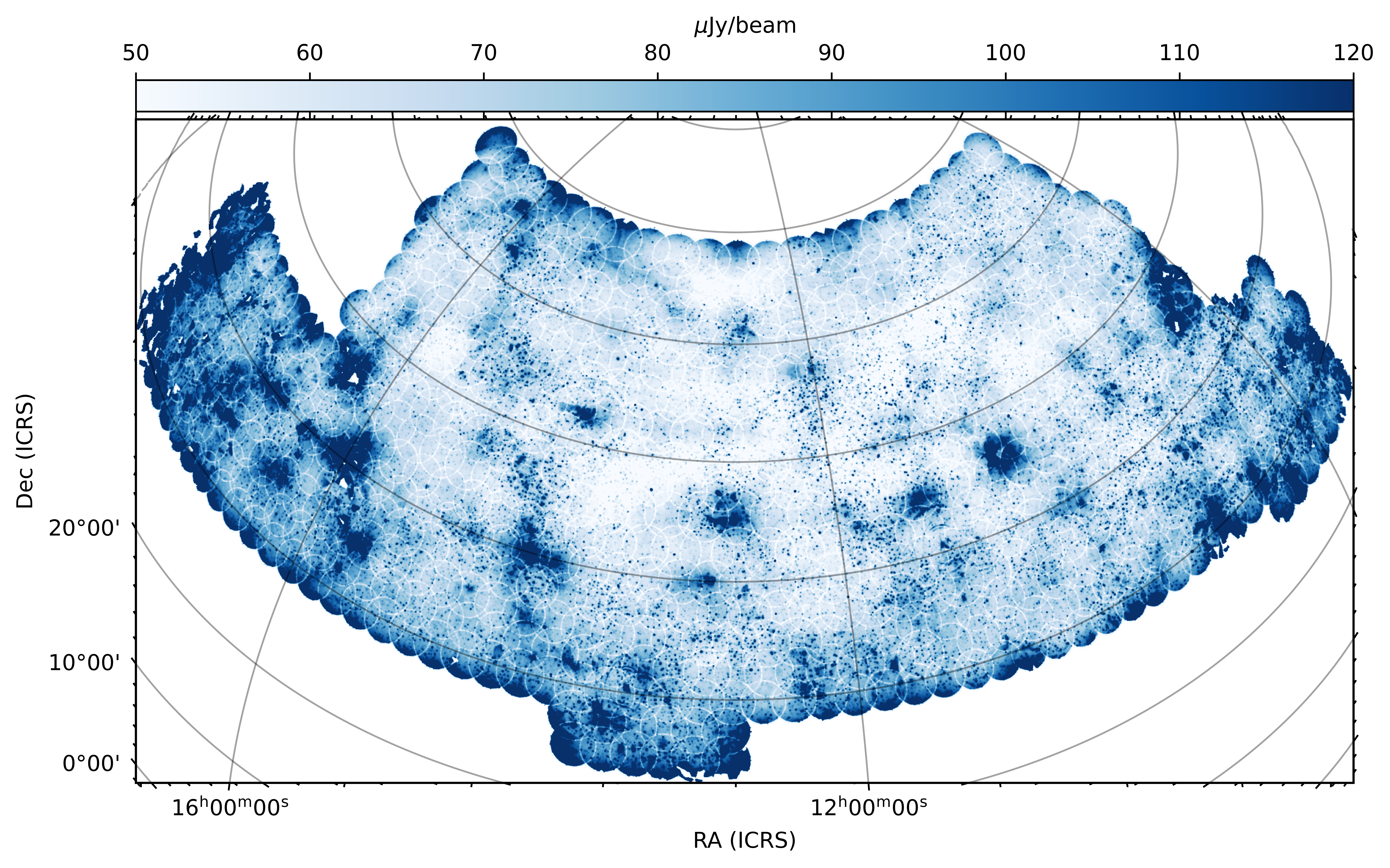}
   \includegraphics[width=0.9\linewidth]{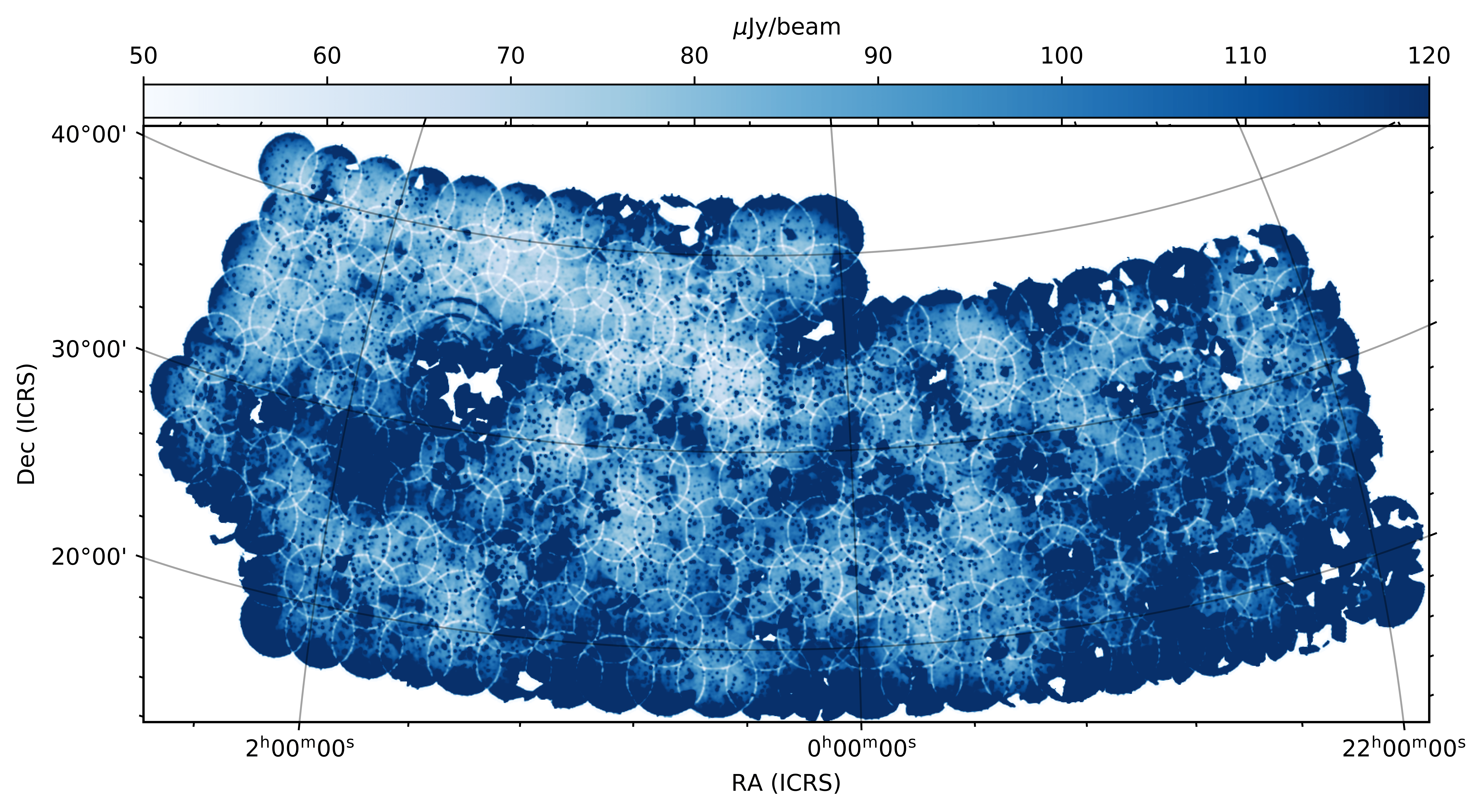}
      \caption{The noise variations in the two regions covered by LoTSS-DR2 with the coverage of the 841 individual pointing outlined. The RA-13 (top) and RA-1 (bottom) regions span 4178 and 1457 square degrees and have median rms values of 74$\mu$Jy/beam and 106$\mu$Jy/beam respectively. Failed facets (white regions) are generally caused by either poor ionospheric conditions or dynamic range issues around bright sources such as 3C\,48 and 3C\, 196.}
   \label{fig:mosaic-noisemap}
\end{figure*}

\begin{figure*}   \centering
   \includegraphics[width=0.42\linewidth]{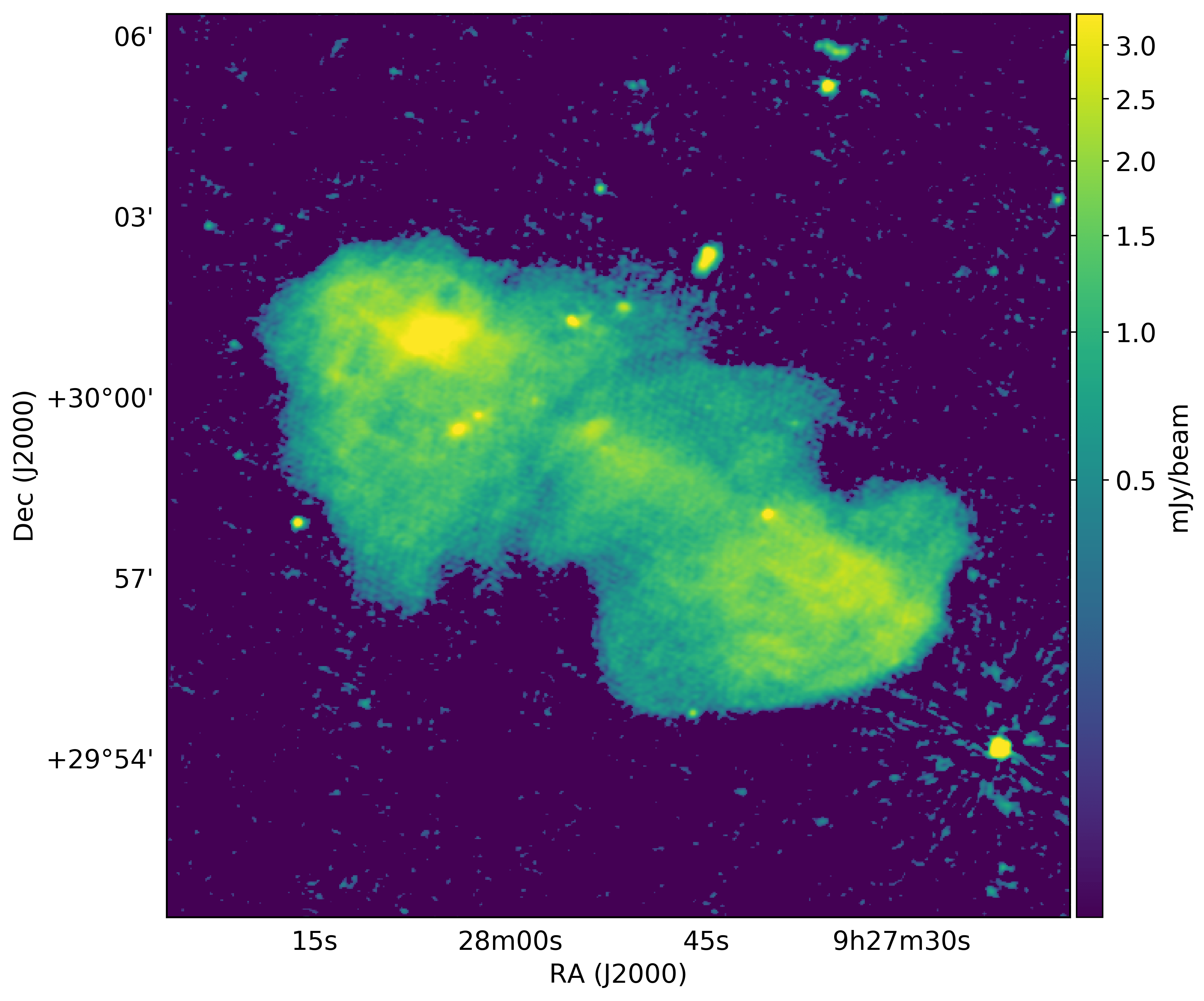}
   \includegraphics[width=0.42\linewidth]{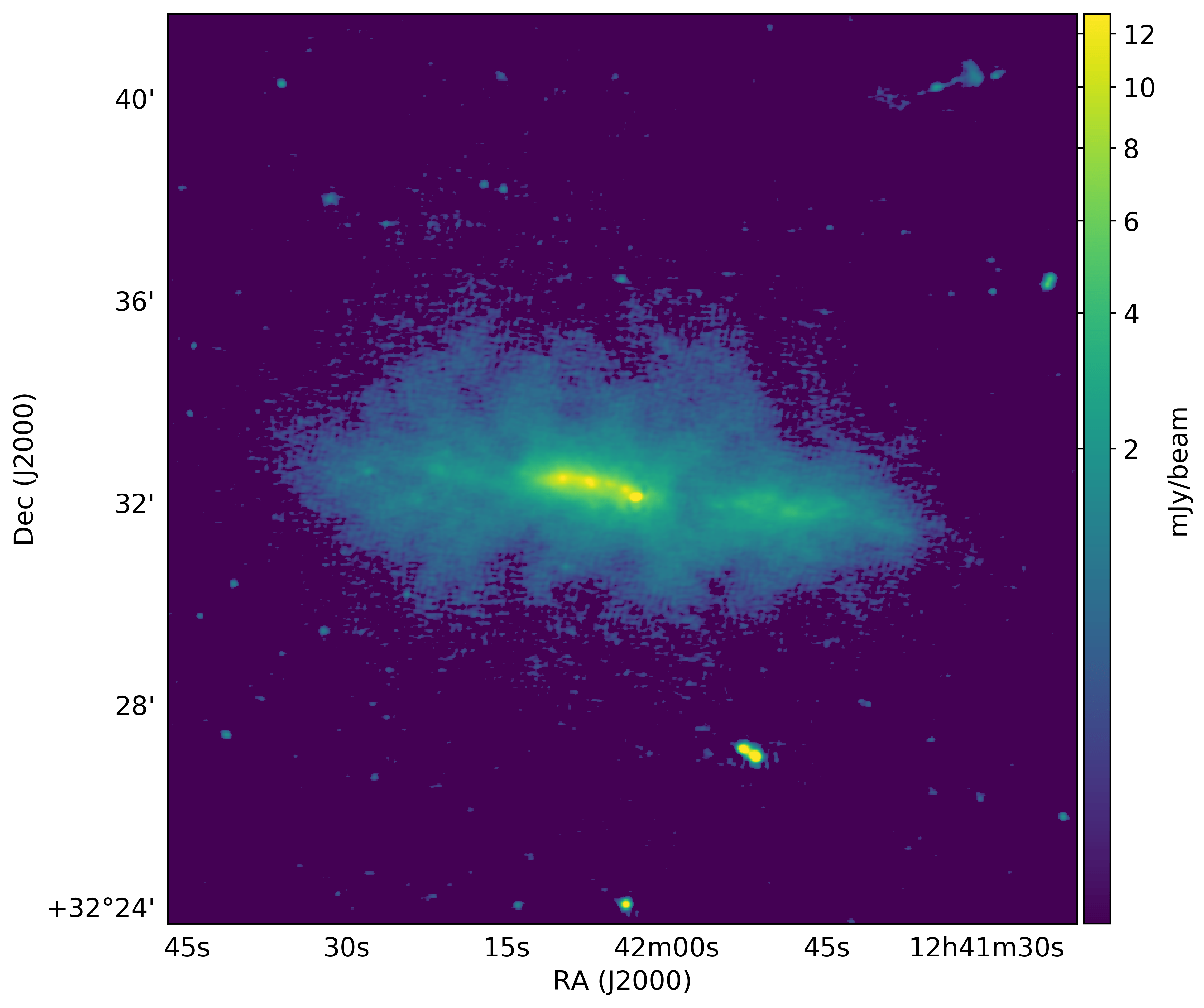}
   \includegraphics[width=0.42\linewidth]{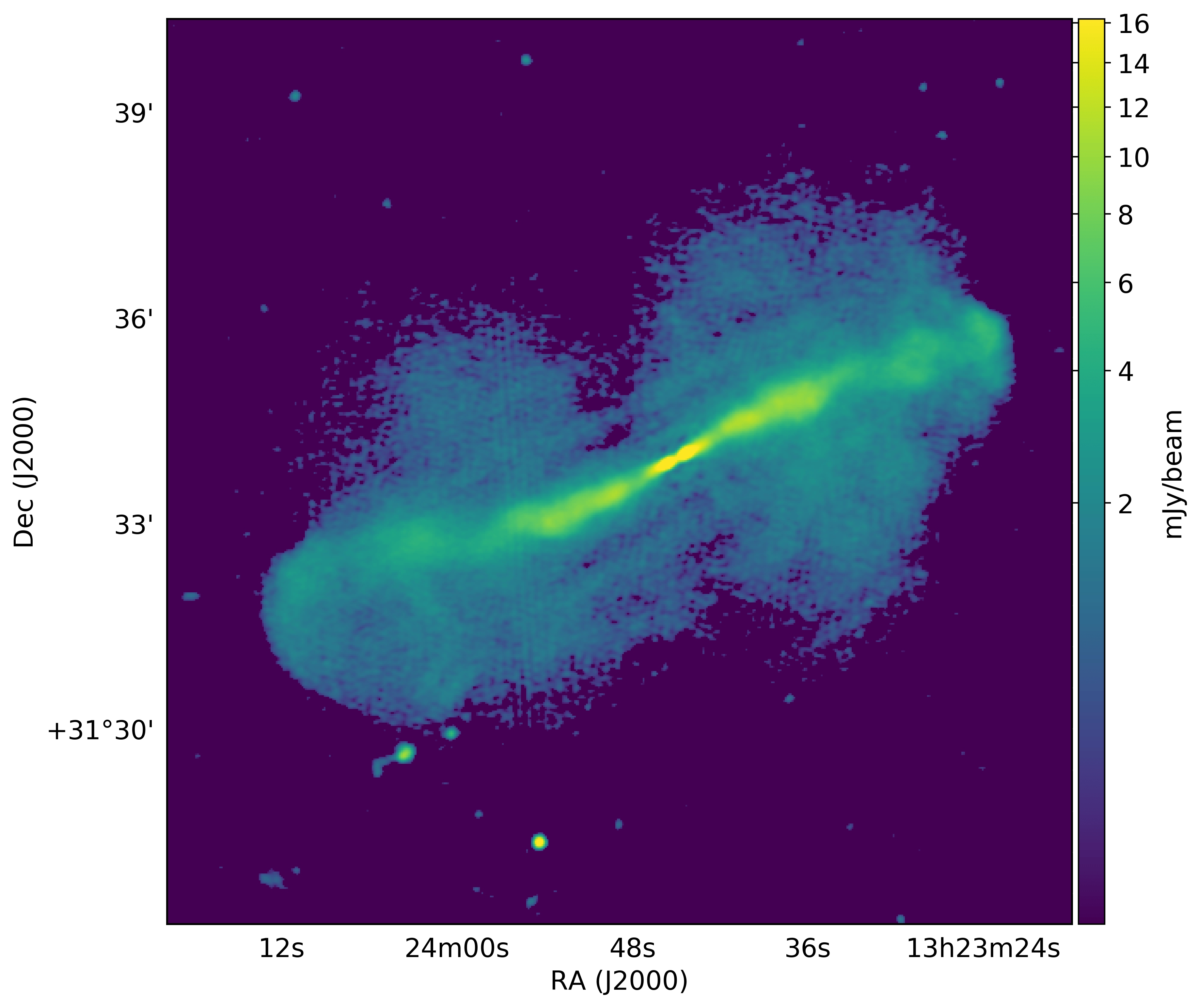}
   \includegraphics[width=0.42\linewidth]{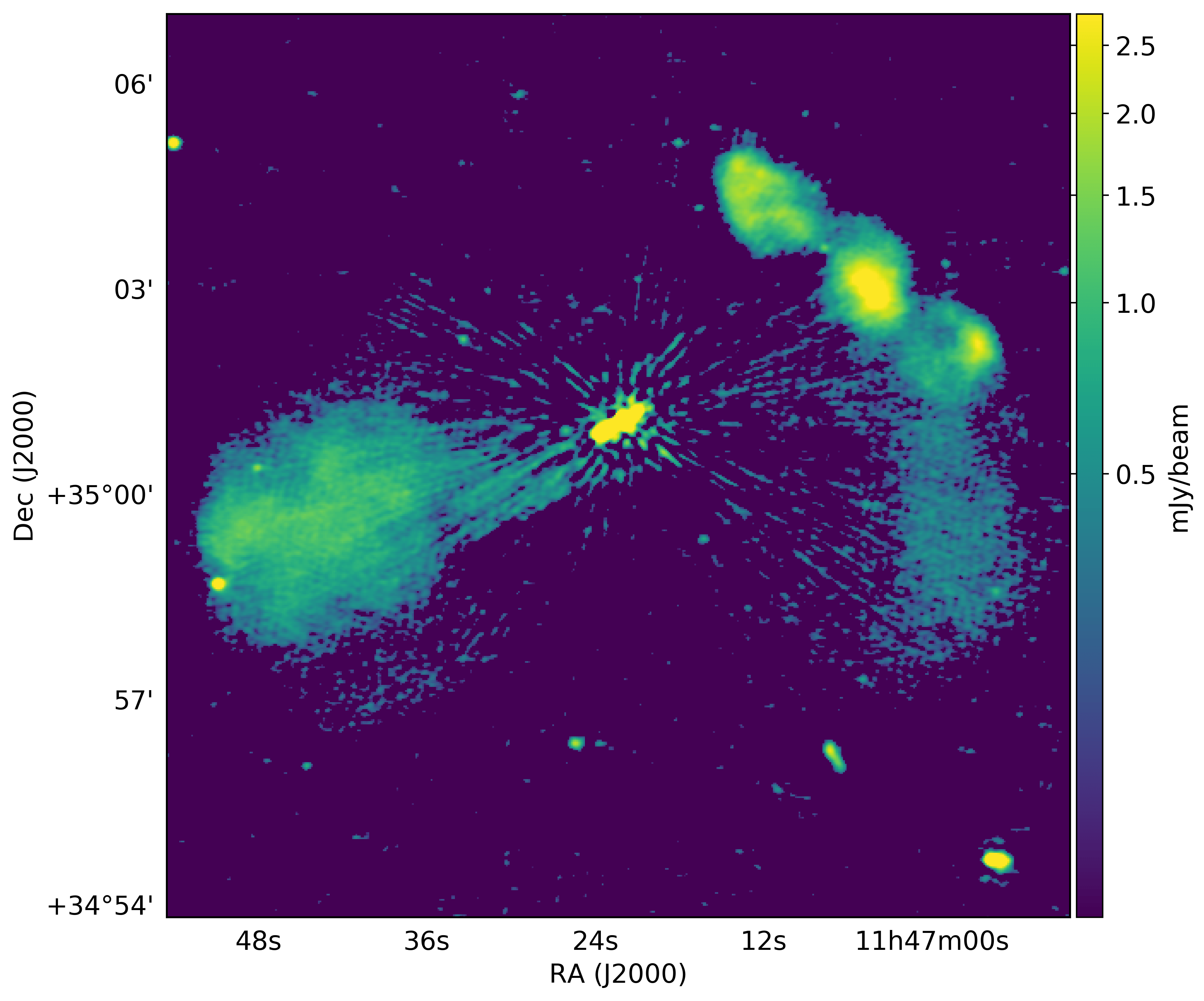}
   \includegraphics[width=0.42\linewidth]{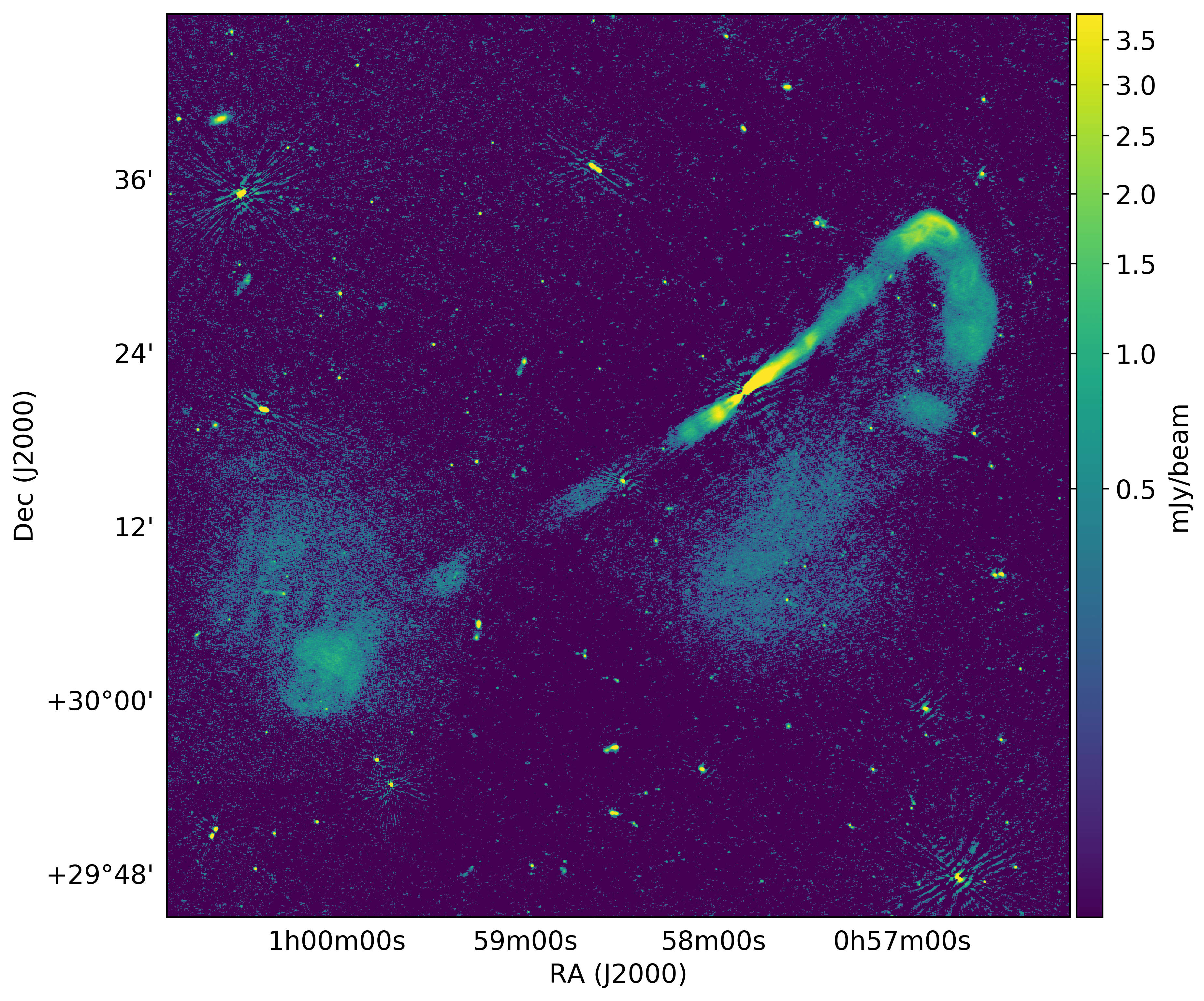}
   \includegraphics[width=0.42\linewidth]{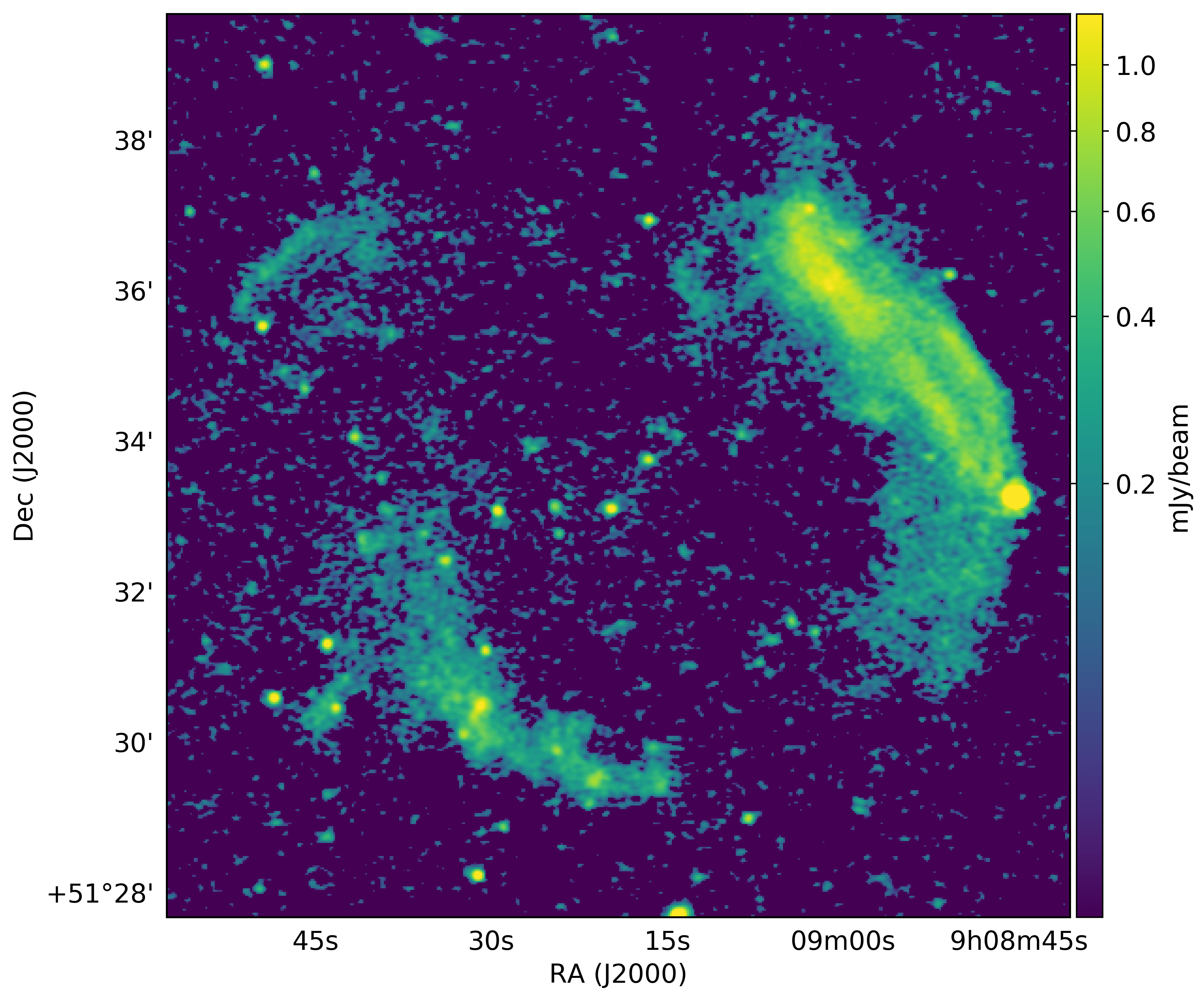}
      \caption{LoTSS-DR2 images of a selection of highly resolved
        sources with the colour scale, contours and image size chosen for display
        purposes. Clockwise from top left, the objects
        depicted are: radio galaxy B2 0924+30, hosted by IC 2476;
        galaxy NGC 4631; BL Lacertae type object B2 1144+35 B, hosted by Z 186-48; cluster of galaxies Abell 746; radio galaxy NGC 315; radio galaxy B2 1321+31,
        hosted by NGC 5127.}
   \label{fig:example-maps}
\end{figure*}

The mosaic images, and catalogues derived from them, have significant overlap so when producing the final full-area catalogue we remove duplicate sources by only keeping those in a given mosaic if they are closest to the centre of that particular mosaic. Our final full-area catalogue consists of 4,396,228 radio sources made up of 5,121,366 Gaussian components. The overall sensitivity distribution is shown in Fig. \ref{fig:mosaic-noisemap} and some example maps from the data release are shown in Fig. \ref{fig:example-maps}.
 
\section{Image quality}
\label{sec:image_quality}

\begin{figure*}[htbp]
   \centering
   \includegraphics[width=0.48\linewidth]{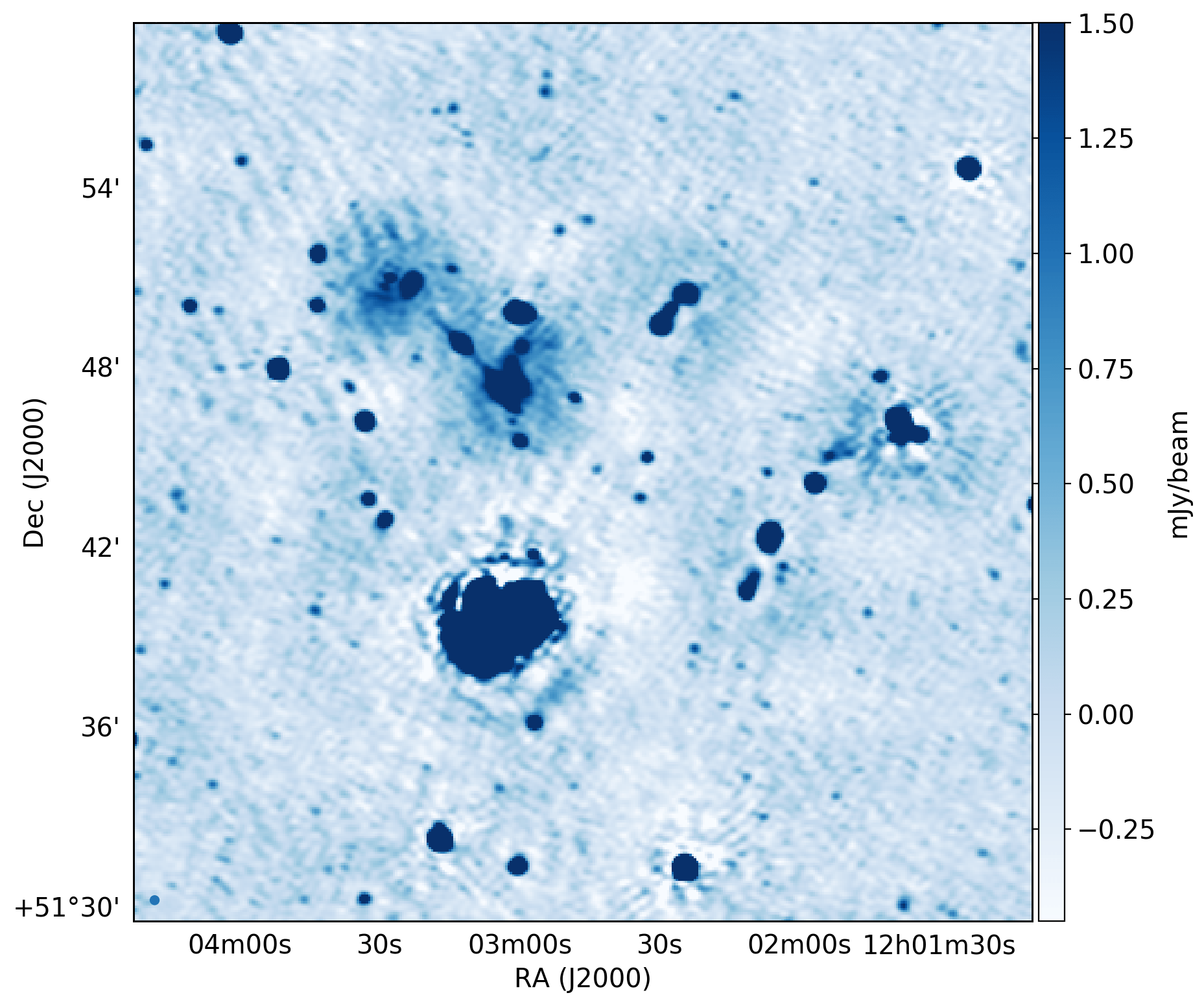} 
    \includegraphics[width=0.48\linewidth]{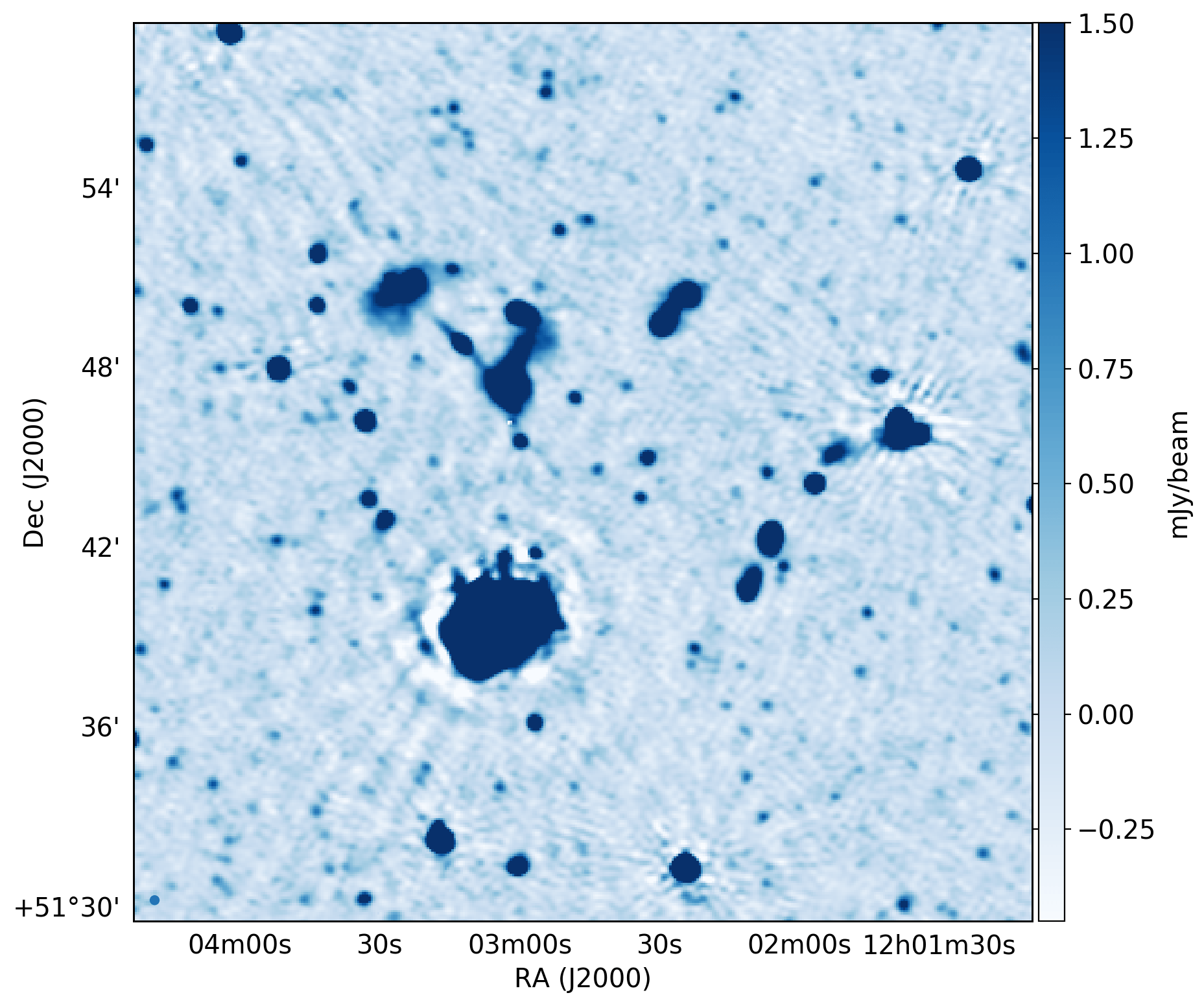} 
   \caption{A demonstration of the improved fidelity of diffuse emission in LoTSS-DR2 compared to LoTSS-DR1. The image on the left is the 20$\arcsec$ resolution LoTSS-DR1 image of a region where it is apparent that low level artificial halos are present in regions surrounding real diffuse emission (most prominent around the radio galaxy centred at 12h03m12s +51$^\circ$48$\arcmin$46$\arcsec$). The image on the right shows the equivalent LoTSS-DR2 image of the region where the artificial halos are no longer present. Another clear difference that is apparent in these panels is the low level structures in the noise in the LoTSS-DR1 images that we have successfully removed from the LoTSS-DR2 images.}
   \label{fig:diffuse_emission}
\end{figure*}

\begin{figure*}[htbp]
   \centering
   \includegraphics[width=0.48\linewidth]{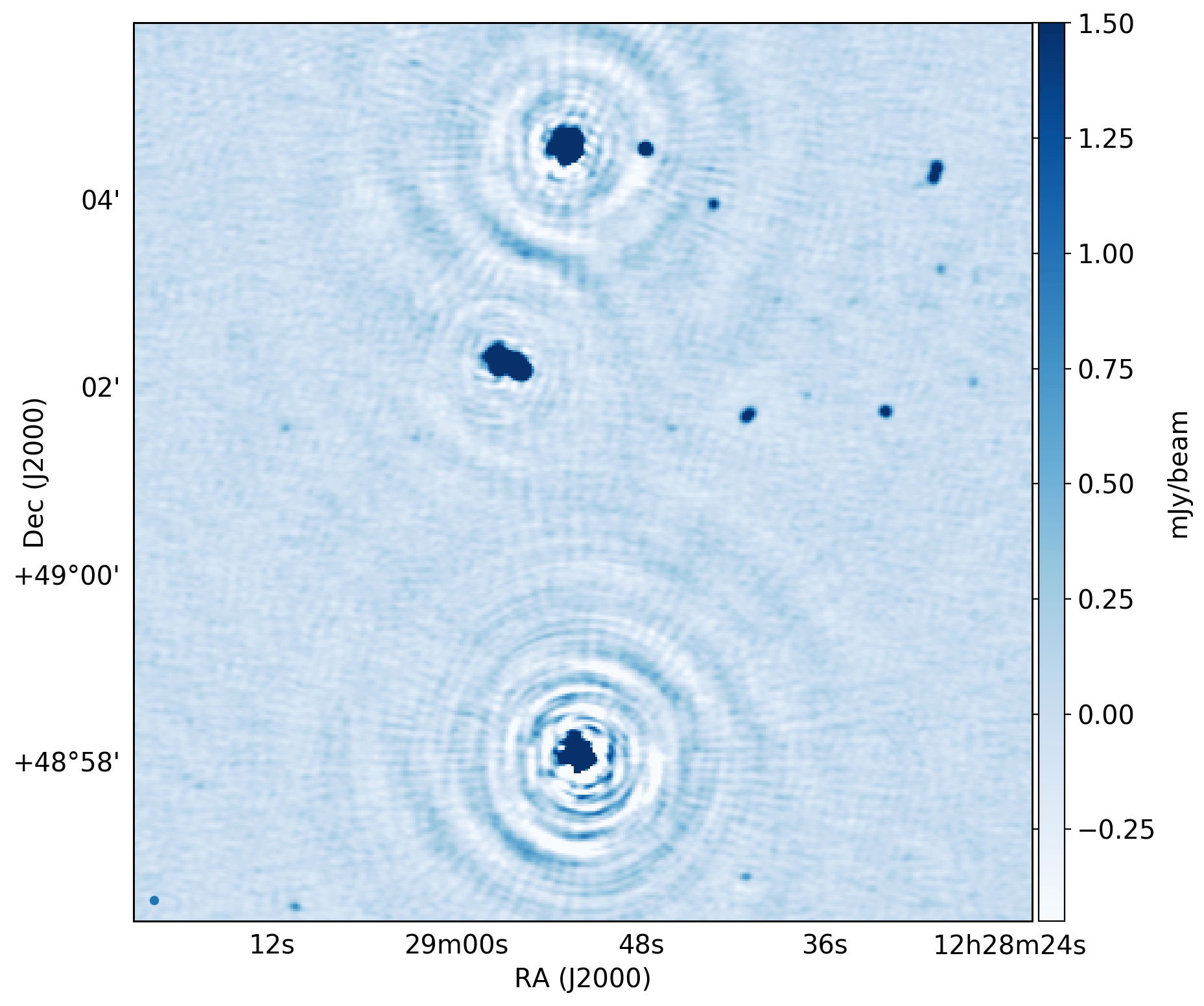} 
    \includegraphics[width=0.48\linewidth]{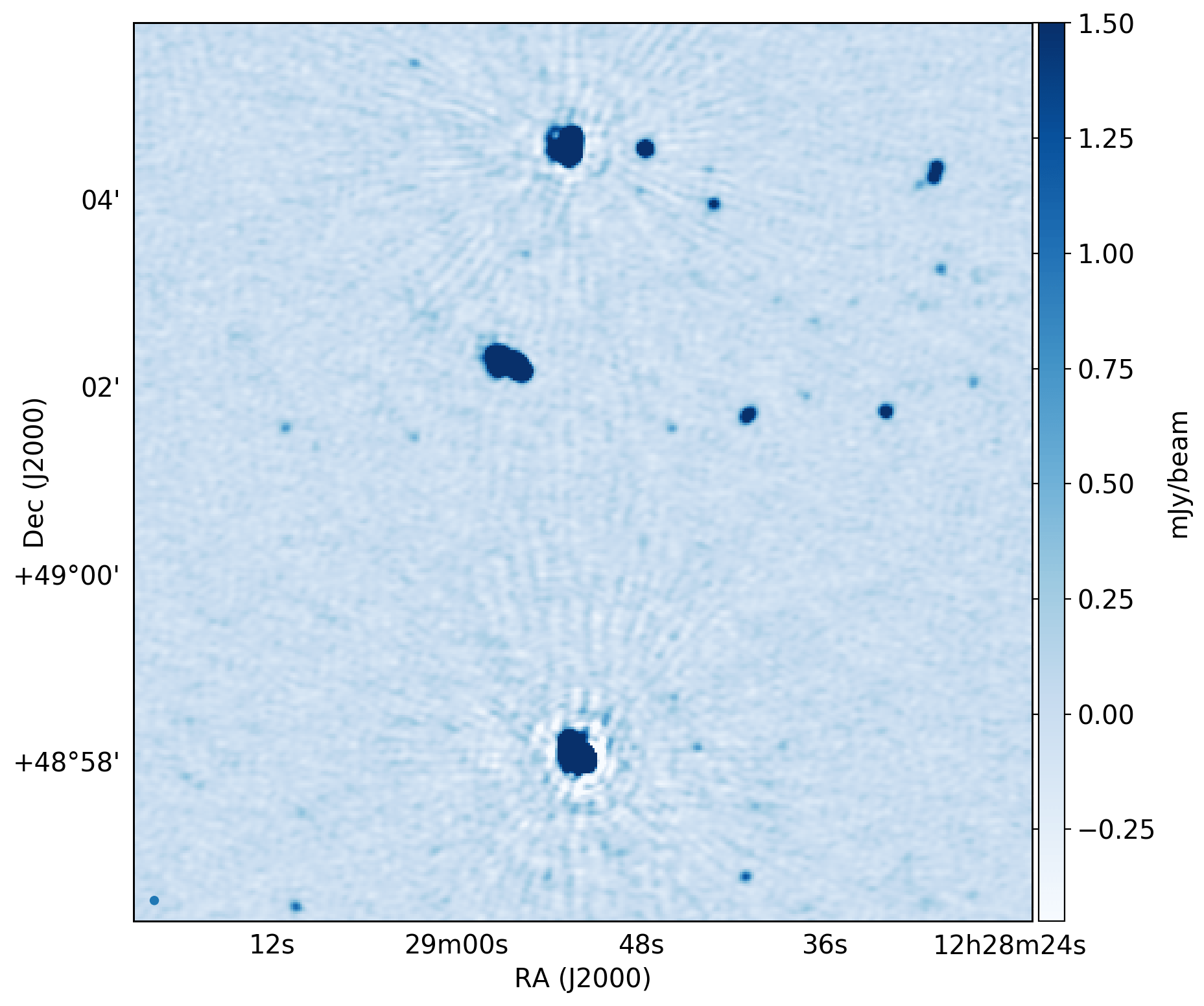} 
   \caption{A demonstration of the improved dynamic range in LoTSS-DR2 compared to LoTSS-DR1 at 6$\arcsec$ resolution. The left and right panels show the LoTSS-DR1 and LoTSS-DR2 images of the same example region respectively.}
   \label{fig:dynamic_range}
\end{figure*}

Several aspects of the image quality have been improved in LoTSS-DR2 compared to LoTSS-DR1. Notably, as described in \cite{Tasse_2021} and demonstrated in Fig. \ref{fig:diffuse_emission} and \ref{fig:dynamic_range}), the recovery of unmodelled emission and the dynamic range have been improved by the revised data processing strategy. In addition we have added a new post processing step to refine the flux density scale. The fully automated processing has allowed us to image a large fraction of the sky that was limited by the area observed to-date. In this section we discuss in detail a number of key aspects of the image quality for the released products and describe the tests we have carried out for this characterisation over the LoTSS-DR2 regions.

\subsection{Source extensions} \label{sec:source_sizes}

Distinguishing point-like from extended sources is important to identify and characterise different source populations. However, as detailed in this subsection, robustly making this separation can be challenging. 

If the data were perfectly calibrated and deconvolved, then point-like sources in the absence of noise would have a ratio of integrated flux density ($S_I$) to peak brightness ($S_P$)  that is unity and a size equal to that of the restoring beam. In a scenario where the errors on the integrated flux density and peak brightness are not correlated and there is no bias in the characteristics of the detected sources then, as described in e.g. \cite{Franzen_2015}, the natural logarithm of the ratio of the integrated flux density ($S_I$) to peak brightness ($S_P$), given by $R=\ln(S_I/S_P)$ will follow a Gaussian distribution centred on zero with a standard deviation equal to:
\begin{equation}
\sigma_R = \sqrt{\left(\frac{\sigma_{S_I}}{S_I}\right)^2 + \left(\frac{\sigma_{S_P}}{S_P}\right)^2}
\end{equation}
However, even in the simple case of detecting simulated point-like sources added to a map of  Gaussian random noise this breaks down because, for example, there is a correlation between $\sigma_{S_I}$ and $\sigma_{S_P}$ which are the statistical errors on the fitted integrated flux density and peak brightness respectively. Additionally, as $S_I$ decreases there are increasing errors on the source sizes and a general over-estimation of these which skews the distribution of $\frac{S_I}{S_P}$. The observed LoTSS-DR2 distribution of source sizes is further complicated, not only by the real distribution of source sizes, but by aspects such as calibration errors, the 1.5$\arcsec$ pixel size, or uncorrected time- and bandwidth-smearing, all of which can artificially blur sources and further impact $\frac{S_I}{S_P}$.

\begin{figure}[htbp]
   \centering
   \includegraphics[width=\linewidth]{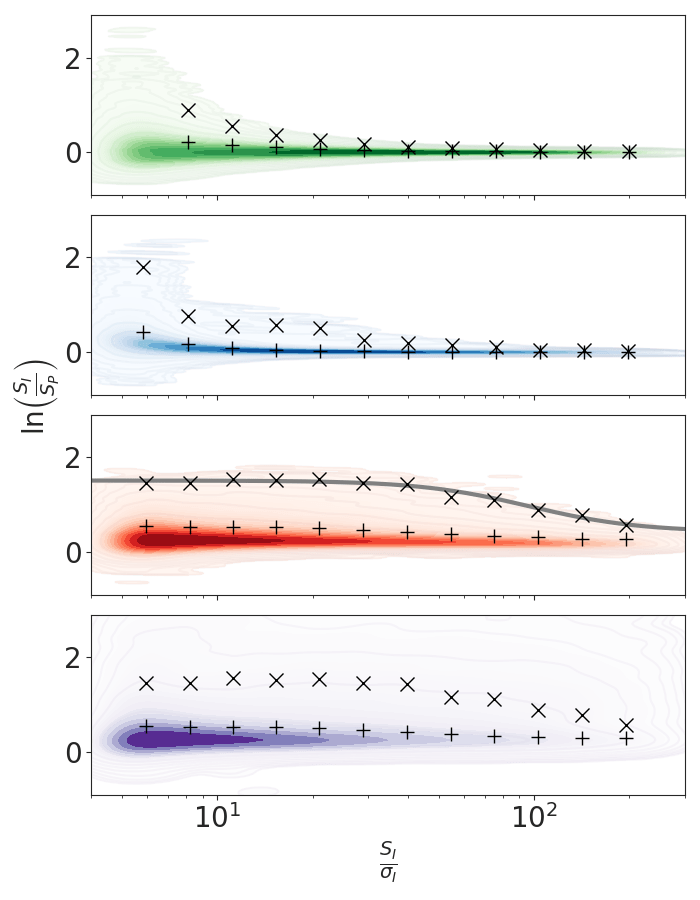}   
   \caption{The ratio of integrated flux density to peak brightness as a function of signal to noise for populations of simulated and real LoTSS-DR2 sources. The green (top) curve shows the idealised distribution where there is no bias in the detections and the noise in the integrated flux density and peak brightness are uncorrelated. The blue (second from top) curve is a population of point-like sources that were injected into images with a Gaussian noise background and catalogued using \textsc{PyBDSF}. The red (second from bottom) and purple (bottom) curves show populations of compact and all LoTSS-DR2 sources respectively. In each of the four distributions the $+$ and $\times$ symbols show the 84.1 and 99.9 percentile levels in different signal to  noise bins. If the data were described by a Gaussian distribution, which is only the case for the green points, these would correspond to the standard deviation and $3\times$ the standard deviation respectively. The sigmoid function we have defined for separating extended and point-like sources is shown with the grey line fitted to the 99.9 percentile level of the population of compact LoTSS-DR2 sources (red points).}
   \label{fig:source-sizes}
\end{figure}

To demonstrate the level at which our real source size distribution deviates from, or mimics, these ideal situations, we have conducted two simple simulations for comparison with our real LoTSS-DR2 catalogues. The first is where point-like sources are drawn from a distribution of flux densities and independent Gaussian random errors are added to the peak brightness and the integrated flux densities separately. The second is where point-like sources are drawn from the same distribution, convolved with a 6$\arcsec$ restoring beam and injected into a map with Gaussian random noise and then catalogued using the same \textsc{PyBDSF} parameters as we have used to create our real source catalogues -- thus characterising the performance of the source identification software to such a population. The distribution of $\frac{S_I}{S_P}$ from these simulations are shown together with real LoTSS-DR2 sources in Fig. \ref{fig:source-sizes}. We note that the flux density distribution of sources in our simulations  mimics the flux density distribution of the real LoTSS-DR2 catalogue by ensuring there are the same number of simulated and real sources in each of the signal-to-noise bins shown in Fig. \ref{fig:source-sizes}.

As is clear from Fig. \ref{fig:source-sizes}, neither of our
simplistic simulations replicates our real LoTSS-DR2 sources.  Hence,
to define a criterion with which extended LoTSS-DR2 sources can be
separated from unresolved sources we make use of the detected source
population itself. First we identify the best candidates for being
genuine point-like sources in the LoTSS-DR2 images which we classify
as being those that are \textsc{PyBDSF} `S'
type sources (i.e. those that are fitted with a single Gaussian), that
are isolated (which, unless otherwise stated, we define throughout as being separated by 45$\arcsec$ or more from a neighbouring source), have a
LoTSS measured major axis size less than 15$\arcsec$, are in the
deconvolution mask for all contributing pointings and have $\frac{S_P}{\sigma_{S_P}} > 5$.
This selection corresponds to just 363,052 sources (8.3\% of the LoTSS-DR2 catalogue) but we use it to identify the much larger population of unresolved sources in our catalogue. 
For these 363,052 sources, the envelope that encompasses the 99.9 percentile of the $\frac{S_I}{S_P}$ distribution is shown in Fig. \ref{fig:source-sizes} and the best fit sigmoid function to this is:
\begin{equation}\label{eq:resolved_sources}
R_{99.9} = 0.42 + \left( \frac{1.08}{1+\left( \frac{S/N}{96.57}\right)^{2.49}} \right)
\end{equation}
where $\textrm{S/N}$ is the signal-to-noise ratio and is defined throughout as $\frac{S_I}{\sigma_I}$. Whilst there is no definitive way of separating unresolved from resolved sources we recommend using this, or comparable, criteria for distinguishing the two populations as the method attempts to account for calibration inaccuracies and signal-to-noise and is more robust than e.g. using just the catalogued source sizes which can be misleading even for unresolved sources (approaches for tackling this issue have been applied in numerous works such as \citealt{Condon_1997}, \citealt{Prandoni_2001} and more recently to LOFAR images in e.g. \citealt{Mahony_2016}, \citealt{Williams_2016} and previous LOFAR surveys data releases). 
In total, by applying this criterion to the LoTSS-DR2 catalogue we find that only 351,153 sources (8.0\% of the full catalogue) have $R$ values outside of this envelope and can be classified as extended at the resolution of LoTSS-DR2, with the remaining 92.0\% unresolved. 
As highlighted though, the separation cannot be done definitively  and a balance must be chosen between the completeness and reliability of the resolved/unresolved classification. Due to the large number of sources that are close to the separation boundary, small differences in the approach can lead to large differences in the outcome (particularly for faint sources) and our conservative 99.9\% percentile separation results in a lower fraction of resolved sources than that found in LoTSS-DR1 (14\%; \citealt{Shimwell_2019}) and LoTSS-deep (between 11.3\% and $\sim$30\% depending on the adopted method - see \citealt{Sabater_2021} and \citealt{Mandal_2021} for details) but a correspondingly higher level of confidence in their genuine extension.

\begin{figure}[htbp]
   \centering
   \includegraphics[width=\linewidth]{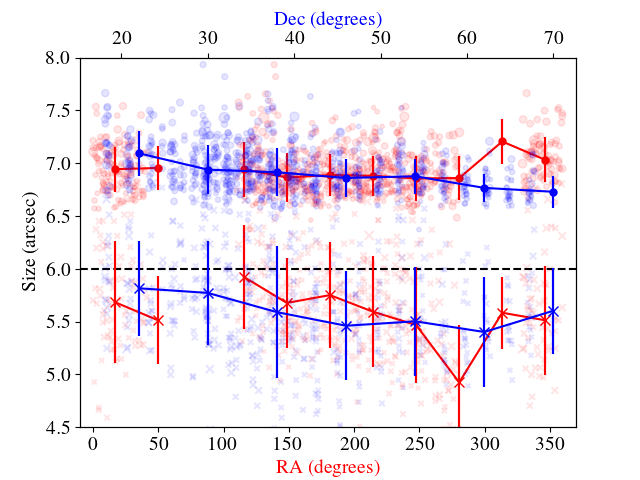}
   \caption{The compact source and synthesized beam size as a function of right ascension (red, lower x-axis) and declination (blue, upper x-axis). The synthesized beam (transparent `$\times$' symbols) size for each LoTSS-DR2 pointing is the average of the major and minor FWHM of a fitted Gaussian. The compact source size (transparent circles) is the median of the average of the major and minor FWHM of the compact sources (according to Eq. \ref{eq:resolved_sources}) with S/N greater than 20 in each mosaic. The marker sizes are proportional to the image rms (accounting for the flagging fraction and integration time) which is a proxy of data quality. The solid lines and larger markers show the median of both the synthesized beam (solid `$\times$' symbols) and compact source (solid circles) sizes as a function of right ascension (red) and declination (blue) and the errors show the corresponding standard deviations. The dashed black line shows the size of the restoring beam which is kept constant over the entirety of LoTSS-DR2 and is generally larger than the synthesized beam.}
   \label{fig:source-sizes-ra-dec}
\end{figure}

We do note that the criterion we have derived is an average over the entire surveyed region and does not reflect field-to-field variations. In Fig. \ref{fig:source-sizes-ra-dec} we show the synthesized beam and the median measured source size for compact (Eq. \ref{eq:resolved_sources}) sources with an S/N exceeding 20 for each of our mosaics as a function of right ascension and declination. Where for consistency, even though the real synthesized beam profile is highly non Gaussian (see Sec. \ref{sec:emission_recovery}), we measured its extent by fitting a two dimensional Gaussian profile using \textsc{PyBDSF}. Similarly to Fig. \ref{fig:source-sizes}, we clearly see  artificial blurring of sources. We also see field-to-field variation that is generally larger or comparable to the underlying trends with right ascension and declination. This blurring effectively decreases our image resolution and systematically makes compact, high S/N sources $\sim 1.0 \arcsec$ larger than our restoring beam which was already  chosen to be larger than the fitted synthesized beam size for the majority (75\%) of fields.  

\subsection{Astrometric precision} \label{sec:astrometry}

As described by \cite{Shimwell_2019} and \cite{Tasse_2021}, during our data processing we perform a facet based astrometric correction to statistically align each facet within our LoTSS-DR2 images with the Pan-STARRS optical catalogue (\citealt{Flewelling_2020}) which is itself thought to be accurate to within 0.05$\arcsec$ (\citealt{Magnier_2020}). Furthermore, during the mosaicing of our individual images we exclude facets that have an uncertainty in the estimated astrometric correction exceeding 0.5$\arcsec$. For LoTSS-DR1, through a comparison with Pan-STARRS, we found an astrometric accuracy of our radio catalogue of approximately 0.2$\arcsec$ for compact sources brighter than 20\,mJy.

To assess the accuracy of the astrometrically corrected LoTSS-DR2 catalogue we make use of a preliminary enriched version of the catalogue where the radio sources have been cross-matched with the DESI Legacy Imaging Surveys (\citealt{Dey_2019}) and ALLWISE data release (\citealt{Cutri_2014}) using a likelihood ratio cross matching technique in addition to a manual visual identification of the optical counterparts through the public LOFAR galaxy zoo project (see Sect. \ref{sec:value_added_cats}). This procedure is similar to that described in \cite{Williams_2019} and as the manual identifications are not yet complete it will be fully detailed in an upcoming publication (see Sect. \ref{sec:value_added_cats}). To minimise the complexities of resolved sources we use the criterion presented in Eq. \ref{eq:resolved_sources} to filter the enriched catalogue to contain only compact LoTSS sources. We also removed sources that are not fully deconvolved in all pointings that make up the mosaic in that specific region and those where we have not identified a counterpart in the DESI Legacy Imaging Surveys through the likelihood ratio. We note that by excluding sources without a DESI counterpart we may inadvertently underestimate our true astrometric errors but given we have already excluded regions with astrometric errors exceeding 0.5$\arcsec$ from the mosaics we believe this impact to be minimal. The filtering results in a catalogue with 375,648 entries that are distributed over the region where LoTSS-DR2 and DESI overlap (93\% of LoTSS-DR2). The DESI Legacy Survey has a high astrometric accuracy: when the recorded  positions of catalogued bright stars are compared to the Gaia DR1 catalogue (\citealt{Gaia_2016}) there is a root mean square (RMS) scatter of approximately 0.02$\arcsec$ (\citealt{Dey_2019}). We can therefore safely assume that any offsets between the DESI Legacy Surveys position and the LoTSS-DR2 positions are dominated by the uncertainty in our radio images.

\begin{figure}[htbp]
   \centering
   \includegraphics[width=0.95\linewidth]{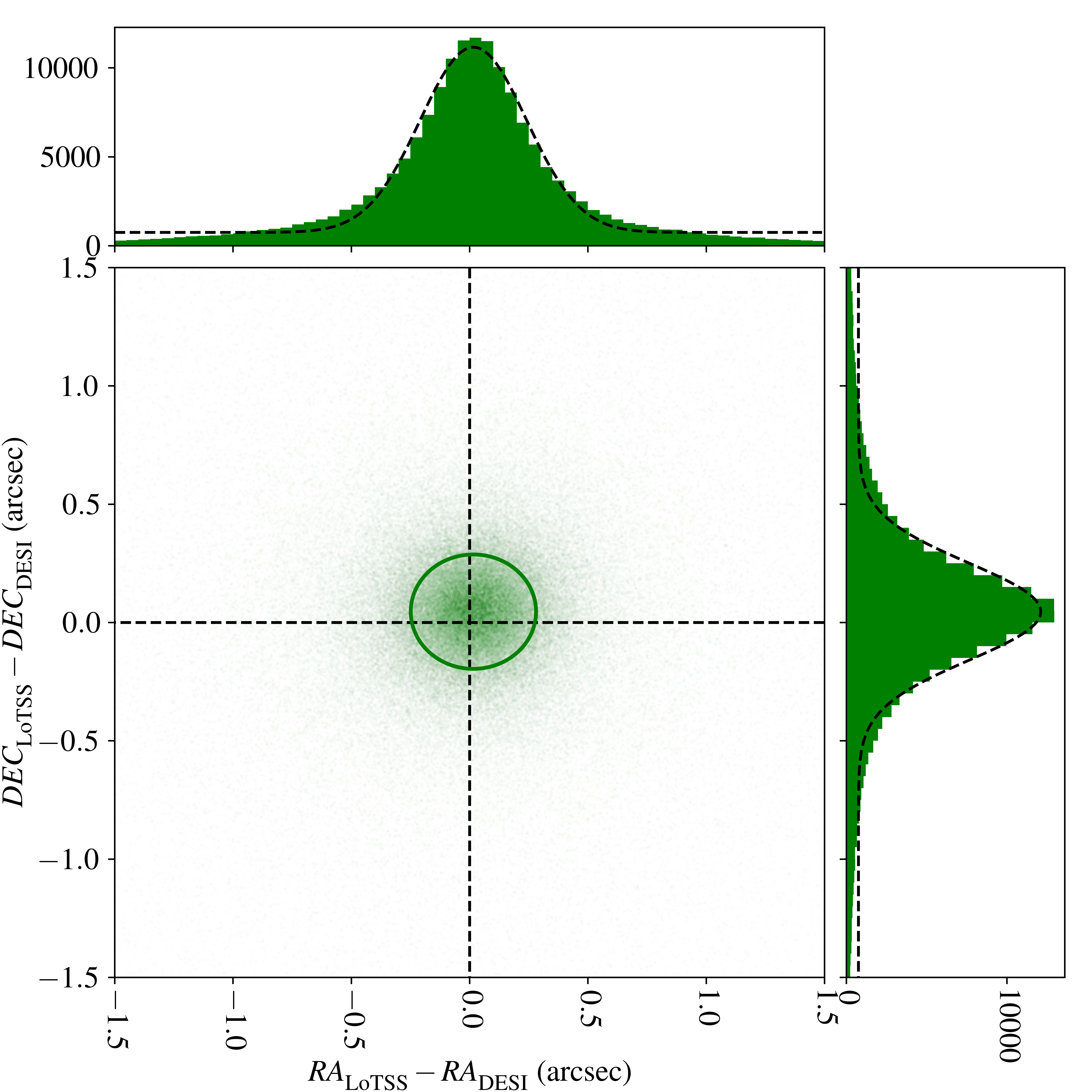}   
   \includegraphics[width=0.95\linewidth]{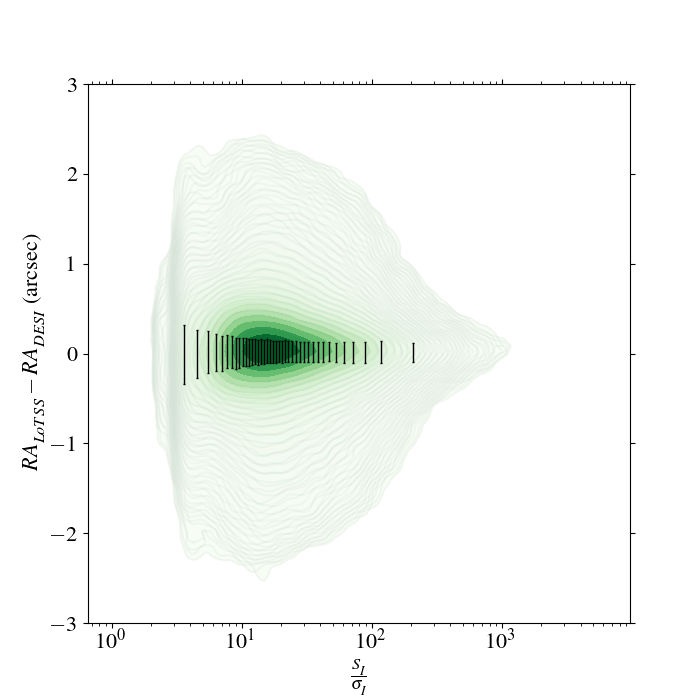} 
   \caption{The astrometric accuracy of the LoTSS-DR2 radio wavelength positions. The top plot shows the offsets in RA and DEC for compact, deconvolved LoTSS sources from their DESI Legacy Imaging Surveys counterpart. The histograms show the distribution of offsets as a function of RA and DEC and the fitted Gaussian (plus a constant) functions have $\sigma_{RA}=0.22\arcsec$ and $\sigma_{DEC}=0.20 \arcsec$ and are centred at offsets of 0.02$\arcsec$ in RA and 0.05$\arcsec$ in DEC. The constant in the fitted function is included to allow for possible mismatches in the association of LoTSS and DESI sources. The bottom plot shows the distribution of RA offsets as a function of signal to noise with the total length of the error bars corresponding to the standard deviation of the Gaussian (plus a constant) function fitted to the sources in that particular signal to noise bin. The signal to noise bins are selected such that they contain an equal number of sources.}
   \label{fig:source-positions}
\end{figure}

In Fig. \ref{fig:source-positions} we show the LoTSS-DR2 offsets in RA and DEC from their DESI Legacy Survey counterparts. Our astrometric precision is $\sigma_{RA}=0.22\arcsec$ and $\sigma_{DEC}=0.20\arcsec$, which we derived from Gaussians fitted to histograms of the offsets of cross-matched, compact sources with S/Ns in excess of 20. The fitted Gaussians are centred on 0.02$\arcsec$ in RA and 0.05$\arcsec$ in DEC demonstrating the low level of systematic positional offset between LoTSS and DESI. The astrometric precision found for LoTSS-DR2 is thus 
comparable to that obtained for LoTSS-DR1 (where both $\sigma_{RA}$ and $\sigma_{DEC}$ were $\sim0.2\arcsec$)

We also explored the positional uncertainties as a function of S/N and found that, as expected, they increase gradually below approximately an S/N of 20 and reach a maximum of 0.5$\arcsec$ at a S/N of 5 - we characterise this behaviour as $\sigma_{RA} = \frac{1.58}{S/N} + 0.17$ and  $\sigma_{DEC} = \frac{1.19}{S/N} + 0.16$. Additionally, we searched for declination dependencies that could not really be explored in LoTSS-DR1 due to its much smaller range of declination. We found that for sources with S/Ns exceeding 20 the astrometric errors at lower declination are marginally larger than at higher declination, with $\sigma_{RA}=0.25\arcsec$ and $\sigma_{DEC}=0.24\arcsec$ at DEC 30$^\circ$ decreasing to $\sigma_{RA}=0.20\arcsec$ and $\sigma_{DEC}=0.19\arcsec$ at DEC 65$^\circ$. Overall, whilst the LoTSS astrometric precision depends on several factors, it is sufficient for accurate cross matching with other surveys and for follow-up spectroscopic programmes and thus meets our aims.

\subsection{Flux density scale} \label{sec:flux_scale} 

As a consequence of uncertainties in the current LOFAR beam model, deriving calibration solutions from observations of calibrator sources (e.g. those in \citealt{Scaife_2012}), using accurate source models,  and then transferring these to the target field, does not 
necessarily result in an accurate flux density scale for the target field. The recovered flux densities of sources are further impacted by ionospheric effects that, as described in Sect. \ref{sec:source_sizes}, can distort sources by an amount that varies depending upon the observing conditions; imperfections in the ionospheric calibration can also scatter flux across the image. To overcome these flux density scale issues, in LoTSS-DR1 a bootstrapping approach, adapted from \cite{Hardcastle_2016}, was used to align the integrated flux densities with VLSSr and WENSS.
A comparison with the 150\,MHz TGSS-ADR1 catalogue 
was used to place a conservative error of 20\% on the LoTSS-DR1 integrated flux density scale. As detailed below, in this data release we have improved the accuracy of our recorded flux densities by adding a post-processing step that utilises the 6th Cambridge survey of radio sources (6C; \citealt{Hales_1988} and \citealt{Hales_1990}) and NVSS
to align the final images of each individual LoTSS pointing with the
flux density scale described in \cite{Roger_1973}, which is commonly used at
low radio frequencies (see e.g. \citealt{Scaife_2012}) and is consistent with \cite{Perley_2017} to within 5\%. This procedure
is the same as the one briefly outlined by \cite{Hardcastle_2021} but
is described in more detail below together with an assessment of the variations in the flux density scale over the LoTSS-DR2 region and an estimation of the overall accuracy. 

We note that our characterisation in this subsection is performed on simple compact sources but is applicable to all catalogued sources. However, for significantly extended sources there are additional considerations such as possible flux density suppression during calibration and the limitations of accurately automatically cataloguing of complex structures; these aspects are outlined in Sect. \ref{sec:emission_recovery} and \ref{sec:value_added_cats} respectively.

\subsubsection{Flux density scale alignment procedure}
\label{sec:flux_density_alignment}

The 151\,MHz 6C survey was used to refine the LoTSS-DR2 flux density scale as it was carefully calibrated to be consistent with the
\cite{Roger_1973} flux density scale to within 5\% (\citealt{Hales_1988}), and it
mapped the majority (approximately 8,600 square degrees) of the sky
north of declination +30$^\circ$ with a source density of around 4
sources per square degree. Unfortunately we are unable to directly
align the entirety of LoTSS-DR2 with 6C, not only because the low
source density of 6C makes this challenging but also, more
fundamentally, because the 6C catalogues do not cover the entirety of
the LoTSS-DR2 area. To overcome this, whilst still making use of the 6C flux density accuracy, we also made use of the NVSS 1.4GHz survey  which encompasses 82\% of the celestial sky including the entirety of the LoTSS-DR2 region, has a source density of approximately 52 sources per square degree and also has a flux density accuracy of approximately 5\% (see e.g. \citealt{White_1997}). 
Under the assumption that 6C, NVSS and the underlying radio source populations have no  systematic effects that impact the ratio between NVSS and 6C source flux densities as a function of their location on the sky we can use a global ratio of NVSS to 6C source flux densities to refine the LoTSS-DR2 flux density scale. 

To do this we first catalogued each individual LoTSS pointing. For
each of these catalogues we performed a simple nearest neighbour
($<1\arcmin$) cross-match with a combined 6C catalogue after filtering
out LoTSS sources that are both highly resolved (defined here as those with measured major axis
size greater than 20$\arcsec$) or faint ($S_I < 0.15$ Jy)
which would not have 6C counterparts. We then discarded all fields where the number of sources
cross-matched was less than 20, which left 529 fields, and we
calculated the median of the ratios of the integrated flux densities for the
matched sources, $F_{6C}$. The median value of the 529 $F_{6C}$ values
is 1.023 and the standard deviation is 0.146. Similarly, we performed
a nearest neighbour cross-match (10$\arcsec$) between the catalogues
of the individual fields with NVSS after again filtering out faint
($S_I<30$ mJy), highly resolved LoTSS sources (defined here as those
with measured major axis sizes greater than 10$\arcsec$) and those
that are not isolated (where the filtering conditions were altered with respect to those applied in the 6C cross-matching to reflect the higher resolution and sensitivity of NVSS).
For the same 529 fields we calculated the median of the ratio of the LoTSS to NVSS 
integrated flux densities of the matched sources in each field, $F_{NVSS}$. The
median of $F_{NVSS}$ is 5.939 and the standard deviation is 0.673. The
derived $F_{6C}$ and $F_{NVSS}$ values are highly correlated and the
median of $F_{NVSS}/F_{6C}$ is 5.724. If we rescale the LoTSS
pointings by $F_{NVSS}/5.724$ and repeat the cross match with 6C then
by definition the median of the 529 newly derived medians of the LoTSS
to 6C integrated flux density ratios is 1.0 but the standard deviation of these
values has decreased significantly from 0.146 before any scaling to
0.074 after scaling. Thus this rescaling makes the flux density scale more
consistent with 6C throughout the LoTSS-DR2 region. Finally, we
account for the 6C survey being at 151\,MHz and LoTSS-DR2 being at
144\,MHz. From $F_{NVSS}/F_{6C} = 5.724$ we know that the median
spectral index of sources in both 6C and NVSS corresponds to $-0.783$ and
using this spectral index we align our LoTSS maps with the 6C flux density
scale by scaling our maps according to the frequencies of LoTSS and 6C such that
$F_{NVSS}=5.724\times(144/151)^{-0.783} =
5.936$. Given that the quoted 6C flux density scale uncertainty is 5\%, the
6C-LoTSS standard deviation of 7.4\% means that a lower limit on the
per-field LoTSS flux density scale uncertainty is around 6\%.

\subsubsection{Positional variations in the flux density scale}
\label{sec:pos_flux_vary}

To more precisely ascertain how well the flux density scale alignment procedure has worked, we examine the variation in flux densities amongst the 841 different LoTSS-DR2 pointings. Using source catalogues of each individual field we perform a simple nearest neighbour cross match (sources within $5\arcsec$) between catalogues of neighbouring pointings (those within 4$^\circ$) which are each filtered to remove sources that are resolved  (according to the criterion presented in Eq. \ref{eq:resolved_sources}), are not in the deconvolution mask, have a nearest neighbour within $45\arcsec$ or have an astrometric uncertainty exceeding $0.5\arcsec$. When we apply an additional filter requiring sources to be within the 60\% level of both pointings primary beams we find a median of 181 matched sources over 2620 overlapping pairs of pointings; if we instead use a 30\% cutoff in the primary beam (which is used for the mosaicing) we find a median of 700 sources in the same overlapping pairs. 

For each pointing pair we quantify the fractional offset in the flux density scale by calculating a linear relationship with zero intercept between the integrated flux densities, which are more robust against ionospheric disturbances and calibration errors than the peak brightness values (see Sect. \ref{sec:source_sizes}), derived from the two separate pointings. As outliers can still exist in these cross matched catalogues we use several different fitting methods (\citealt{Huber_1981}, \citealt{Sen_1968}) that are available within the scikit-learn package\footnote{https://scikit-learn.org/stable/} as well as simple linear regression. These different methods have different outlier mitigation criteria and we select the fit that provides the lowest mean absolute error in the fit residuals.

\begin{figure}[htbp]
   \centering
   \includegraphics[width=\linewidth]{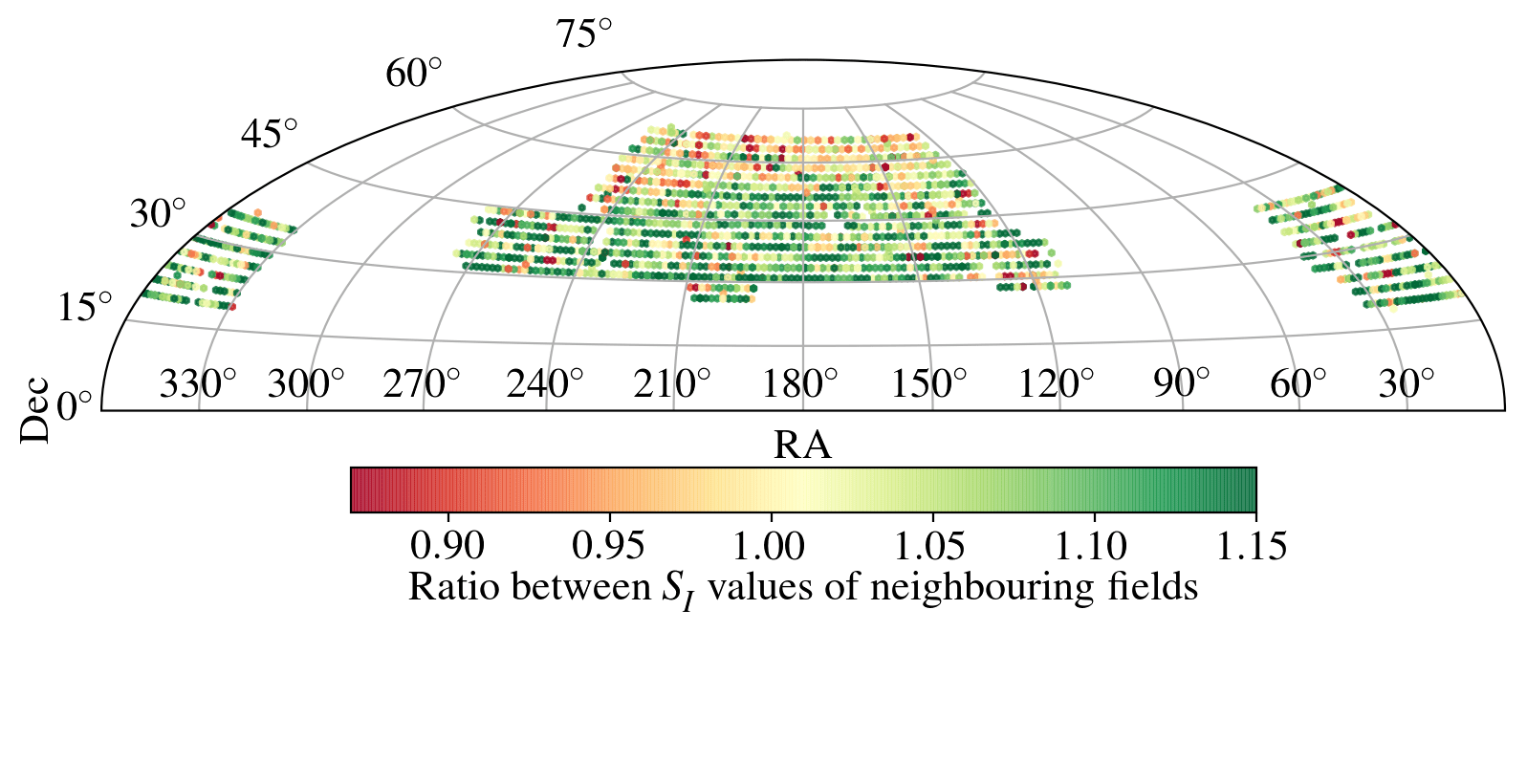} 
   \includegraphics[width=\linewidth]{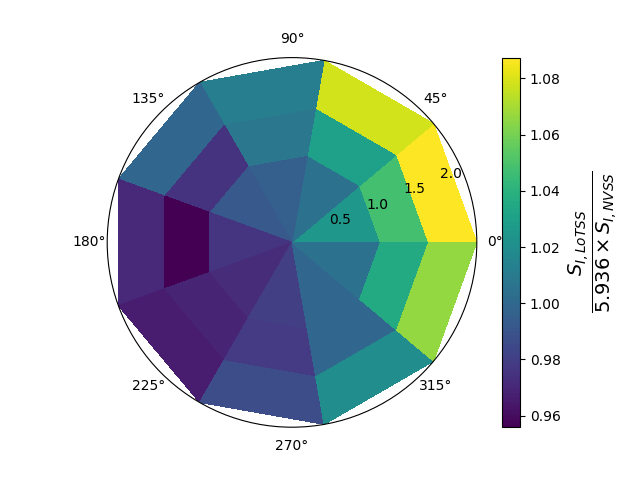}
   \caption{The top panel shows the ratios between $S_I$
     measurements of isolated, compact, deconvolved sources, within
     30\% of the primary beam power level in
     overlapping LoTSS-DR2 pointings that are predominantly separated in declination. These are projected onto the sky with the position being the mean position
     of the overlapping pointings and the colour scale indicates the
     best fit ratio between the $S_I$ values. The bottom panel shows the median LoTSS to NVSS $S_I$ ratios (divided by 5.936 to centre on unity) for different segments of individual fields averaged together for all LoTSS-DR2 fields with a declination less than 50$^\circ$. Here 0$^\circ$ is North, the angle increases from North through West and the segments extend up to 2$^\circ$ from the pointing  centres.}
   \label{fig:overlapping_flux_ratios}
\end{figure}

Applying the same scaling factors that were derived as described in Sect. \ref{sec:flux_density_alignment} to each individual pointing decreases the standard
deviation of a Gaussian fitted to the distribution of the integrated flux density ratios
derived from overlapping pointings from 0.13 to 0.10 (when
applying either the 30\% or 60\% primary beam cuts). The width of this fitted Gaussian can be further decreased if we only consider the overlapping regions with large numbers of sources, which is a proxy of good data quality. For example, the standard deviation drops to 0.07 when considering only the 20\% of the overlapping regions with the largest numbers of cross-matched sources. However, even after applying scaling factors, the best fit ratios between integrated flux density values derived from
neighbouring pointings vary as a function of position on the sky and on the separation in right ascension and declination between the two pointings. In the upper panel of Fig. \ref{fig:overlapping_flux_ratios} we show these ratios only for neighbouring pointings that are offset predominantly in declination, and we see that in a given region the flux densities of sources derived from the pointing centred at lower declination are generally higher than those derived for the same sources but from a pointing centred at higher declination. A similar plot for overlapping regions for pointings that are offset in right ascension does not reveal such clear trends. 

To examine this further we take the regions where the effect is most prominent (south of declination 50$^\circ$) and group sources within each pointing into segments and we examine the median ratio of bright ($S_I > 30$ mJy) LoTSS to NVSS integrated flux densities ($F_{NVSS}$) in each of the segments combined over LoTSS-DR2 pointings. The results of this analysis are shown in the lower panel of Fig. \ref{fig:overlapping_flux_ratios} and provide further evidence that the LoTSS-DR2 flux density scale varies across a given pointing, with the northern region generally having excess ($\approx$12.5\%) flux density compared to the southern region. To better understand the cause of these flux density scale drifts we are examining the application of the LOFAR element beam (that of a single dipole) and array factor (which accounts for the effect of combining dipoles into stations and of the analog tile beam former) in our data processing and the beam models themselves are also undergoing a refinement.  

The effect of mosaicing significantly dilutes the flux density scale variations over a given pointing because pointings with high noise are downweighted, as are pixels further from pointing centres and the overestimation of flux density in the northern regions of a pointing is somewhat reduced by the underestimation in the southern regions of a neighbouring pointing. If $F_{NVSS}$ is calculated over different segments for the mosaiced maps instead of the individual pointing images, the maximum variations between the segments are measured to be 3\% compared to the 13\% that is measured from the individual pointing catalogues. Despite mosaicing significantly mitigating the severity of the variations we still conservatively place a 10\% flux density scale accuracy on LoTSS-DR2 catalogues which, as described above, is the standard deviation of a Gaussian fitted to the distribution of the integrated flux density ratios
derived from overlapping pointings. We adopt this uncertainty as it reflects the observed behaviour in the individual pointings which we consider a worst case scenario. We note that this just accounts for
variations across the surveyed area and the impact of, for example, the inaccuracy in the modelling or application of the LOFAR beam at large distances from the pointing centre, but it does not account for any systematic bias of the entire flux density scale, which, however, should be tied to that of 6C as described above. 

To assess the scope/limitations for further improving the alignment of
the flux density scale of our individual pointings we make use of the deep
field dataset presented in \cite{Sabater_2021} where 22 epochs
totaling over 160\,hrs of data were synthesized together using the
same data processing pipeline as used for LoTSS-DR2 to form a single
deep image of the European Large-Area ISO Survey-North 1 (ELAIS-N1)
region. As part of that work, maps were also made of the individual
epochs which were all independently calibrated from the same sky model and with the
same calibration parameters (i.e. facet layout, calibration cadence and other options); thus, in an ideal scenario, after
appropriate scaling, we would hope that sources in each individual
epoch have equivalent flux densities. In reality there will be some
scatter in the recovered flux densities due to imperfections in the
calibration and varying ionospheric conditions, and even though the pointings are all centred in the same location and observed at comparable times, there will be small differences in the LOFAR beams. 

A very careful
analysis of the flux density scale and variations between the different epochs
was conducted by \cite{Sabater_2021} who explored the dependence on
S/N, distance from pointing centre, source size and ionospheric
conditions. Here we perform a complementary analysis where we use the
22 epochs to test the alignment scheme we have used for LoTSS-DR2. We
thus follow the same procedure that we described above to explore flux density
scale difference in overlapping LoTSS pointings. We scale each
ELAIS-N1 map according to our flux-alignment method, cross match the
sources between each of the maps, filter the cross matched catalogue
and fit for the ratio of the integrated flux densities of each epoch against
each other epoch (231 cross matched catalogues). The best-fitting Gaussian to the histogram
of ratios is very narrow. When we consider just sources between 0.6
and 0.7 of the primary beam power level (so similar to those sources
we examine in the LoTSS-DR2 analysis), the standard deviation of the
fitted ratios is just 0.02 after scaling compared to 0.045 before. Hence, with different ionospheric conditions but a similar LOFAR beam configuration, and the same facet layout and sky model for calibration, we are able to get a much tighter distribution of flux densities between different epochs than we see for overlapping LoTSS-DR2 pointings.
This 
highlights that there is still
significant room for more precision in the LoTSS flux density scale via improved calibration and beam correction.

\subsubsection{Overall flux density scale accuracy}

Finally, we attempt to assess the overall accuracy of the LoTSS-DR2
flux density scale rather than variations throughout the surveyed region. To
do this we first filter the LoTSS-DR2 catalogue so that it contains
just isolated 
sources. We then cross match these sources with NVSS (10$\arcsec$ matching
radius), WENSS (6$\arcsec$ matching radius), 6C (1$\arcmin$ matching
radius) and VLSSr (10$\arcsec$ matching radius); the WENSS catalogue
has had the flux densities scaled by a factor of 0.9 (see
\citealt{Scaife_2012}) to align it with the flux density scale we use here. We
then filter the cross matched catalogues based on the
\cite{Sabater_2021} flux density product thresholds which attempt to account for the sensitivity limitations of the various surveys (see Tab. 3. of that
paper) and this results in 1952 sources that are matched between all
5 surveys under consideration. 

From these sources we randomly select
100 sources and use the bootstrapping procedure outlined by
\cite{Hardcastle_2016} to ascertain the factor by which the LoTSS-DR2 sources need to be scaled by in order to align the flux density scale of the
LoTSS-DR2 catalogue with the flux density scales of other surveys.
This approach makes use of the {\tt emcee} Markov chain Monte Carlo
(MCMC) library (\citealt{Foreman-Mackey_2013}) to derive  LoTSS-DR2
scaling factors where the normalisation and power-law spectral index
for each source are free parameters that are determined via
Levenberg-Marquart $\chi^2$ minimisation, and the likelihood is calculated from the total 
$\chi^2$ of all the sources in the field.
Whilst we make the
assumption that sources have power-law spectra between 74\,MHz and
1.4\,GHz the procedure does allow us to exclude sources that are
poorly described by a power law (these may include e.g. incorrect
cross matches or sources with significant spectral curvature) by rejecting those
with high  $\chi^2$ before refitting (on average only a few sources
are removed, with $93\pm 3$ out of 100 remaining). We repeat this procedure 500 times, each with just 100 randomly selected sources, and examine the distribution of derived LoTSS-DR2 scaling factors which is found to be approximately Gaussian and centred on 1.01 with a standard deviation of 0.03. 

We note that whilst the results of this analysis are  encouraging there is still some uncertainty about the level of systematic error in the flux density measurements in the LoTSS-DR2 catalogues. If for example we repeat the above analysis but just making use of VLSSr, LoTSS-DR2 and NVSS then from the 6102 sources matched between the 3 surveys we derive LoTSS-DR2 scaling factors that are again approximately Gaussian but are this time centred on 0.91 which is in agreement with the findings by \cite{deGasperin_2021} who examined the accuracy of the LoLSS flux density scale. The fact that these comparisons with different surveys produce different results is likely a consequence of  low level systematic errors in flux density scales combined with our assumption that the cross-matched source populations have genuinely power-law spectra without any curvature. Either way the results suggest the systematic overall flux density scale error of LoTSS-DR2 is less than 10\% and we have therefore not applied further scaling. For the total error on the flux density of a given source this overall scaling uncertainty should be added to the 10\% error due to positional variations that are described in Sect. \ref{sec:pos_flux_vary}.

\subsection{Recovery of diffuse emission}
\label{sec:emission_recovery}

\begin{figure}[htbp]
   \centering
   \includegraphics[width=\linewidth]{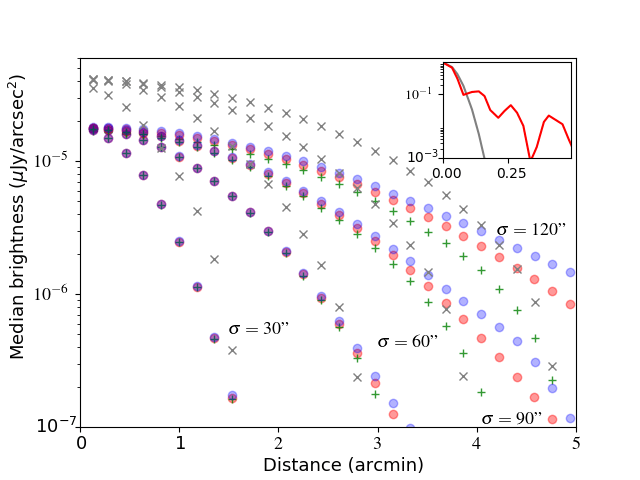} 
   \caption{The recovery of simulated Gaussian skymodels of varying width and brightness that are injected into calibrated $uv$-data. The red points show the brightness as a function of distance from the centroid of the simulated skymodels convolved by a Gaussian with a size that matches the restoring beam dimensions. The blue points show the profile of the emission recovered in the restored image created from the simulated $uv$-data. The  green `+' symbols show the flux density in the corresponding model image created during the deconvolution of the simulated $uv$-data that has been convolved with a Gaussian with a size that matches the restoring beam. The `$\times$' symbols show the flux density in the undeconvolved image of the $uv$-data. The different profiles show the different values of standard deviations of the injected Gaussian profile. When the emission is deconvolved the model injected into the data matches the profile and flux density of the emission recovered in the restored image. However, when the emission is not fully deconvolved (e.g. in the undeconvolved image or in the restored image where the emission becomes very faint) the apparent flux density can be either severely overestimated or underestimated and can be highly misleading due to the mismatch in area of the synthesised (red) and restoring beams (grey) which are shown in the inset.}
   \label{fig:lofar-perfectuv-injection}
\end{figure}

Extended, several arcminute scale emission can be accurately recovered in LOFAR images 
 due to the excellent $uv$-coverage provided by the large number of short baselines. This has been quantified in several studies such as \cite{Hoang_2018} and \cite{Botteon_2020} who make use of LoTSS data and have shown that, for example, they are able to recover 95\% of the flux density for a large (up to $\sim9\arcmin$) simulated galaxy cluster radio halo. For completeness we again demonstrate this by injecting two-dimensional Gaussian profiles of different widths (a standard deviation ranging from 30$\arcsec$ to 120$\arcsec$ with equal minor and major axis) into real LOFAR $uv$-data but for clarity in the interpretation we remove noise and contaminating sources. As shown in Fig. \ref{fig:lofar-perfectuv-injection}, in these idealised simulations, which do not account for any effects of calibration, when imaging the simulated $uv$-data we essentially recover all of the flux density even for the broadest profiles that we have injected (within a separation of 5 times the standard deviation of the injected Gaussian from its centre at least 97\% of the flux density is recovered).

\begin{figure*}[htbp]
   \centering
   \includegraphics[width=0.31\linewidth]{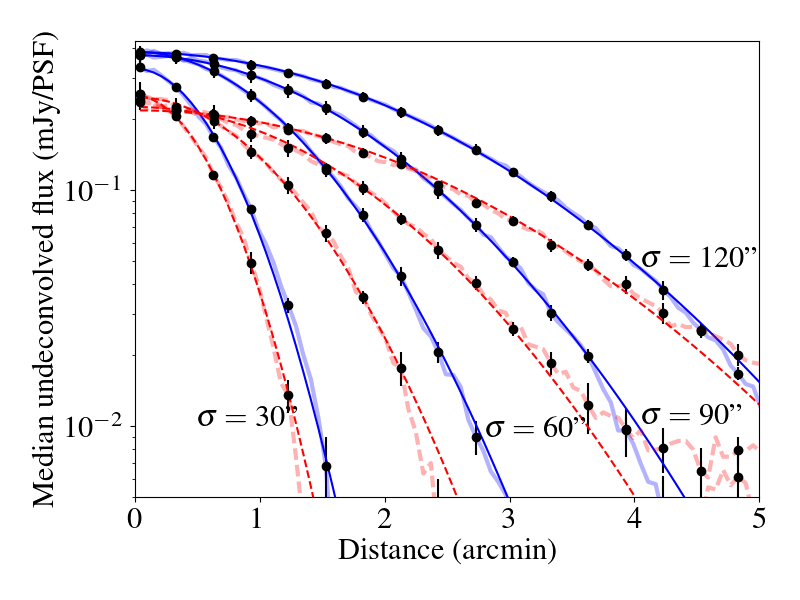}
    \includegraphics[width=0.31\linewidth]{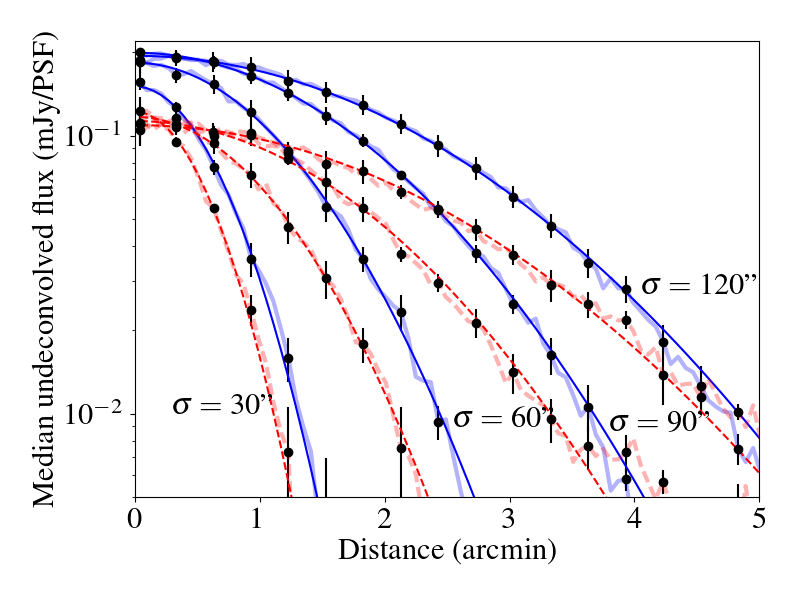}
    \includegraphics[width=0.31\linewidth]{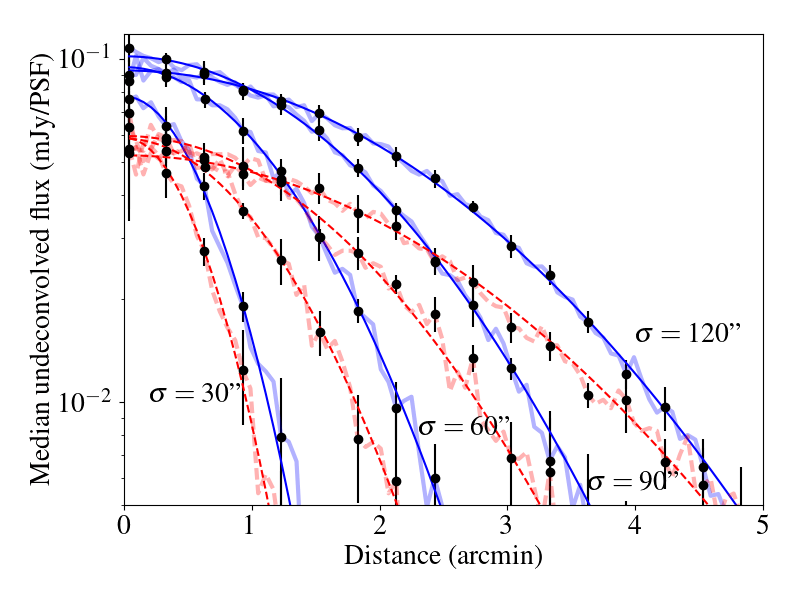}
   \caption{The recovery of unmodelled flux density in the LoTSS-DR2 pipeline for highly extended sources. Four Gaussian profiles (with standard deviations of 30, 60, 90, 120$\arcsec$) with different peak brightness levels (10$\mu$Jy/pixel, $5\mu$Jy/pixel and 2.5$\mu$Jy/pixel from left to right) are each injected 5 times into LoTSS-DR2 $uv$-data prior to a direction dependent calibration step using a skymodel that does not include their emission. The black error bars and thicker lines show the median and the standard deviation of the image brightness values as a function of distance for each of the different injected profiles. The thin solid lines are the best fit profiles to these median values. Blue lines correspond to values derived from the undeconvolved pre direction dependent calibrated image (i.e. image from \ref{pipeA:3} or the centre panel in Fig. \ref{fig:lofar-unmodelleduv-injection-images}) and the red lines correspond to the undeconvolved post direction dependent calibrated images (i.e. image from \ref{pipeA:5} or the right panel in Fig. \ref{fig:lofar-unmodelleduv-injection-images}). Overall, by integrating the simulated and recovered signal out to five standard deviations from the Gaussian centoid we find that in this worst-case scenario typically $60 \pm 6\%$ of injected but completely unmodelled flux density is recovered after direction dependent calibration.}
   \label{fig:lofar-unmodelleduv-injection}
\end{figure*}

\begin{figure*}[htbp]
   \centering
   \includegraphics[width=0.32\linewidth]{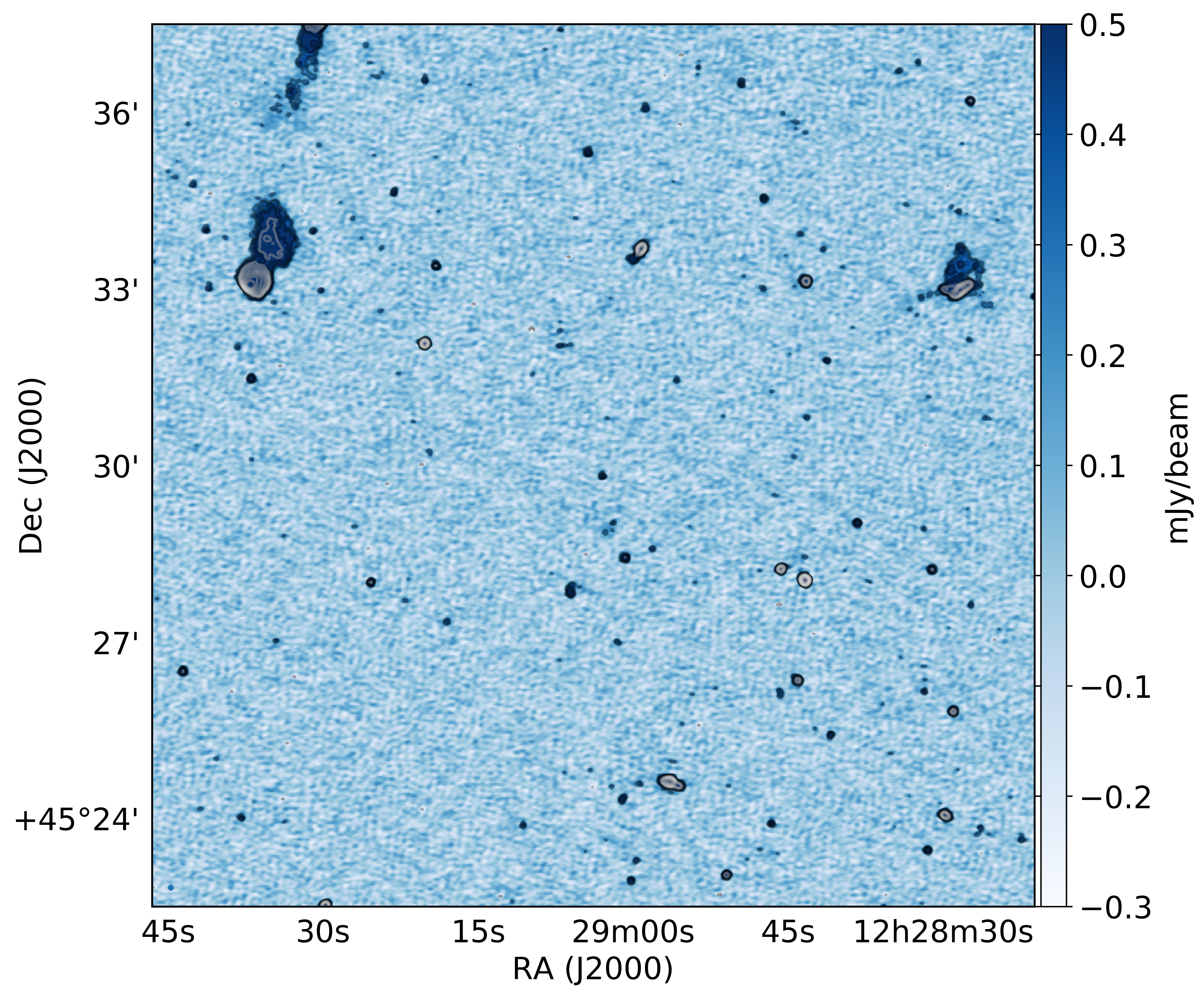}
      \includegraphics[width=0.32\linewidth]{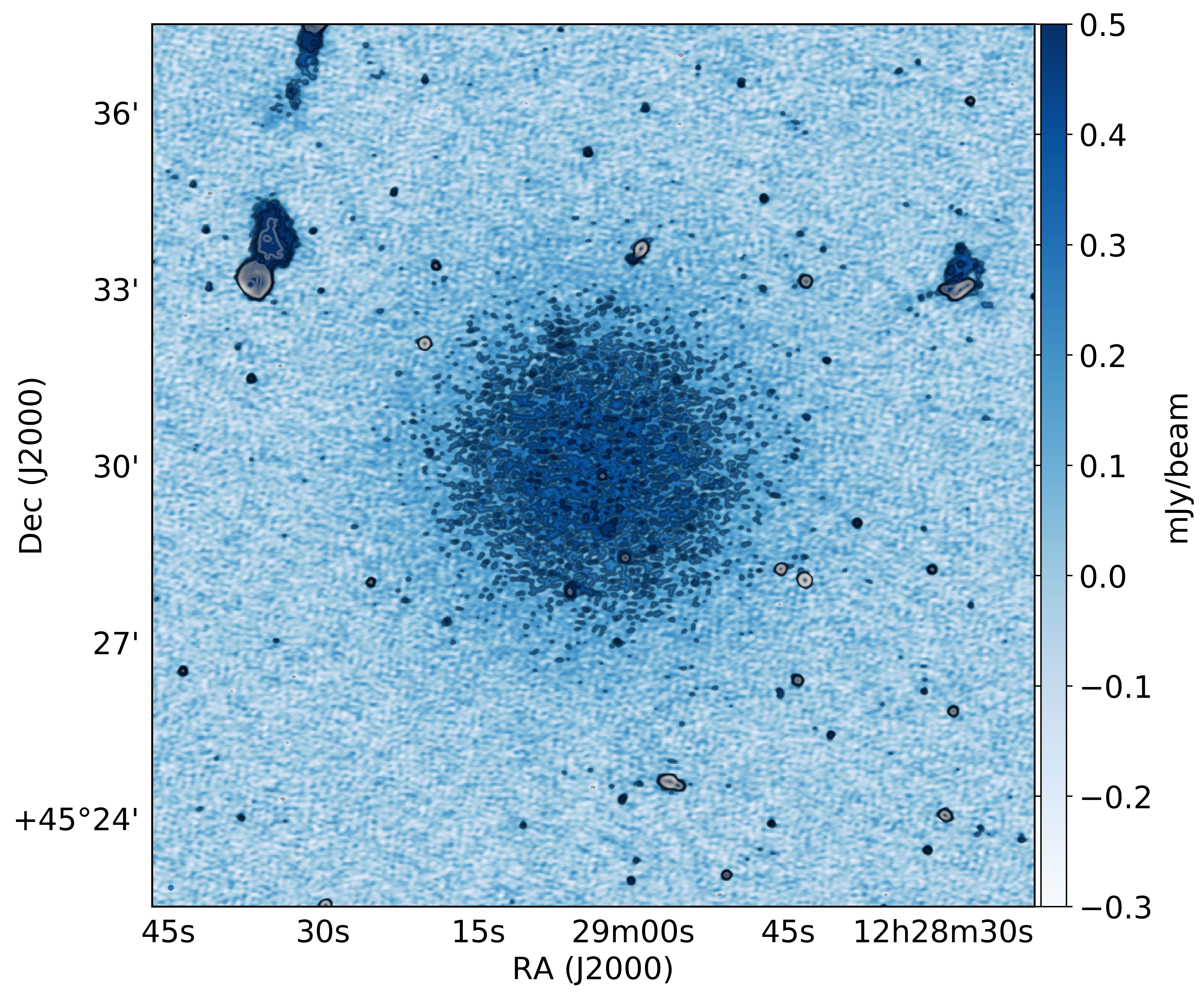}
      \includegraphics[width=0.32\linewidth]{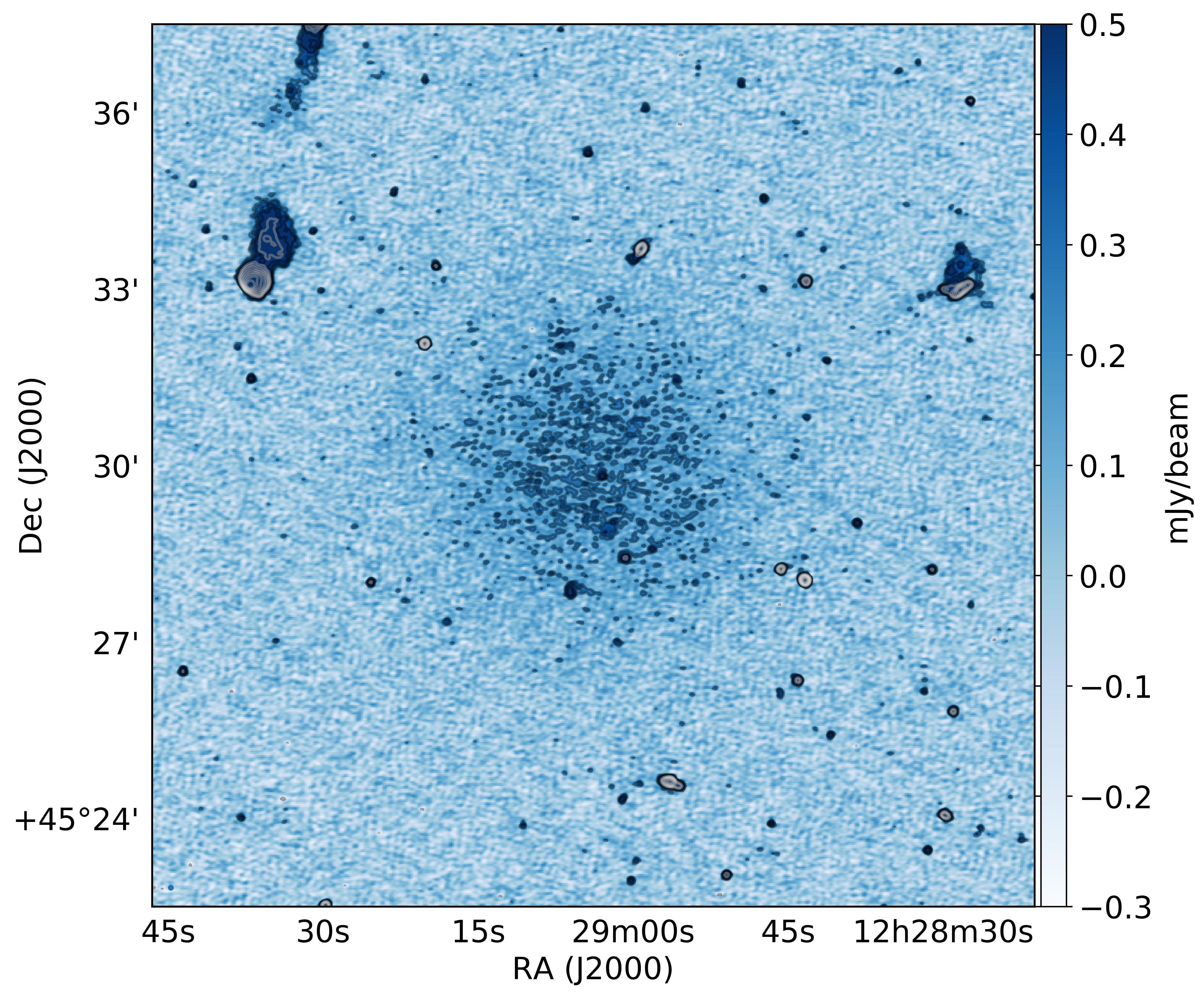}
   \caption{An example of an injected profile, in this case the injected profile has $\sigma=120\arcsec$ and a peak brightness of  10$\mu$Jy/pixel. The left shows the original data. The centre shows the original data plus the injected profile that is added but not deconvolved. The right shows the resulting image after it is calibrated again but without the profile in the model and hence shows the suppression of completely unmodelled emission.}
   \label{fig:lofar-unmodelleduv-injection-images}
\end{figure*}

As described in detail by \cite{Tasse_2021} and demonstrated qualitatively in Fig. \ref{fig:diffuse_emission}, we have significantly improved the LoTSS processing pipeline by improving the regularisation and conditioning of our calibration to  more accurately recover diffuse emission. However, a lingering concern has been that in the calibration of LoTSS data there are a large number of degrees of freedom and, whilst we believe point-like sources are accurately recovered (see \citealt{Shimwell_2019} for some simulations), the level at which unmodelled diffuse emission is suppressed in the LoTSS-DR2 direction dependent calibration pipeline was yet to be thoroughly quantified. This is particularly important because the pipeline products are routinely used for studies of highly extended low surface brightness sources such as galaxy clusters,  nearby galaxies or the faint lobes of active galactic nuclei which can be hard to accurately include in the sky models used for calibration. \cite{Tasse_2021} presented an initial assessment of the performance of the LoTSS-DR2 pipeline and here we build upon this by performing the following simulations:
\begin{enumerate}[label=Step.\arabic*,ref=Step.\arabic*,leftmargin=1.9cm]
\item Process a LOFAR dataset with the LoTSS-DR2 pipeline (see \citealt{Tasse_2021} for a full description of the pipeline or Sect. \ref{sec:observations} for a summary)  \label{pipeA:1}
\item Inject simulated Gaussian profiles of various sizes, flux-densities and positions into the $uv$-data using the final solutions derived from  \ref{pipeA:1} which corrupt the profiles by direction dependent effects. \label{pipeA:2}
\item Re-image the data, which now includes the simulated profiles, with the solutions derived from the  LoTSS-DR2 pipeline run. \label{pipeA:3}
\item Re-calibrate the data, which now includes the simulated profiles using the final skymodel from \ref{pipeA:1} pipeline which does not include the simulated profiles. \label{pipeA:4}
\item Re-image the data with the newly derived solutions. \label{pipeA:5}
\end{enumerate}
In this procedure the images from  Step.1, 3 and 5 are made from the real data, the real data plus the idealised
recovery of the profile based on the LOFAR $uv$-coverage and finally
the real data plus the recovered profile after it has been completely
excluded from the sky model used for direction dependent calibration.
Hence, by subtracting the output of \ref{pipeA:1} from the output images of both \ref{pipeA:3} and \ref{pipeA:5} we can remove contamination from discrete sources and compare how much flux density is recovered in the idealised case relative to the completely unmodelled case which is the worst-case scenario and is where some of the emission is suppressed by the calibration process. The results for this simulation are shown in Fig. \ref{fig:lofar-unmodelleduv-injection} and \ref{fig:lofar-unmodelleduv-injection-images}. The fraction of recovered flux density varies not only with the profile but also with the distance from the profile centroid, but we find that typically, by integrating pixel values from regions where the flux density exceeds the image noise and lies within five standard deviations of the Gaussian centroid, the recovered integrated flux density is $60\pm6\%$ of that injected. There is some dependence on both the size of the injected Gaussian and the flux density,  with poorer recovery of fainter and larger Gaussians. When averaged over the different simulations peak brightness values we recover $69\pm7\%$, $58\pm4\%$, $56\pm2\%$ and $57\pm4\%$ of the injected flux density for injected Gaussians with standard deviations of 30$\arcsec$, 60$\arcsec$, 90$\arcsec$ and 120$\arcsec$ respectively. When averaged over the Gaussian width in the different simulations we find that the recovered integrated flux density is  $54\pm6\%$, $58\pm6\%$ and $59\pm6\%$  for simulations with injected peak brightness levels of 2.5$\mu$Jy/pixel, $5\mu$Jy/pixel and 10$\mu$Jy/pixel respectively.

A more realistic scenario for many of the detectable faint diffuse structures in the LoTSS-DR2 images is that they are partly deconvolved during the LoTSS-DR2 pipeline processing and are thus not completely absent from the models used during calibration steps. The precise level and accuracy of the deconvolution will depend upon aspects such as the brightness, extent, complexity and local environment but can be assessed by inspecting the deconvolution residual images. 
To simulate this scenario we incorporate additional deconvolution and masking steps into the simulations between Step.3 and  Step.4. These additional steps mimic the masking and deconvolution that is performed in the LoTSS-DR2 pipeline and allow for the simulated Gaussians to be partly incorporated into a model which is then used for the calibration in Step.4. As previously, in these new simulations the \ref{pipeA:3} and \ref{pipeA:5} images contain the undeconvolved simulated emission before and after calibration respectively (where the only difference is that now the calibration is done off a model that partly includes the simulated emission rather than completely excludes it) and by comparing these then we can again examine how much flux density is lost during the calibration. As before, we subtract the image from \ref{pipeA:1} from both the \ref{pipeA:3} and \ref{pipeA:5} images to remove contaminating sources from the analysis. We find that in this more realistic scenario we are able to recover significantly more flux density if the injected Gaussians are prominent enough to be picked up during the masking and deconvolution steps. For example, we find that when averaged over the different simulations peak brightness values, we recover $75\pm12\%$, $70\pm15\%$, $93\pm20\%$ and $93\pm17\%$ of the injected flux density for injected Gaussians with standard deviations of 30$\arcsec$, 60$\arcsec$, 90$\arcsec$ and 120$\arcsec$, respectively. When averaged over the different simulations Gaussian width, we find that the recovered integrated flux density is  $54\pm9\%$, $85\pm11\%$ and $93\pm4\%$  for simulations with injected peak brightness levels of 2.5$\mu$Jy/pixel, $5\mu$Jy/pixel and 10$\mu$Jy/pixel, respectively. As expected for brighter Gaussians we approach the theoretical recovery of LOFAR (i.e. comparable to injecting into $uv$-data and not accounting for calibration effects) whilst for fainter Gaussians, where much of the emission remains largely undeconvolved, we approach the situation outlined in the worst-case scenario simulations where calibration is performed off sky models that completely exclude the emission and a substantially larger fraction of the flux density is suppressed. Similarly, wider Gaussians are more likely to be picked up during the masking and the emission is typically better recovered than that from the narrower simulated Gaussians.

A further important caveat for studies of diffuse emission is that if the emission is not deconvolved then, even considering the possible flux-density suppression described above, its apparent brightness in the images can be significantly over estimated or otherwise misleading. This is because the integral of the synthesised beam in a moderately sized region around its peak far exceeds that of the 6$\arcsec$ restoring beam and also has a very different shape with substantial sidelobes. The severity of this issue is shown in Fig. \ref{fig:lofar-perfectuv-injection} which demonstrates that for various sized Gaussian sources injected into $uv$-data the apparent emission at the centroid is overestimated by a factor $\sim 2.3$ if none of the emission is deconvolved. This overestimation is due to the synthesized beam sidelobes associated with the surrounding emission artificially enhancing the apparent levels of emission at the centroid. Furthermore, the difference between the deconvolved and undeconvolved images changes significantly as a function of distance from the centroid and these changes reflect the total contribution of the synthesized beam sidelobes from the surrounding emission at a given point. 

Through various masking and imaging steps in the LoTSS-DR2 pipeline we try to automatically detect diffuse emission and ensure that it is deconvolved and enters the skymodel used for calibration and in such cases the recovery of the emission is expected to be close to the theoretical limits from the $uv$-coverage of the observation. Unfortunately, it is inevitable that low level emission is sometimes missed and our simulations reflect the challenges in precisely interpreting this emission. Reassuringly, even in the worst-case scenario of completely undeconvolved emission that is missing from our sky models, a reasonable fraction is  still present in our final images. However, we do urge particular caution when conducting studies of diffuse emission (or placing limits on the lack of diffuse emission) and suggest a careful examination of deconvolution residual images to ensure that sources have been included in the skymodel and fully deconvolved. The effects of non-deconvolution of diffuse sources at the full resolution of LoTSS-DR2 can be significantly mitigated by making use of the low-resolution (20$\arcsec$) images that are also provided as a standard DR2 data product. Alternatively, the released $uv$-data can be reimaged with tailored masking and deconvolution or even further processed to optimise the image fidelity and deconvolution parameters for a particular target (see \citealt{vanWeeren_2021}).  Finally, we note that the analysis presented in this subsection has focused on the recovery of faint diffuse emission in the images and not on the accuracy of the \textsc{PyBDSF} characterisation of such objects. For this purpose we have released the \textsc{PyBDSF} residual mosaics which allow for an assessment of the catalogued Gaussian components and we refer the reader to \cite{Mohan_2015} and \cite{Hopkins_2015} as well as our ongoing efforts to refine the \textsc{PyBDSF} catalogues that are outlined in Sect. \ref{sec:value_added_cats}.

\subsection{Dynamic range}
\label{sec:dynamic_range}

The dynamic range  of the LoTSS-DR2 images, which here we define as the image noise increase due to the proximity of a source, varies depending on aspects such as the ionospheric conditions during the observations and the complexity and brightness distribution of sources, particularly those that are challenging to precisely model on fine scales. However, as described by \cite{Tasse_2021}, we have significantly improved the dynamic range in LoTSS-DR2 compared to LoTSS-DR1. This improvement is due to the inclusion of additional steps in our pipeline that perform a direction independent calibration of the data using a direction dependently produced sky model that has been distorted by the appropriate calibration solutions to predict its direction independent appearance. 

\begin{figure}[htbp]
   \centering
   \includegraphics[width=\linewidth]{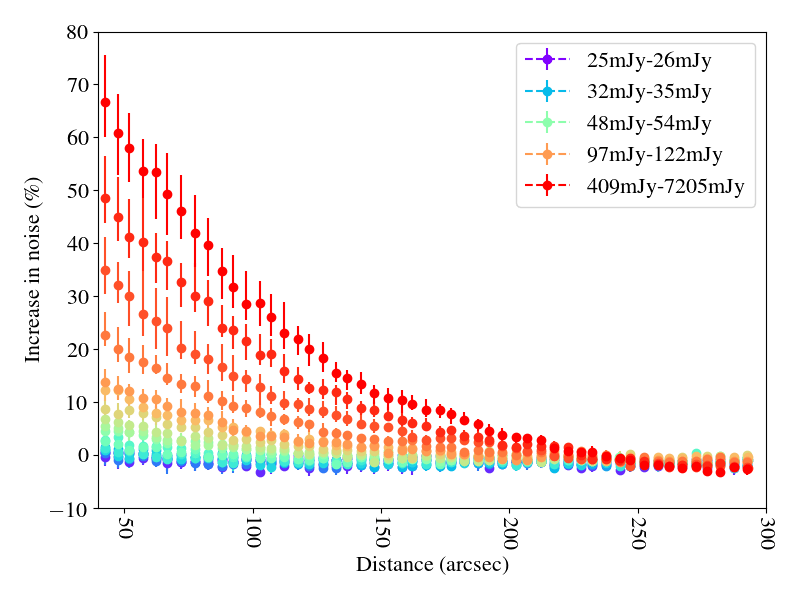}  
   \caption{Percentage increase in the  local noise level as a function of distance from the centres of sources of various integrated flux density levels. The  increase in local noise level is defined as the percentage at which the image pixel standard deviation in regions close to sources is higher than the standard deviation of pixels away from  contaminating sources. Here statistics are determined for compact and isolated LoTSS-DR2 sources that are grouped into bins of different integrated flux density levels (for direct comparison with LoTSS-DR1 through Fig. 11 in \citealt{Shimwell_2019}) which are shown with different colour markers. The integrated flux density bins were selected to each contain the same number of sources (239) and the errors are 95\% bootstrap confidence intervals.}
   \label{fig:dynamic-range-profiles}
\end{figure}

To quantify the dynamic range in LoTSS-DR2 we examined the noise levels around all compact (Eq. \ref{eq:resolved_sources}), isolated (45$\arcsec$ or more from a neighbouring source and no sources with $S_I>5$\,mJy within 300$\arcsec$ to reduce contamination in our measurements) sources with an integrated flux density greater than 25\,mJy. Grouping sources by source flux density we then ascertain the average factor by which sources of a particular flux density increase the noise level in the surrounding region and how this varies as a function of distance from the source. The results are shown in Fig. \ref{fig:dynamic-range-profiles}. 

In comparison to LoTSS-DR1, where the local noise in regions 50$\arcsec$ from sources of e.g. 0.12-0.15\,Jy was on average approximately 100\% higher than in regions without contaminating sources, the dynamic range is typically at least a factor of four better in LoTSS-DR2 (see Fig. \ref{fig:dynamic-range-profiles}). Furthermore, in LoTSS-DR1 we deduced that 8\% of the area of the data release had an increased noise ($>15\%$ above the local thermal noise) and performing the same analysis for LoTSS-DR2 indicates that in this data release only 2.5\% of the area is now impacted in this way. At only 5\% and 10\% levels of increased noise we measure that 6.6\% and 3.7\% of the surveyed area are impacted respectively. Note that these calculations only account for contamination of the released images and do not include failed facets which result in gaps in our mosaic coverage (see Fig. \ref{fig:mosaic-noisemap}). However, in LoTSS-DR1 the processing of 6\% of the fields failed due to dynamic range limitations and could not be included in that release, whereas in LoTSS-DR2 we processed over 13 times more fields and, whilst individual facets within fields have failed (e.g. around the bright sources 3C\, 196 and 3C\,48), not a single field has had to be entirely removed from this data release due to dynamic range limitations.  A qualitative comparison highlighting the improvement between LoTSS-DR1 and LoTSS-DR2 is shown in Fig. \ref{fig:dynamic_range} and for a careful assessment of the noise level around targets of interest we have included LoTSS-DR2 noise maps in this data release.

\subsection{Sensitivity and completeness}
\label{sec:sensitivty}

\begin{figure}[htbp]
    \centering
   \includegraphics[width=\linewidth]{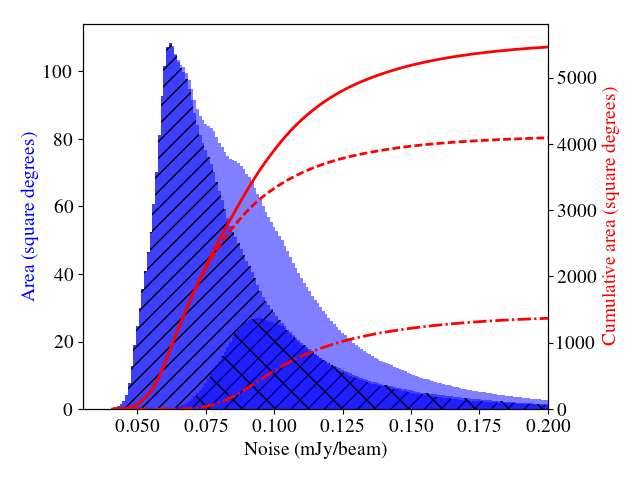}
    \caption{The overall rms distribution of  LoTSS-DR2 is shown in the light blue histogram (y-axis on the left). The contributions from the RA-13 and RA-1 regions are shown by the darker blue histograms with $\setminus$ and $/$ hatchings respectively. The  rms distributions of the two regions are approximately equal above 0.1\,mJy/beam  but otherwise the RA-13 region is on average higher declination and therefore generally more sensitive than the  RA-1 region. The median and mean of the rms are 83$\mu$Jy/beam and 95$\mu$Jy/beam over the entire LoTSS-DR2 region where 90\% and 95\% is below 140 and 170$\mu$Jy/beam respectively. The solid red line shows the cumulative area of the entire LoTSS-DR2 region with an rms noise below a given value with the dashed and dashdot lines showing the same but for the RA-13 and RA-1 regions (y-axis on the right).}
    \label{fig:mosaic-depth}
\end{figure}

\begin{figure}[htbp] 
     \centering
   \includegraphics[width=\linewidth]{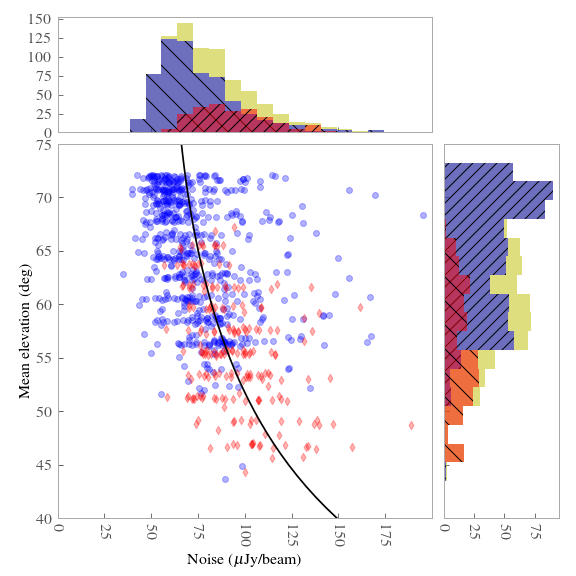}
    \caption{The sensitivity of each individual pointing that makes up
      LoTSS-DR2 as a function of the mean elevation during the
      observation. The sensitivities are scaled to their equivalent for an 8hr run by taking into account the exact integration time and the fraction of data that are flagged. The blue points and blue histograms with $\setminus$ hatchings show pointings in the RA-13 region and the red points and red histograms with $/$ hatchings show the pointings in the lower average declination RA-1 region. The yellow histograms correspond to all pointings in both regions. The black line shows the best fit A$\times$cos(90-elevation)$^{-2.0}$ curve where A is found to be 62$\mu$Jy/beam and the dependence on elevation is fixed according to the projected size of the LOFAR stations.}
    \label{fig:noise-elevation}
\end{figure}

The aim of LoTSS is to image the northern sky to a sensitivity of 100$\mu$Jy/beam at optimal declinations (i.e those close to the latitude of LOFAR). As shown in Fig. \ref{fig:mosaic-depth} we meet this objective in over 70\% of the region covered by LoTSS-DR2 with 50\%, 90\% and 95\% of the region having sensitivities below 83, 139 and 171$\mu$Jy/beam respectively. The regions typically at higher noise level are those impacted by dynamic range limitations (see Sect. \ref{sec:dynamic_range}), those observed at low elevation 
(Fig. \ref{fig:noise-elevation}) and those on the edge of the LoTSS-DR2 coverage where only a single pointing contributes to the image and the effective integration time is significantly reduced (see e.g. Fig. \ref{fig:mosaic-noisemap} where edge effects are clear). We note that 
once the survey is complete we can expect a lower fraction of regions of enhanced noise due to edge effects as the only edges in the coverage that will remain are those at the $0^\circ$ declination limit of the survey.  If for example we exclude the LoTSS-DR2 edge regions (24\% of the total LoTSS-DR2 region) from our sensitivity calculations we find that our sensitivity is better than 100$\mu$Jy/beam in 78\% of the region with 50\%, 90\% and 95\% of the region having sensitivities below 76, 122 and 148$\mu$Jy/beam respectively. However, sensitivity variations with the elevation of the target field will remain a feature of the final LoTSS sensitivity pattern. This is because as shown in Fig. \ref{fig:noise-elevation} even though our calibration procedures can effectively calibrate a thicker ionosphere and approximately achieve the thermal noise at low elevation, the projection of the LOFAR stations substantially reduces the effective collecting area. The field of view is however also increased when observing at low elevation whilst our mosaic grid spacings remain equal and thus the additional overlap between pointings does partly mitigate the sensitivity variations (see Fig. \ref{fig:DR2-region}). Additionally, we are adopting 12\,hr (rather than the usual 8\,hr) duration observations for pointings at declinations less than $20^\circ$ which will further improve our sensitivity for these fields that are generally those observed at the lowest elevations.

To assess the completeness of the survey we inject populations of sources into random positions of the \textsc{PyBDSF} residual mosaic images (i.e. our final maps but with sources removed using their catalogued models) and ascertain how many injected sources are recovered as a function of the injected source flux density. A simple approach to quantify the point-source completeness is to inject delta-functions convolved to the survey resolution into the residual images. However, as described in Sect. \ref{sec:source_sizes} we know that genuinely point-like sources in LoTSS  are blurred due to imperfections in the calibration. To account for this we also perform simulations where we inject real deconvolution components of the skymodels of sources that we classified to be point-like according to the criterion given in Sect. \ref{sec:source_sizes} into the residual images, after again convolving the models to the survey resolution. 

In both cases we inject a population of sources described by the 7-th order polynomial fit to the deep 150\,MHz source counts presented by \cite{Mandal_2021}, namely that in an area of $A_{sr}$ steradians and a narrow flux density bin of size $\Delta S_{p,mJy}$ mJy, there are $\Delta N_{\Delta{S_p,mJy}}$ sources according to:
\begin{multline*}
\Delta N_{\Delta{S_p,mJy}} = A_{sr}\Delta S_{p,mJy} 10^{(6.155 - 2.615 X + 0.227 X^2 + 0.518 X^3} \\
^{ - 0.450 X^4 + 0.160 X^5 - 0.0285 X^6 + 0.002 X^7)},
\tag{3}
\end{multline*}
where $X$ is $\log_{10}S_{p,mJy}$ and $S_{p,mJy}$ is the central flux density of the bin in mJy. We randomly draw sources between a flux density of 0.2\,mJy and 2.0\,mJy from the distribution 
and inject these into the image. These flux density limits correspond to a source density of approximately 3070 sources per square degree. When using the realistic deconvolution component models we scale the sum of the model to equal the desired flux density and when injecting delta functions we use the desired flux density value but with all emission in a single deconvolution component. The two situations are equal if the source is genuinely point-like and modelled by a single component in the deconvolution. 

\begin{figure}   \centering
   \includegraphics[width=\linewidth]{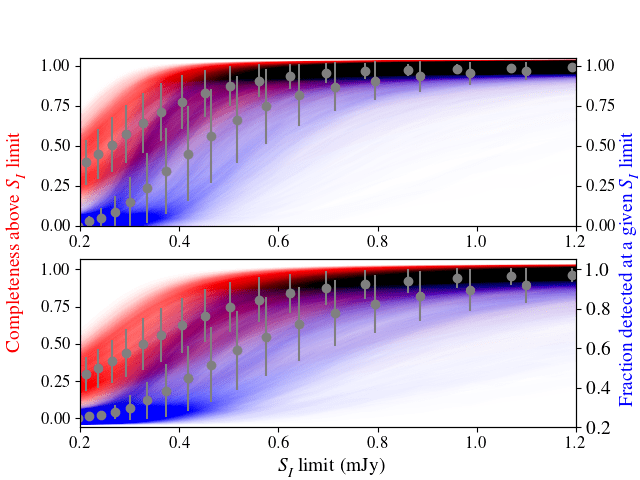}
   \caption{The point source completeness, or fraction of simulated sources detected above some $S_I$ limit, of LoTSS-DR2 for genuinely point-like sources (top) and for real point-like sources as defined from their recovered deconvolution models (bottom). Each red line shows the completeness derived from approximately 370,000 sources that are injected into each of the 841 DR2 mosaics through a series of 10 simulations with realistic source density and flux density distributions. The overlaid grey points represent the mean between the different mosaics weighted by the area covered by the mosaics and the error bars show $\pm$ the standard deviation of the different mosaic values. Due to the number of sources injected into the maps the Poisson errors are negligible by comparison. The fraction of sources detected at a given flux density are also shown with blue lines where grey points again show the corresponding weighted mean values as a function of flux density. }
   \label{fig:DR2-completeness}
\end{figure}

In both cases, for each of the 841 \textsc{PyBDSF} residual mosaic images we perform 10 simulations where point-like or deconvolution model source populations are injected. Using the same \textsc{PyBDSF} settings as for the real LoTSS catalogue (except that, for efficiency, the wavelet source characterisation is not used as we are only injecting simple sources) we then classify sources as detected if we recover them to be within 10$\arcsec$ of their injected location and with a difference in recovered and injected integrated flux of less than 10 times the statistical error on the recovered integrated flux.
Our results are shown in Fig. \ref{fig:DR2-completeness} and for perfect point-like sources LoTSS-DR2 is 50\%, 90\% and 95\% complete at 0.27\,mJy, 0.58\,mJy, 0.71\,mJy whereas in our more realistic simulations using real deconvolution models the completeness is  50\%, 90\% and 95\% at 0.34\,mJy, 0.8\,mJy, 1.1\,mJy.

\subsection{In-band spectra}

\label{sec:inband_spectrum}

\begin{figure}   \centering
   \includegraphics[width=\linewidth]{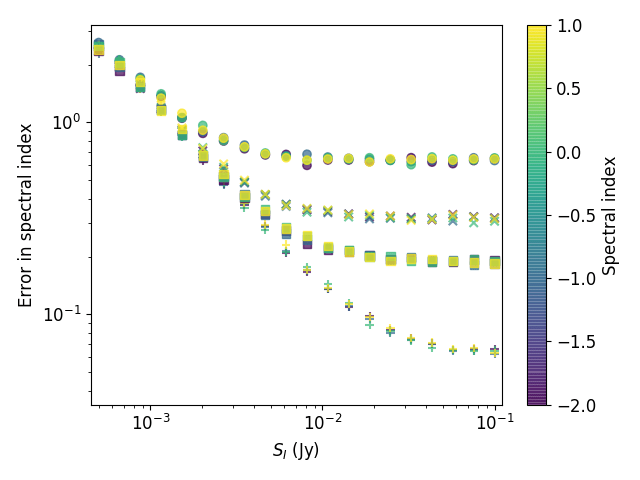}
    \caption{The simulated measurement error in spectral index as a function of $S_I$ where the marker colour indicates the simulated spectral index ($\alpha$). The simulation assumes an independent 10\% (circles), 5\% ($\times$'s), 3\% (boxes), or 1\% ($+$'s) flux density alignment uncertainty on each in-band image which are added in quadrature to the image noise values of 220, 195 and 190\,$\mu$Jy/beam for the 128.0 143.7 and 160.2\,MHz in-band images respectively. The spectral index errors were calculated by simulating 1000 sources at each flux density, spectral index and flux density alignment uncertainty and computing the standard deviation of the recovered spectral index values.} 
    \label{Inband_simple_simulation} 
\end{figure}

LoTSS observations are conducted with a frequency range of 120-168\,MHz. We create three 16\,MHz bandwidth images with central frequencies of 128, 144 and 160\,MHz to explore the spectral properties of sources across the 48\,MHz band. However, as shown in Fig. \ref{Inband_simple_simulation}, through a simple simulation of a source population with a range of spectral properties and flux-densities as well as realistic thermal noise, it is clear that the narrow frequency range, combined with non-negligible uncertainties in the alignment of the flux density scale of the in-band images, hinders our ability to derive accurate in-band measurements even for very high signal to noise simple sources over a wide range of spectral indices. For the brightest sources, where the thermal noise impact is negligible, the flux density alignment uncertainty limits the accuracy, while at lower flux density the thermal noise plays an increasingly large role in the spectral index uncertainty. Nevertheless, our in-band images still offer the opportunity to constrain the spectrum of many sources that are yet to be detected at other frequencies and identify extremes of the distribution, such as inverted-spectrum sources. 

To derive the in-band spectra, we first attempt to minimise the uncertainty in the alignment of the in-band flux density scale. To do this we filter the full  bandwidth LoTSS-DR2 catalogue so that it contains  only compact (according to Eq. \ref{eq:resolved_sources}), isolated and fully deconvolved sources. The filtered LoTSS-DR2 catalogue is then cross matched with catalogues derived from each of the LoTSS  in-band images for each pointing. Using these cross matched catalogues we scale the in-band images such that the median spectral index between each in-band image and the filtered full bandwidth LoTSS-DR2 catalogue is $\alpha=-0.783$, where this spectral index was chosen for consistency with the flux density scale corrections we applied to our continuum images (see Sect. \ref{sec:flux_scale}) but we note the effect of altering it later in this subsection.  The resulting aligned 16\,MHz bandwidth images have median noise levels of 220$\pm$100, 180$\pm$85, 160$\pm$105$\mu$Jy/beam at 128, 144 and 160\,MHz. Within 2$^\circ$ of the pointing centres typically 660$\pm$50 sources are detected per pointing in each band.

To assess the accuracy of our flux density scale alignment we examine the in-band spectral index distribution using the catalogues derived during the alignment procedure but with the alignment corrections applied. For each source we determine the in-band spectral index by fitting the three 16\,MHz in-band integrated flux density values with a power law function. 
Where possible, for the same sources, we also derive a spectral index between the full-bandwidth LoTSS-DR2 integrated flux densities and corresponding values in the 1.4\,GHz NVSS catalogue (assuming a 10\% flux density scale uncertainty on the LoTSS measurements and a 5\% flux density scale uncertainty on NVSS). 

\begin{figure}   \centering
   \includegraphics[width=\linewidth]{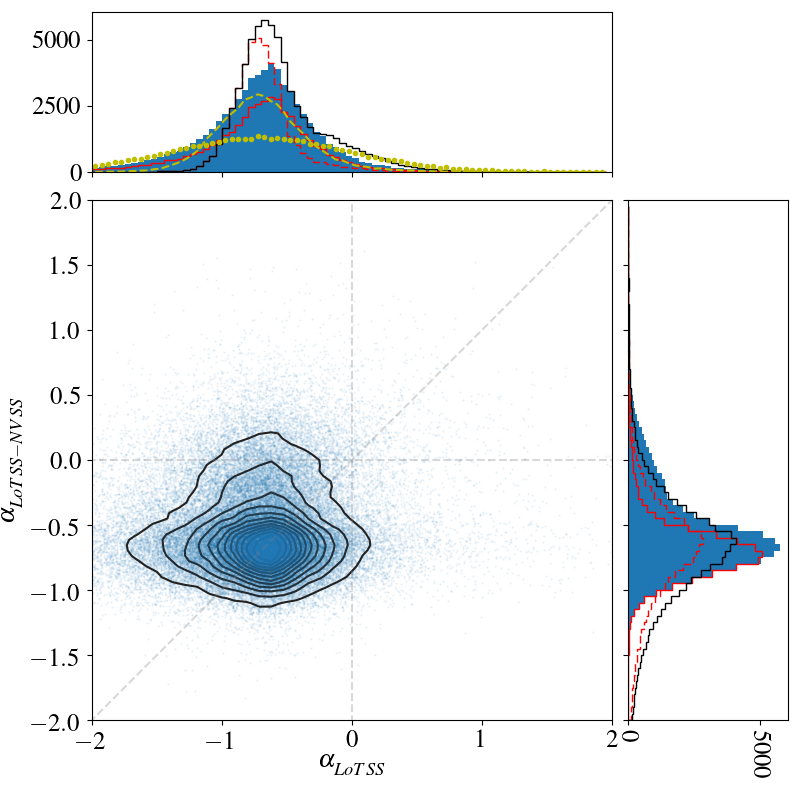}
          \caption{The derived LoTSS-NVSS and in-band LoTSS spectral index measurements for isolated, compact, fully deconvolved LoTSS-DR2 sources overlaid with contours showing the density of points. The solid blue histograms show the in-band LoTSS spectral index distribution (top) and the LoTSS-NVSS spectral index distribution (right) for all sources where we have both measurements. The histograms outlined in black show the LoTSS-NVSS spectral index distribution of these sources overlaid on the in-band LoTSS spectral index distribution or vice versa. The solid and dashed red lines show the spectral index values derived from only bright (LoTSS $S_I>30$\,mJy/beam) LoTSS sources for in-band and LoTSS-NVSS (top) and LoTSS-NVSS and in-band (right) respectively.  The dotted and the dashed yellow histograms (top) show the results from  simulations where the LoTSS-NVSS spectral index values of bright (LoTSS $S_I>30$\,mJy/beam) sources, and associated errors, are used to simulate LoTSS in-band flux densities which are then fitted with a power law to give a spectral index. Here flux densities for each band and source are extrapolated from the LoTSS-NVSS spectral index and LoTSS continuum measurement plus a random error drawn from a Gaussian with a standard deviation equal to 10\% (dotted) or 3\% (dashed) of the flux density (mimicking a flux density scale alignment uncertainty) added in quadrature to the statistical uncertainty from the \textsc{PyBDSF} estimation of noise local to the source. }
    \label{Inband_v_NVSS} 
\end{figure}

By assuming that the genuine scatter in the in-band spectral indices is comparable to that of the LoTSS-NVSS spectral indices we can approximate the accuracy of our alignment. To make this comparison we conduct a simulation where we take LoTSS-NVSS spectral indices and associated errors and simulate flux densities at the frequencies of our in-band LoTSS images. These simulated in-band flux densities are then altered by adding different levels of random error associated with the flux density scale alignment uncertainty in quadrature with the local noise. The simulated in-band values and associated errors are then fitted with a power law. The resulting distribution of in-band spectral indices derived from the simulation is similar in both width and height to that found from the real measurements when the level of uncertainty in the flux density scale alignment is 3\% which we hereafter adopt. This is demonstrated in Fig. \ref{Inband_v_NVSS} which shows our in-band  spectral index measurements compared to those we have derived between LoTSS and NVSS for all compact, isolated, fully deconvolved LoTSS-DR2 sources that are cross matched between the two surveys. 
 
 We note that there is an apparent offset between the peak of the spectral index distribution of the simulated population and the in-band population (and more generally between the median LoTSS-NVSS spectral index and the in-band spectral index of the same sources). This offset is associated with differences in the alignment procedures used for the LoTSS continuum and the LoTSS in-band measurements which both assumed median values of $\alpha=-0.783$ but for different source selections. As LoTSS is significantly deeper than NVSS for typical negative spectrum, and even flat spectrum sources (the sensitivities being comparable for an inverted $\alpha \approx 0.7$ spectrum source), we imposed a 30\,mJy limit on the LoTSS integrated flux densities during the alignment of the continuum measurements -- such sources are  detectable in NVSS unless they are unusually steep spectrum ($\alpha \lessapprox -1.2$). However, when aligning our in-band spectrum we did not impose such a flux density cut because the large number of sources at lower flux densities in the in-band images aids with the alignment. Hence, the offset implies that our in-band measurements for sources with integrated flux densities fainter than 30\,mJy are preferentially steeper spectrum than brighter sources.

This apparent variation in spectral index as a function of flux density can be largely accounted for by biases (e.g. due to the varying noise levels in the different bands resulting in different levels of \cite{Eddington_1913} bias and incompleteness) that artificially steepen the in-band spectrum of fainter sources. However, these effects are also likely entangled with real spectral index variations as a function of flux density and source population (see e.g. \citealt{Williams_2021} and \citealt{deGasperin_2018}). Furthermore, these issues clearly highlight the degree of uncertainty in the most appropriate spectral index to assume for the alignment process (as previously stated we assume $\alpha=-0.783$) and thus a further significant limitation of our present in-band spectral index measurements -- altering the assumed spectral index has the effect of shifting our derived in-band spectral index distribution in $\alpha$. These limitations will remain until either alignment is no longer required due to better flux density scale calibration or 
critical astrophysical related aspects such as the typical spectral index variations with flux density and curvature levels for the selected sources at the observed frequencies are better known. Even then the impact of e.g. ionospheric blurring, beam model errors, incompleteness and \cite{Eddington_1913} bias will remain substantial and in some cases require accurate knowledge of the underlying undetected source population and its spectral properties. The size of these effects will vary both within a particular pointing as well as between pointings, and thus require extensive modelling to isolate from genuine properties of the observed sources. 

\begin{figure}   \centering
   \includegraphics[width=\linewidth]{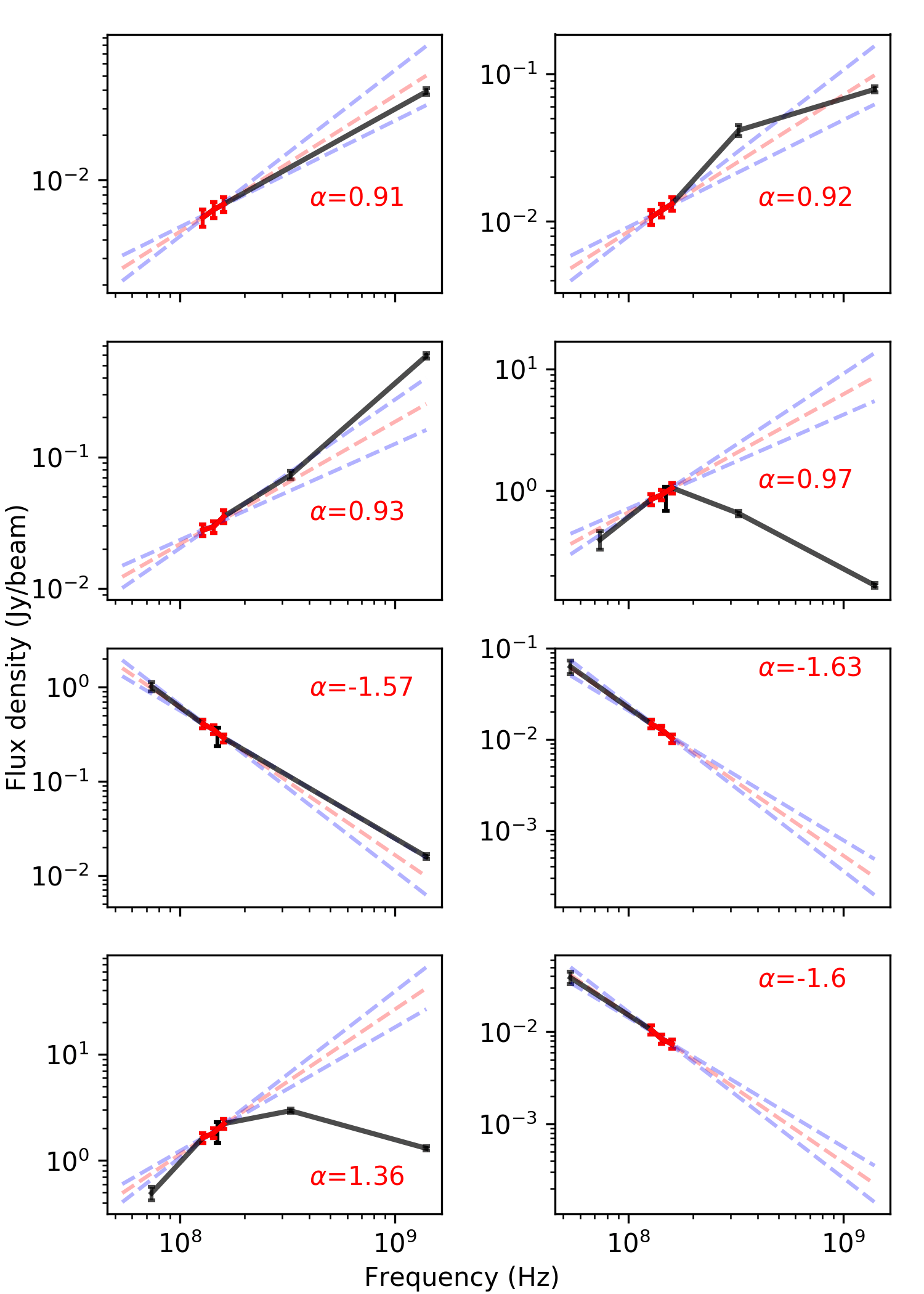}
       \caption{Examples of sources with extreme in-band LoTSS spectral indices. The in-band LoTSS measurements are shown in red where the red dotted line shows the fitted $\alpha$ (values of $\alpha$ are displayed on the panels) and the blue dotted lines show the fitted $\alpha\pm0.2$. The black points show measurements from LoLSS, VLSSr, TGSS, WENSS and NVSS where available. Clockwise from top left, the sources shown are: ILTJ004259.17+233517.1, ILTJ084911.00+403231.2, ILTJ101731.34+434312.5, ILTJ113903.61+583040.5, ILTJ125903.65 +594635.9, ILTJ082236.93+511230.0, ILTJ011755.08+320622.5 and ILTJ004259.17+233517.1.}
    \label{Spec_selection} 
\end{figure}

Despite the limitations, the distribution of the derived in-band spectral index values is largely as expected (also compared with earlier investigations of this issue by \citealt{Hardcastle_2016}), with LoTSS-NVSS providing more accurate measurements whilst our in-band measurements are hindered by the $\sim$3\% uncertainty in our flux density scale alignment which broadens the apparent spectral index distribution. Furthermore, even though sensitivity limitations impact the LoTSS-NVSS cross matched source populations for sources with $\alpha \lessapprox 0.7$, the LoTSS-NVSS spectral indices do reveal a large population of  flat and rising spectrum (e.g. $\alpha \gtrapprox -0.4$) sources although steep spectrum (e.g. $\alpha \lessapprox -1.0$) sources are largely absent (the large flat spectrum population was not identified in the GLEAM survey by \citealt{Callingham_2017} but this may be due to LoTSS accessing a fainter population than GLEAM). The in-band LoTSS measurements suffer less from the  sensitivity limitations that affect the recovered LoTSS-NVSS spectral index distribution and they can also be made for large numbers of sources that are presently only detected in LoTSS. As a demonstration, in Fig. \ref{Spec_selection} we show that interesting sources, or populations of sources, can be readily identified from the in-band spectra alone. 

We do emphasise that given the substantial measurement errors ($>0.2$) and systematic errors (due to bias and effects of the alignment procedure) on the LoTSS in-band measurements, the in-band measurements should be used with utmost caution. 
For example, when interpreting the in-band spectral index of a particular target of interest, a careful statistical analysis of the in-band spectrum of nearby sources (preferably also making use of measurements from auxiliary data) is valuable to characterise the local measurement errors and typical in-band spectral indices values in addition to the biases in that particular region. Furthermore, for resolved sources even more care is required as other aspects discussed in Sect. \ref{sec:emission_recovery} and \ref{sec:value_added_cats} such as the deconvolution, flux density suppression and \textsc{PyBDSF} characterisation add additional complications. There is scope to further improve the alignment of in-band measurements and provide more precise and valuable spectral indices (see  Fig. \ref{Inband_simple_simulation}) with reduced systematic uncertainty. New northern hemisphere surveys such as VLASS, APERTIF, LoLSS and LoDSS are underway and these will be critical for this and will supplement in-band LoTSS measurements by allowing for the spectral properties of many more LoTSS detected sources to be constrained over wide frequency and flux density ranges. 

\subsection{Source counts}
\label{sec:source_counts}

The source populations being recovered in our images span a wide range of flux densities (with substantial numbers of sources from $0.2$\,mJy$<S_I<5000$\,mJy), and as shown by e.g. \cite{Williams_2016} and \cite{Hale_2019}, even an individual LoTSS-depth pointing has sufficient  area and sensitivity to  accurately constrain radio source counts over several orders of magnitude in flux density (from approximately 1\,mJy to 1\,Jy).  However, recent longer duration and more sensitive observations at low-frequency (LoTSS-deep) allow for better statistics to constrain fainter populations that are largely beyond the reach of LoTSS depth observations (\citealt{Mandal_2021} present counts down to 0.25\,mJy). Thus, as shown in \cite{Hardcastle_2021}, the combination of the area and depth achieved by combining LoTSS-DR2 and the LoTSS-deep fields provides excellent statistics over an exceptionally wide range of flux densities.

\begin{figure}[htbp]
   \centering
   \includegraphics[width=\linewidth]{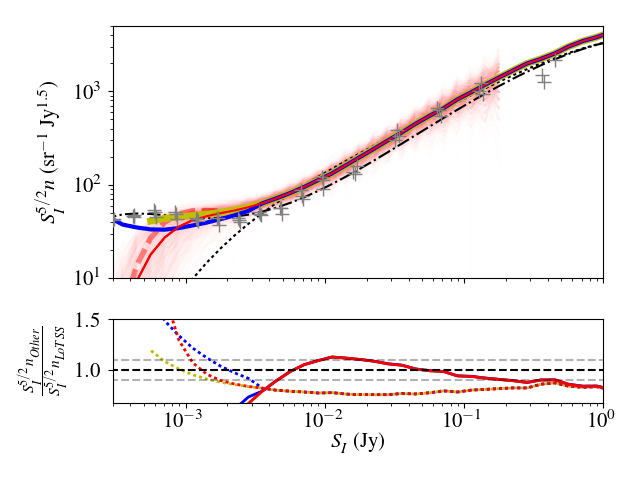}
   \caption{The 120-168\,MHz Euclidean normalized differential source counts derived from LoTSS-DR2  \textsc{PyBDSF} catalogues. The completely uncorrected source counts are shown in the top panel with transparent red solid lines showing those derived from individual pointings for flux density bins where at least 90\% of the pointings contain more than 10 sources. The opaque red solid line shows the combined uncorrected counts from all mosaics thus spanning the entirety of LoTSS-DR2. The solid yellow line shows the counts produced from only highly complete (at least 90\% of sources detected at a given threshold) LoTSS-DR2 regions (Method One in Sect. \ref{sec:source_counts}). The solid blue line shows counts that are corrected for incompleteness using the simulations that are detailed in Sect. \ref{sec:sensitivty} (Method Two in Sect. \ref{sec:source_counts}). The simulations where sources are injected according to the \cite{Mandal_2021} source count fit and recovered with \textsc{PyBDSF} are shown by the thick red dashed line.  The black dashdot line shows the \cite{Mandal_2021} fitted counts, the black dotted line shows the \cite{Intema_2017} fitted counts and the grey $+$'s show the sources counts of the three different fields analysed by \cite{Mandal_2021} that were combined in their fitting. Poisson errors on our counts are too small to be seen but the variation amongst the different mosaics (transparent red lines) reflects the level of agreement between different regions in LoTSS-DR2. In the bottom panel we show the ratio of the \cite{Intema_2017} (solid lines) and \cite{Mandal_2021} (dotted lines) source counts compared to those from LoTSS-DR2 (colour scheme the same as the top panel). The black and grey horizontal dashed lines mark a source counts ratio of 1.0 and 1.0$\pm$0.1 respectively. }
   \label{fig:sourcecounts}
\end{figure}

 The observed 120-168\,MHz Euclidean normalized LoTSS-DR2 differential source counts are shown in Fig. \ref{fig:sourcecounts} where there is good agreement (within 10\%) between our counts and the fit derived from TGSS-ADR by \cite{Intema_2017} above approximately 10\,mJy and with the fit derived from LoTSS-deep by \cite{Mandal_2021} from 1\,mJy to $\sim$5\,mJy. The overall shape of the source counts at higher flux densities ($>5$\,mJy) is also in reasonable agreement with the polynomial description derived by \cite{Mandal_2021} but there is a substantial scaling discrepancy. We note that this discrepancy is partly explained by the 7-th order polynomial fit by \cite{Mandal_2021} which does not fully capture the curvature of their measured counts at intermediate flux densities and systematically underestimates them between 10\,mJy and 100\,mJy. At these intermediate flux densities the \cite{Mandal_2021} polynomial fit is least well constrained because at lower flux densities there are good source statistics in the LoTSS-deep fields whilst at higher flux densities the polynomial fit takes into account the TGSS-ADR counts which steer it. Furthermore, if we consider the LoTSS-deep counts (crosses in Fig. \ref{fig:sourcecounts}) rather than the corresponding polynomial fit, we find that there is significantly better agreement with both LoTSS-DR2 and TGSS-ADR at flux densities above $\sim$30\,mJy.

 Other factors may contribute to the apparent discrepancy between LoTSS-DR2 and LoTSS-deep at flux densities below $\sim$30\,mJy, for example,  \cite{Mandal_2021} examined three LoTSS-deep fields of areas between 7 and 10 square degrees and found (see Fig. \ref{fig:sourcecounts}) comparable but slightly smaller (although up to 20\% in number of sources in a given flux density bin) overall scaling discrepancies to that seen here. \cite{Mandal_2021} attributed these to sample variance which may again contribute the LoTSS-deep and LoTSS-DR2 discrepancy.  Additional effects could also play a role though: for example, an overall flux density scale offset between LoTSS-DR2 and the LoTSS-deep fields could also contribute significantly, but if the effect is solely due to this, then the flux density scale offset is $\sim25\%$ which is larger than our estimated overall systematic error on the LoTSS-DR2 ($<10\%$; Sect. \ref{sec:flux_scale}) and the LoTSS-deep fields ($<10\%$; \citealt{Sabater_2021}) flux density scales. A further effect may come from the careful combination of \textsc{PyBDSF} radio components into single radio sources that was done for the LoTSS-deep fields (\citealt{Kondapally_2021}) but is yet to be completed for LoTSS-DR2 (see Sect. \ref{sec:value_added_cats}). However, a decrease in LoTSS-DR2 counts in a particular flux density bin of $\sim20\%$ would be required if this were the only contributing effect but given we found only a 2\% effect in LoTSS-DR1 (\citealt{Williams_2019}) it seems unlikely that this can explain the full discrepancy.  We also note that, at some flux densities (e.g. 10\,mJy to 30\,mJy), improving the agreement between the LoTSS-DR2  and LoTSS-deep source counts will increase the discrepancy between those derived from LoTSS-DR2 and TGSS-ADR.
 
As shown in Sect. \ref{sec:sensitivty}, LoTSS-DR2 begins to significantly suffer from incompleteness below $\sim$1\,mJy, although some regions of the survey are more sensitive than others (see Fig. \ref{fig:mosaic-noisemap}), and the catalogue contains 2,284,168 $S_I<$1\,mJy sources (i.e. 52\% of the total number of catalogued sources). As a consequence the 
 raw LoTSS-DR2 source counts significantly underestimate the true population at flux densities below $\sim$1\,mJy (where the counts are dominated by low-excitation radio galaxies and star forming galaxies - see Best et al. in prep). To assess incompleteness we follow two different approaches, both of which make use of the same deconvolution model simulations we presented in Sect. \ref{sec:sensitivty} where we injected large numbers (over 180 million) of $0.2<S_I<2.0$\,mJy sources into residual images across the entire LoTSS-DR2 region.

In the first approach, which we refer to as Method One, for each mosaic we use our simulations to examine the completeness for that particular mosaic. When counting our real sources as a function of flux density we then remove mosaics where the fraction of simulated sources detected at that particular flux density is less than 90\%. This approach thus mitigates incompleteness by focusing just on mosaics with good completeness at a particular flux density. 

In the second approach, referred to as Method Two, for each mosaic, we count the number of simulated sources that are detected as a  
function of flux density and only remove mosaics where the fraction of simulated sources detected at a particular flux density is below 20\%. As a mosaic begins to become incomplete, the recovered simulated source counts (derived using the flux density measurements from \textsc{PyBDSF}) begin to differ significantly from the injected ones, but we derive scaling factors for each flux density bin to correct the recovered simulated source counts so that they match the injected simulated ones. These same scaling factors are then applied to our real source counts. This approach therefore attempts to account for incompleteness as well as the characteristics of the image noise and source finding algorithm but is limited by the assumption that our simulations (our deconvolution model simulations that we presented in Sect. \ref{sec:sensitivty} which use the \citealt{Mandal_2021} source counts) accurately model the real source population (e.g. we do not account for the source sizes and occurrence as a function of flux density).

The results of these analyses are shown in Fig.  \ref{fig:sourcecounts} where it is apparent that both approaches somewhat correct our source counts below $\sim$1\,mJy. However, neither method brings our counts quite in line with expectations from \cite{Mandal_2021} which better probe the low flux density end due to more sensitivity and correspondingly lower levels of incompleteness (Method One results in the closest agreement but a $\sim$15\% discrepancy remains). We note that our simulations to correct for incompleteness only use point-like sources (with calibration errors accounted for) and future simulations that include a more realistic population of faint source morphologies may be required to bring the LoTSS-DR2 and LoTSS-deep field counts into better agreement. Furthermore, as previously mentioned, our source counts are derived from \textsc{PyBDSF} radio catalogues where real sources (of any flux density) may be split into several different components but efforts  
to combine LoTSS-DR2 components into final source catalogues with optical and infrared counterparts and redshift estimates are ongoing (Sect. \ref{sec:value_added_cats}). The outcome of this will   
enhance the accuracy of our source counts and permit further studies of various different source populations as well as their evolution and distribution but this is beyond the aim of this paper.

\subsection{Polarisation image properties}
\label{sec:pol_properties}

For each of the 841 fields in LoTSS-DR2 we create Stokes Q and U image
cubes with spatial resolutions of  $\sim 20\arcsec$ (maximum $uv$ distance of 25.75\,km) and $\sim4\arcmin$ (maximum $uv$ distance
of 1.6\,km) and with a frequency resolution of 97.6\,kHz. This provides image
cubes with 480 planes that span the LoTSS frequency coverage for each
pointing and each resolution. In addition we made a 120-168\,MHz
continuum Stokes V image at 20$\arcsec$ resolution for all pointings.
Detailed scientific analysis of these products will be presented in
forthcoming publications with O'Sullivan et al. (in prep) and
\cite{Erceg_inprep} exploring the Rotation Measure (RM) of emission in
the linearly polarised cubes (including 2,461 discrete, primarily extra-Galactic, sources and diffuse
Galactic structures) whilst Callingham et al. (in prep) characterise the
circularly polarised emission (including detections of approximately 100 sources which are mainly stellar systems or pulsars). We also refer the reader to these publications for discussions regarding the lack of an absolute polarisation angle calibration and the handedness of Stokes V.

\begin{figure*}   \centering
   \includegraphics[width=0.32\linewidth]{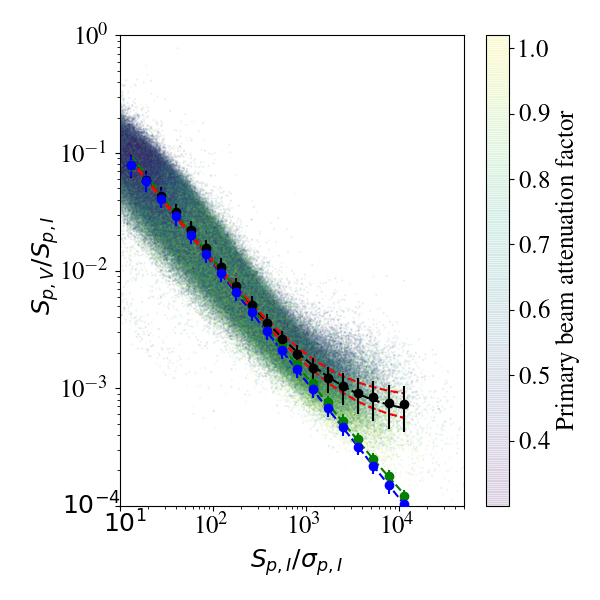}
   \includegraphics[width=0.32\linewidth]{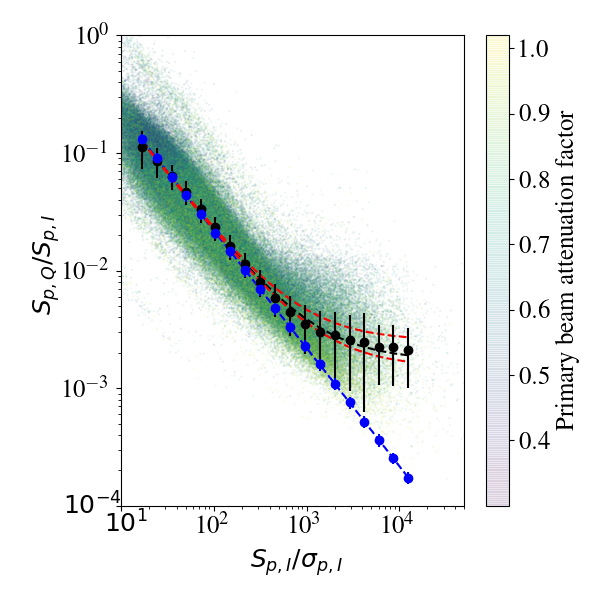}
   \includegraphics[width=0.32\linewidth]{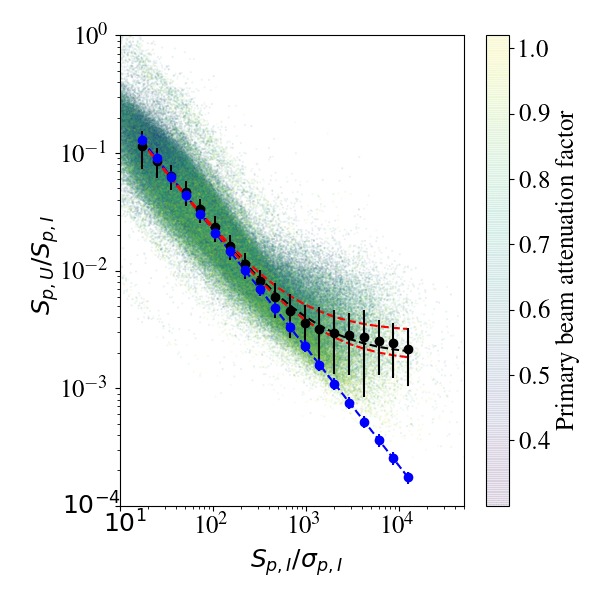}
    \caption{The observed Stokes V (left), Q (centre) and U (right) to Stokes I maximum absolute image pixel ratios as a function of the Stokes I signal-to-noise ratio.  The small points show measurements obtained from all compact, isolated LoTSS-DR2 sources with $S_I>$10\,mJy. The large black points show the median values in $S_{p,I}/\sigma_{p,I}$ bins and the error bars show the standard deviation of the sources within each bin. The dashed black fitted line shows the best fit through the points which are:  $S_{p,V}/S_{p,I} = 1.16 \times \sigma_{p,I}/S_{p,I} + 5.7\times10^{-4}$; $ S_{p,Q}/S_{p,I} = 2.12 \times \sigma_{p,I}/S_{p,I} + 1.7\times10^{-3}$; and $ S_{p,U}/S_{p,I} = 2.1 \times \sigma_{p,I}/S_{p,I} + 1.9\times10^{-3}$.
    The red dashed lines above and below the best fit line show the best fits derived when considering sources with a primary beam attenuation factor below 0.65 and above 0.65 respectively.  The large blue dashed line fitted with blue dots shows the situation if the Stokes V, Q or U signals are Gaussian random noise (with pixels chosen from a region size to align the observed and simulated distributions at low $S_{p,I}/\sigma_{p,I}$). In the Stokes V panel, the green dashed line fitted with green dots represents when the aperture is offset from the target and, as expected, this follows the Gaussian random noise simulation and demonstrates its validity. The real observations begin to differ significantly from the leakage-free simulations at around $S_{p,I}/\sigma_{p,I}=1,000$ -- above this $S_{p,I}/\sigma_{p,I}$ we have over 23,000 measurements in Q and U and 30,000 in V (where our maps are slightly wider in area) that constrain our leakage estimates at high $S_{p,I}/\sigma_{p,I}$.}
    \label{fig:pol-leakage}
\end{figure*}

Our Stokes V images, once flux density scaled by applying the scaling factors derived from the continuum LoTSS-DR2 maps, have a median sensitivity of 95\,$\mu$Jy/beam with a standard deviation of 30\,$\mu$Jy/beam. However, the noise in the Stokes V maps is not completely Gaussian, with noticeable artefacts around prominent Stokes I sources. To characterise this leakage from Stokes I into Stokes V we examined all our Stokes V maps at the locations of isolated LoTSS-DR2 sources with $S_I >10$\,mJy and assume that there is no real Stokes V signal in the images. For each source we find the maximum Stokes I ($S_{p,I}$) and the maximum absolute value of Stokes V  ($S_{p,V}$) in a $10\times10$ pixel aperture around the catalogued source position. We also find the local Stokes I noise level ($\sigma_{p,I}$). As shown in Fig \ref{fig:pol-leakage} the distribution of the recovered Stokes V to Stokes I ratio as a function of signal-to-noise of the Stokes I detection differs from what we would expect if the Stokes V image were purely Gaussian random noise. Instead, we can describe the observed Stokes V signal as $S_{p,V}/S_{p,I} = A \times \sigma_{p,I}/S_{p,I} + B$ where B describes the leakage from Stokes I into Stokes V and our best fitting parameters are A is 1.16 and B is $5.7\times10^{-4}$. Here A deviates from unity because we choose the maximum pixels within a finite aperture (so Stokes I corresponds to the peak of the source but Stokes V corresponds to the highest noise peak in that aperture) and B reflects the leakage level of 0.057\%. In Fig.  \ref{fig:pol-leakage} we also show that the leakage depends upon distance from the pointing centre, or primary beam attenuation, with sources further from the centre (or with more severe primary beam attenuation) exhibiting larger levels of leakage. 

\begin{figure}   \centering
   \includegraphics[width=\linewidth]{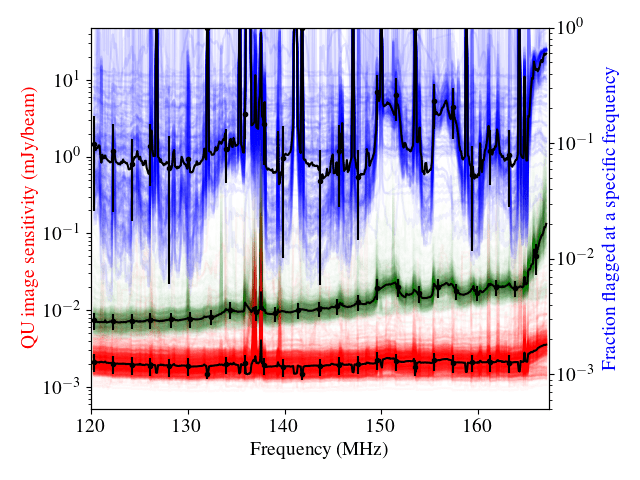}
    \caption{The sensitivity of our Stokes Q and U image cubes as a function of frequency is shown in green (4$\arcmin$ resolution) and red (20$\arcsec$ resolution). The fraction of data flagged in our LoTSS-DR2 datasets as a function of frequency is the key contributing factor to the variation in noise levels and is shown in blue. For each plot the individual coloured lines show the properties of each image cube or dataset where the black line, and corresponding errors, show the median values at a particular frequency plus/minus the standard deviation amongst the 841 different LoTSS pointings. }
    \label{fig:QU-noise}
\end{figure}

Excluding significant outliers and again scaling in the same way that we scaled our Stokes I maps, our Stokes Q and U images have a median sensitivity of 10.8\,mJy/beam and 2.2\,mJy/beam in each 97.6\,kHz bandwidth image with standard deviations of 5\,mJy/beam and 0.6\,mJy/beam for the $\sim4\arcmin$ and $\sim 20\arcsec$ image cubes respectively. However, as shown in Fig. \ref{fig:QU-noise}, the noise levels in the planes of the image cubes varies across the 120-168\,MHz band which is largely due to the amount of data that is flagged either as a result of it being contaminated with radio frequency interference (which is typically at the 9\% level) or otherwise being low quality data. We assess the Stokes I to Stokes Q and Stokes U leakage using the same procedure described above to examine the Stokes I to Stokes V leakage (again assuming that all Stokes Q and Stokes U signals are due to a combination of noise and leakage). Using individual Q and U image cube planes does not give sufficient sensitivity to make the assessment, so we instead use 120-168\,MHz band-averaged 20$\arcsec$ resolution Q and U images after RM synthesis which are described in O'Sullivan et al. (in prep). As shown in Fig. \ref{fig:pol-leakage}, we find leakage from Stokes I to Stokes Q is 0.17\% and from Stokes I to Stokes U is 0.19\%. 

There are several caveats to our analysis. Firstly, none of the V, Q and U images are deconvolved as the version of DDFacet used in the pipeline has only Stokes I deconvolution functionality (see \citealt{Tasse_2018} for more details). Furthermore, in the DI calibration steps in our direction dependent calibration pipeline we make the assumption that $Q = U = V = 0$\,Jy which has the effect of suppressing the instrumental polarisation leakage but the consequence of producing spurious polarised signals when bright real polarised signals ($>10$\,mJy) are present (see \citealt{Tasse_2021} and O'Sullivan et al. in prep for further details). However, encouragingly, we do note that both the leakage levels we have derived, as well as the observed increase in leakage as a function of distance from the pointing centre, are comparable with the results presented in  \cite{Asad_2016} who thoroughly examined leakage through real measurements and simulations that probed the accuracy and limitations of the current LOFAR beam model.

\section{Public data release} \label{sec:data_release}

To enable thorough scientific exploitation of the data, in LoTSS-DR2 a wide variety of different data products are being made publicly available with DOI:10.25606/SURF.LoTSS-DR2 and can be accessed via the LOFAR surveys webpage\footnote{\url{https://lofar-surveys.org/}}, the ASTRON Virtual Observatory\footnote{\url{https://vo.astron.nl}} and the SURF Data Repository\footnote{\url{https://repository.surfsara.nl/}}. More specifically, this data release includes the following Stokes I products:
\begin{enumerate}[label=\textbf{Product.A\arabic*},ref=\textbf{Step.A\arabic*},leftmargin=1.9cm]
\item Mosaiced astrometric-corrected Stokes I 6$\arcsec$  resolution 120-168\,MHz restored images with associated \textsc{PyBDSF} residual and noise maps.
\item Mosaiced Stokes I 20$\arcsec$ resolution 120-168\,MHz restored images.
\item Individual Stokes I 6$\arcsec$ (with and without astrometric correction) and 20$\arcsec$ resolution 120-168\,MHz restored images with associated DDFacet model, residual and mask images.
\item Individual Stokes I 6$\arcsec$ (with astrometric correction)  resolution 16\,MHz bandwidth images with central frequencies of 128, 144 and 160\,MHz.
\item Catalogue of 4,396,228 radio sources and the 5,121,366 Gaussian components that describe them (example entries are shown in Tab. 1).
\item Hierarchical Progressive Surveys (see \citealt{Fernique_2015}) images to visualise the mosaiced Stokes I 6$\arcsec$ and 20$\arcsec$ resolution images.
\end{enumerate}
In addition, the following polarisation products are included in this data release:
\begin{enumerate}[label=\textbf{Product.B\arabic*},ref=\textbf{Step.A\arabic*},leftmargin=1.9cm]
\item   Individual observation Stokes QU 20$\arcsec$ resolution undeconvolved 480-plane image cubes with a frequency resolution of 97.6\,kHz.
\item  Individual observation Stokes QU 4$\arcmin$ resolution undeconvolved 480-plane image cubes with a frequency resolution of  97.6\,kHz.
\item  Individual observation Stokes V 20$\arcsec$ resolution 120-168\,MHZ continuum undeconvolved images.
\end{enumerate}
And finally the following $uv$-data and calibration products are included for each pointing:
\begin{enumerate}[label=\textbf{Product.C\arabic*},ref=\textbf{Step.A\arabic*},leftmargin=1.9cm]
\item DI calibrated visibilities and DD calibration solutions.
\item Facet layout and astrometric corrections 
\end{enumerate}

 We emphasise that the released products do not incorporate the range of limitations that are described in Sect. \ref{sec:image_quality} as these can vary with aspects such as observing conditions and source properties and are not always applicable. For example, the catalogues and noise images contain only the \textsc{PyBDSF} statistical uncertainties from the source characterisation. The extensive variety of products do, however, allow for a robust assessment of the data and image accuracy and, together with the description in this publication, we hope that they can allow users to ascertain which, if any, limitations are appropriate to consider in any given study.
 
\begin{sidewaystable*}
\centering
{{Table 1:} An example of entries in the LoTSS-DR2 source catalogue which contains a total of 4,396,228 sources and is derived from 120-168\,MHz images (central frequency of 144\,MHz) with a resolution of 6$\arcsec$. The columns are: source identifier (ID), J2000 right ascension (RA), J2000 declination (Dec), peak brightness ($\rm{S_{P}}$), integrated flux density ($\rm{S_{I}}$), major axis (Maj), minor axis (Min), deconvolved major axis (DC Maj), deconvolved minor axis (DC Min), position angle (PA),   local noise at the source location (RMS), type of source as classified by \textsc{PyBDSF} (Type -- here `S' indicates an isolated source fit with a single Gaussian; `C' represents sources fit by a single Gaussian but within an island of emission that contains other sources; and `M' is used for sources that are extended and fitted with multiple Gaussians), the mosaic identifier (Mosaic), the number of pointings that are mosaiced at the position of the source (Number pointings), and the primary beam attenuation weighted average of the deconvolution mask images at the source location (Masked fraction). The errors in the catalogue are the uncertainties obtained from the \textsc{PyBDSF} source fitting. Additional uncertainties on the source extensions, astrometry, flux density scale as well as descriptions of completeness, dynamic range and diffuse emission recovery limitations are described in Sect. \ref{sec:image_quality}.  }\\

\label{tab:catalogue-example}
\begin{tabular}{lccccccccccccccc}
\hline
Source ID & RA  & DEC  & $S_{P}$  & $S_{I}$ & Maj & Min & DC\_Maj    & DC\_Min  & PA & RMS  & Type & Mosaic & Number & Masked \\  
 &   &   & (mJy/  & (mJy) & ($\arcsec$) & ($\arcsec$)  &  ($\arcsec$)   &  ($\arcsec$)& ($^\circ$)  & (mJy/  &  & & pointings & fraction  \\ 
& & & beam) &   & & & & & & beam) \\  \hline
ILTJ075321.87 & 118.3411$^\circ$ & 27.9383$^\circ$ & 2.22 & 2.75  & 6.9 & 6.5  & 3.4  & 2.4  & 71 & 0.09 & S & P117+27 & 3 & 1.00  \\
+275617.7 & $\pm$0.1$\arcsec$ & $\pm$0.1$\arcsec$ & $\pm$0.09 & $\pm$0.19 & $\pm$0.3 & $\pm$0.3 & $\pm$0.3 & $\pm$0.3 & $\pm$28 \\
ILTJ133639.64 & 204.1652$^\circ$ & 52.8310$^\circ$ & 0.51 & 0.76  & 7.5 & 7.2  & 4.5  & 3.9  & 114 & 0.06 & S & P5Hetdex41 & 3 & 0.50  \\
+524951.5 & $\pm$0.4$\arcsec$ & $\pm$0.4$\arcsec$ & $\pm$0.06 & $\pm$0.15 & $\pm$1.0 & $\pm$0.9 & $\pm$1.0 & $\pm$0.9 & $\pm$115 \\
ILTJ105955.28 & 164.9803$^\circ$ & 55.0005$^\circ$ & 0.28 & 2.13  & 21.9 & 12.7  & 21.1  & 11.1  & 92 & 0.07 & S & P164+55 & 4 & 0.00  \\
+550001.9 & $\pm$2.1$\arcsec$ & $\pm$1.0$\arcsec$ & $\pm$0.06 & $\pm$0.50 & $\pm$4.9 & $\pm$2.5 & $\pm$4.9 & $\pm$2.5 & $\pm$24 \\
ILTJ105319.51 & 163.3313$^\circ$ & 45.7868$^\circ$ & 0.62 & 0.77  & 7.3 & 6.2  & 4.1  & 1.5  & 6 & 0.06 & S & P163+45 & 4 & 0.87  \\
+454712.5 & $\pm$0.3$\arcsec$ & $\pm$0.3$\arcsec$ & $\pm$0.07 & $\pm$0.13 & $\pm$0.8 & $\pm$0.6 & $\pm$0.8 & $\pm$0.6 & $\pm$29 \\
ILTJ125532.50 & 193.8854$^\circ$ & 45.0397$^\circ$ & 0.48 & 0.78  & 8.8 & 6.6  & 6.5  & 2.8  & 83 & 0.13 & S & P32Hetdex08 & 3 & 0.00  \\
+450223.0 & $\pm$1.2$\arcsec$ & $\pm$0.7$\arcsec$ & $\pm$0.13 & $\pm$0.33 & $\pm$2.7 & $\pm$1.6 & $\pm$2.7 & $\pm$1.6 & $\pm$45 \\
ILTJ085443.03 & 133.6793$^\circ$ & 46.2299$^\circ$ & 0.76 & 1.17  & 8.0 & 6.9  & 5.3  & 3.4  & 124 & 0.08 & S & P133+47 & 4 & 0.50  \\
+461347.4 & $\pm$0.4$\arcsec$ & $\pm$0.4$\arcsec$ & $\pm$0.09 & $\pm$0.21 & $\pm$1.0 & $\pm$0.8 & $\pm$1.0 & $\pm$0.8 & $\pm$34 \\
ILTJ022403.60 & 36.0150$^\circ$ & 30.0882$^\circ$ & 13.00 & 30.94  & 9.2 & 8.9  & 7.0  & 6.6  & 102 & 0.10 & M & P036+31 & 3 & 1.00  \\
+300517.5 & $\pm$0.0$\arcsec$ & $\pm$0.0$\arcsec$ & $\pm$0.10 & $\pm$0.53 & $\pm$0.1 & $\pm$0.1 & $\pm$0.1 & $\pm$0.1 & $\pm$10 \\
ILTJ144426.07 & 221.1086$^\circ$ & 36.7295$^\circ$ & 0.62 & 0.79  & 7.1 & 6.5  & 3.9  & 2.4  & 161 & 0.11 & S & P219+37 & 3 & 0.00  \\
+364346.3 & $\pm$0.5$\arcsec$ & $\pm$0.6$\arcsec$ & $\pm$0.11 & $\pm$0.23 & $\pm$1.3 & $\pm$1.1 & $\pm$1.3 & $\pm$1.1 & $\pm$80 \\
ILTJ143748.86 & 219.4536$^\circ$ & 36.7022$^\circ$ & 2.21 & 2.93  & 7.1 & 6.7  & 3.8  & 3.0  & 170 & 0.07 & S & P219+37 & 3 & 1.00  \\
+364207.9 & $\pm$0.1$\arcsec$ & $\pm$0.1$\arcsec$ & $\pm$0.08 & $\pm$0.16 & $\pm$0.2 & $\pm$0.2 & $\pm$0.2 & $\pm$0.2 & $\pm$28 \\
ILTJ100259.85 & 150.7494$^\circ$ & 59.4202$^\circ$ & 0.66 & 0.98  & 7.4 & 7.2  & 4.4  & 4.1  & 90 & 0.06 & S & P150+60 & 4 & 0.81  \\
+592512.5 & $\pm$0.3$\arcsec$ & $\pm$0.3$\arcsec$ & $\pm$0.06 & $\pm$0.14 & $\pm$0.7 & $\pm$0.6 & $\pm$0.7 & $\pm$0.6 & $\pm$138 \\
ILTJ100138.17 & 150.4091$^\circ$ & 59.2853$^\circ$ & 0.61 & 0.92  & 7.7 & 7.0  & 4.9  & 3.5  & 177 & 0.06 & S & P150+60 & 3 & 0.86  \\
+591707.0 & $\pm$0.3$\arcsec$ & $\pm$0.3$\arcsec$ & $\pm$0.06 & $\pm$0.14 & $\pm$0.8 & $\pm$0.7 & $\pm$0.8 & $\pm$0.7 & $\pm$40 \\
ILTJ215336.49 & 328.4020$^\circ$ & 27.4076$^\circ$ & 0.95 & 1.45  & 8.2 & 6.7  & 5.6  & 3.0  & 64 & 0.15 & S & P330+28 & 2 & 0.75  \\
+272427.3 & $\pm$0.6$\arcsec$ & $\pm$0.5$\arcsec$ & $\pm$0.15 & $\pm$0.36 & $\pm$1.5 & $\pm$1.0 & $\pm$1.5 & $\pm$1.0 & $\pm$36 \\
ILTJ224657.98 & 341.7416$^\circ$ & 26.3063$^\circ$ & 0.25 & 1.93  & 18.3 & 15.3  & 17.3  & 14.1  & 25 & 0.10 & S & P341+26 & 2 & 0.00  \\
+261822.6 & $\pm$2.4$\arcsec$ & $\pm$2.7$\arcsec$ & $\pm$0.09 & $\pm$0.77 & $\pm$6.6 & $\pm$5.3 & $\pm$6.6 & $\pm$5.3 & $\pm$93 \\
ILTJ121006.34 & 182.5264$^\circ$ & 29.5820$^\circ$ & 17.91 & 69.08  & 30.0 & 8.0  & 29.4  & 5.3  & 85 & 0.11 & M & P183+30 & 3 & 1.00  \\
+293455.0 & $\pm$0.0$\arcsec$ & $\pm$0.2$\arcsec$ & $\pm$0.11 & $\pm$0.96 & $\pm$0.6 & $\pm$0.1 & $\pm$0.6 & $\pm$0.1 & $\pm$1 \\
ILTJ221407.92 & 333.5330$^\circ$ & 34.7886$^\circ$ & 8.90 & 15.92  & 9.1 & 7.1  & 6.8  & 3.7  & 97 & 0.10 & S & P334+36 & 2 & 1.00  \\
+344719.0 & $\pm$0.0$\arcsec$ & $\pm$0.0$\arcsec$ & $\pm$0.10 & $\pm$0.27 & $\pm$0.1 & $\pm$0.1 & $\pm$0.1 & $\pm$0.1 & $\pm$2 \\
ILTJ083500.48 & 128.7520$^\circ$ & 54.4843$^\circ$ & 0.74 & 2.31  & 12.3 & 9.1  & 10.8  & 6.9  & 50 & 0.12 & S & P128+54 & 5 & 0.62  \\
+542903.3 & $\pm$0.8$\arcsec$ & $\pm$0.7$\arcsec$ & $\pm$0.12 & $\pm$0.48 & $\pm$2.2 & $\pm$1.4 & $\pm$2.2 & $\pm$1.4 & $\pm$27 \\
ILTJ113243.58 & 173.1816$^\circ$ & 51.8272$^\circ$ & 0.54 & 0.57  & 7.0 & 5.5  & 0.0  & 0.0  & 83 & 0.07 & S & P12Hetdex11 & 4 & 0.00  \\
+514938.0 & $\pm$0.4$\arcsec$ & $\pm$0.3$\arcsec$ & $\pm$0.07 & $\pm$0.13 & $\pm$1.1 & $\pm$0.7 & $\pm$1.1 & $\pm$0.7 & $\pm$27 \\
 \hline  
 \end{tabular}
\end{sidewaystable*}

\section{Future prospects}
\label{sec:future_prospects}

There are still significant technical challenges to address as we build from LoTSS-DR2 to a complete high quality northern hemisphere survey. Furthermore, there is substantial auxiliary information that can be added to the LoTSS-DR2 catalogues themselves to greatly enrich their scientific value. Several avenues we are currently pursuing are outlined below.

\subsection{Value-added catalogue}
\label{sec:value_added_cats}

Our aim, as with LoTSS DR1 \citep{Williams_2019}, is to produce a
value-added catalogue with reliable optical counterparts for as many
radio sources as possible, together with photometric redshifts
\citep{Duncan_2019} and, eventually, spectroscopic redshifts for
bright radio sources to be generated by the WEAVE-LOFAR project
\citep{Smith_2016}. DR2's sky coverage is well matched to the optical
coverage of the DESI Legacy Survey and therefore we use the Legacy
Survey in searches for optical counterparts, together with {\it WISE}
all-sky coverage in the near and mid-infrared.

The process of determining optical identifications for LoTSS DR2
sources is complex and will follow the basic approach laid out for DR1
by \cite{Williams_2019}. Likelihood ratio crossmatching will be used
for compact, isolated radio sources, but for larger sources a range of
heuristics will be used to try to ensure that optical counterparts are  not missed for
e.g. slightly extended single component sources or small double sources where the counterpart may not be coincident
with the radio components or the flux-weighted position. For the largest sources,
and for others selected for human visual inspection as part of the
processing, source `association' (the grouping together of multiple
{\sc PyBDSF} radio components into a single physical source) and
identification are done together as part of the Radio Galaxy Zoo
(LOFAR) Zooniverse project\footnote{\url{https://www.zooniverse.org/projects/chrismrp/radio-galaxy-zoo-lofar}}, with extensive involvement from many
volunteers from the general public as well as astronomers. At the time
of writing, this project is 80\% complete for sources with at least
one large, bright component (major axis $>15\arcsec$ and integrated flux density $>4$ mJy), but
significantly further behind for smaller and/or fainter sources that
may be part of a larger association. Once a particular area of sky is
returned from Radio Galaxy Zoo, there is then further manual
processing to deal with objects flagged by volunteers as needing
further inspection in various ways. There is also some manual inspection
of large sources, and extended (associated) sources without an optical
ID that then are passed through a version of the code of \cite{Barkus_2021} which uses the radio ridge line (pathway of connected highest intensity points) to make a refined estimate of
the maximum likelihood optical counterpart. About a quarter of the
final area of DR2 has complete optical identifications as the end
product of these processes at the time of writing, and the overall
optical identification fraction is 87\%, with 97.7\% of these having optical IDs
from the likelihood ratio process, 0.7\% from the ridge line code and
1.5\% from visual inspection in Radio Galaxy Zoo or the other manual
processes; thus it seems likely that the optical ID fraction in DR2,
with the improved optical depth of the Legacy survey, will be
significantly higher than the 73\% achieved using Pan-STARRS with LoTSS-DR1.
A final value-added catalogue for the DR2 radio sources, with photometry and estimated photometric redshifts, will be produced in the manner described
by \cite{Kondapally_2021} and \cite{Duncan_2021}, and we hope to be able to release such a
catalogue publicly by the end of 2022. A detailed description of the process will be given by Hardcastle et al. (in prep).

\subsection{Improved calibration and imaging}
\label{sec:improved_imaging}

The size of the facets in LoTSS-DR2 is the main source of phase errors in the images. This is clear from \cite{vanWeeren_2021} where a post processing scheme has been presented and was demonstrated to improve the quality of the calibration towards particular targets of interest by calibrating the sky in a small area around a particular target. Such a scheme has already been applied to study the faint diffuse emission from all 309 Sunyaev Zel'dovich detected clusters in the  \cite{Planck_2016} catalogue that lie in the LoTSS-DR2 region, thus highlighting its feasibility even for moderately sized samples (Botteon et al. in prep). The effect can also be demonstrated by simply increasing the number of facets used when processing with our existing pipeline, although at the cost of some stability as facets contain less emission. A promising approach to refine wide-field imaging whilst maintaining stability was outlined by \cite{Albert_2020} who demonstrated that ionospheric phase screens derived from LoTSS-DR2 solutions are able to significantly reduce calibration artefacts in some situations. This procedure is now being more widely tested using LoTSS-DR2 products to ascertain whether we are able to use it to routinely improve our images.

\begin{figure*}   \centering
   \includegraphics[width=0.47\linewidth]{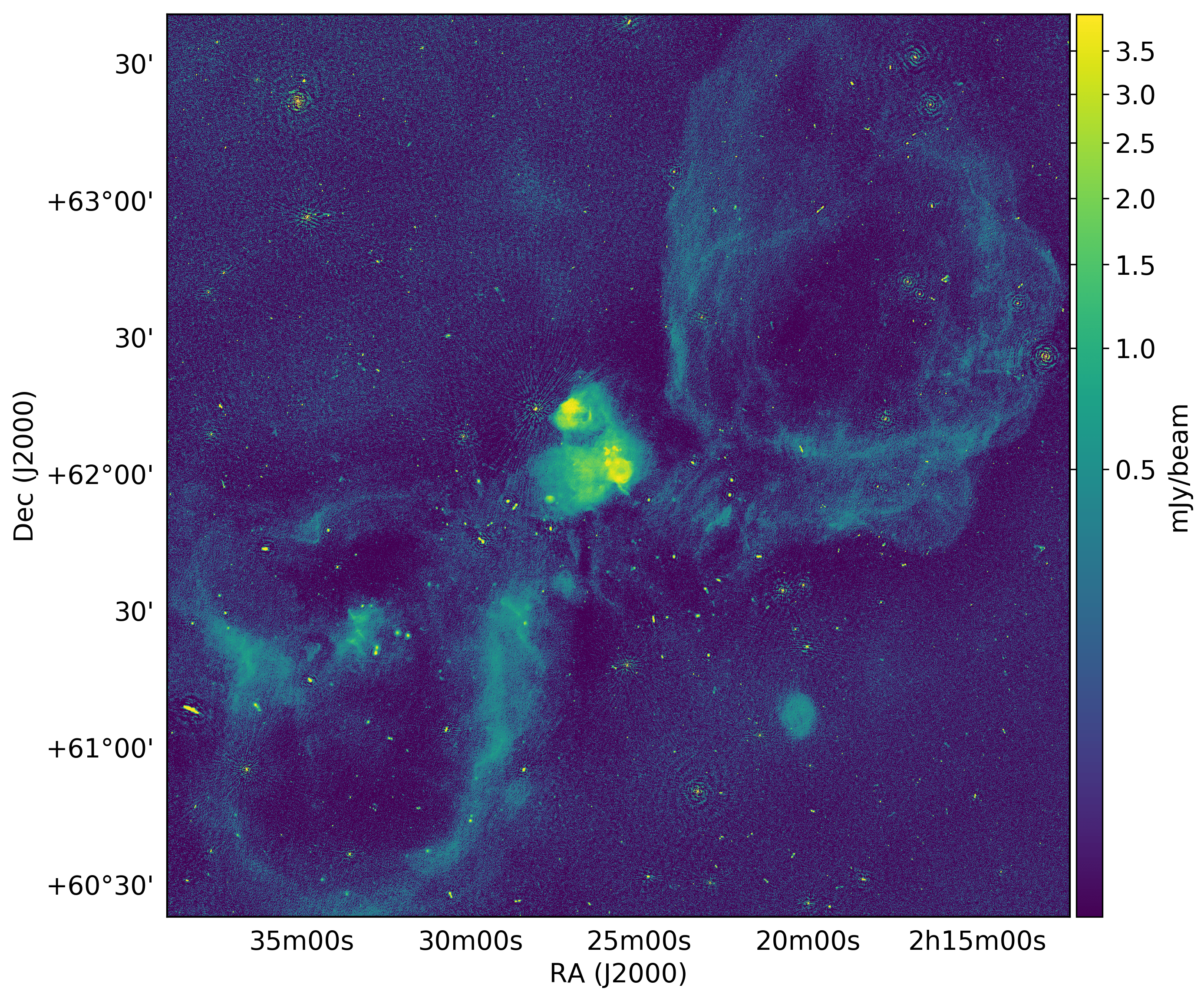}
   \includegraphics[width=0.47\linewidth]{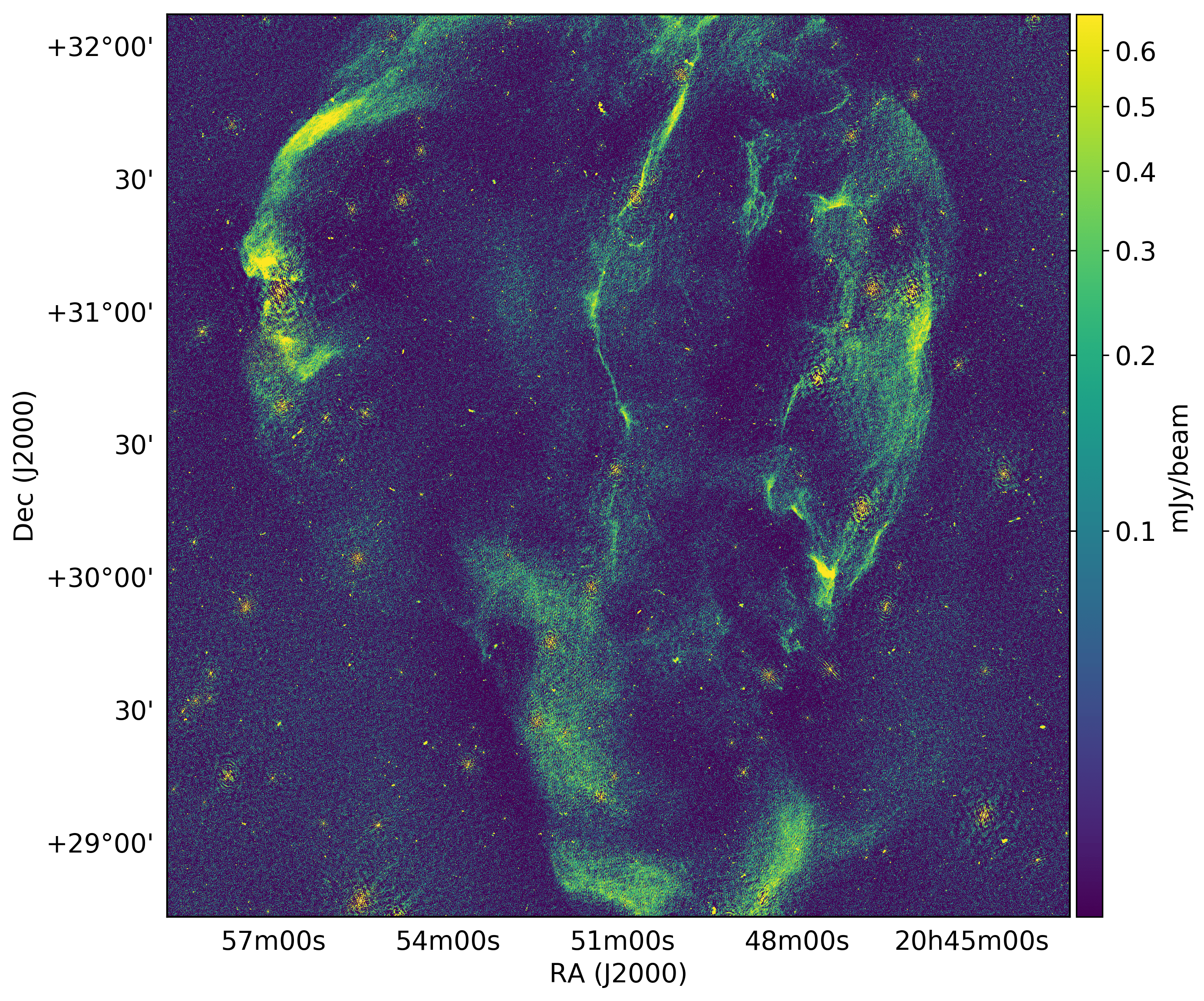}
    \caption{LoTSS mosaics showing the W3/W4/HB3 star forming region (left) and the Cygnus loop supernova remnant at an angular resolution of 6$\arcsec$. The mosaics have sensitivities of approximately 120$\mu$Jy/beam and the emission, which is deconvolved, from each of these Galactic regions spans over 12\,million  1.5$\arcsec$ $\times$ 1.5$\arcsec$ pixels. The calibration of these complex regions was done with a modified version of the pipeline used for LoTSS-DR2 which provided more stability whilst still allowing us to calibrate in 45 different directions for each individual pointing that makes up these mosaics (see Sect. \ref{sec:improved_imaging} for details).}
    \label{fig:cygnus_loop}
\end{figure*}

To date, in both LoTSS-DR1 and LoTSS-DR2 we have focused on imaging
regions at high Galactic latitude. However, the ambition of LoTSS is
to also image the part of the Galactic plane in the northern
hemisphere. Whilst we have presently only accumulated a small amount
of data in this region (see Fig. \ref{fig:DR2-region})  we have used
these data to test the performance of our pipeline in Galactic regions
that contain much more extended emission than extra-Galactic regions.
The two demonstration fields we have used for this are the Cygnus loop
and W3/W4/HB3 region. To image these regions we modified our pipeline by
removing all direction independent calibration steps and rather than performing direction dependent amplitude calibration we performed further direction dependent phase
calibration instead. DDFacet was also modified to allow us to image
large islands of contiguous emission and a final post-processing deconvolution step with a manually drawn deconvolution mask was conducted. These modifications provided more stability in the pipeline and allowed us to image the complex emission in these fields, which is challenging to fully reproduce in the skymodels that are derived during imaging. As in regular LoTSS-DR2 processing, the data from each pointing was imaged over an $8^\circ \times 8^\circ$ region and was calibrated in 45 different directions. In the processing of the 11 pointings that make up the mosaics the calibration did not diverge significantly for any facets. Furthermore the pipeline runtime only increased due to the additional imaging step which typically took 50\,hrs to complete on a compute node with
512 GB RAM and two Intel Xeon E5 v4 processors (16 cores each).
The results of these
demonstrations are shown in Fig. \ref{fig:cygnus_loop} and whilst the
image fidelity is encouraging, we plan to conduct further tests and make
further refinements to ensure that we can image the whole of the Galactic
plane in the final northern-hemisphere survey.

\subsection{Other uses of the data}

As described in Sect. \ref{sec:data_release} we have released a large
variety of products to allow for studies of continuum emission,
polarisation and further processing. There are, however, still many possibilities
to further exploit the data which can be achieved by reprocessing the
LoTSS datasets in the LTA to make use of the high time and frequency
resolution of the data before it is averaged for standard LoTSS
processing. For example, the data have sufficient time and frequency
resolution to facilitate wide-field subarcsecond resolution imaging allowing us to resolve much more than the 8.0\% of sources we resolve at 6$\arcsec$ resolution (see e.g. \citealt{Morabito_prep} and \citealt{Sweijen_2022}). In
addition, the frequency resolution is high enough for wide area
searches for spectral lines such as carbon radio recombination lines
(e.g.  \citealt{Emig_2020}, \citealt{Salas_2019}). There are also
other novel uses such as examining the time-variability of the data to search exoplanet signatures  
(e.g. \citealt{Vedantham_2020}) or transient signals but at even higher time resolution than has been done to-date. The LoTSS datasets can also be reprocessed to, for example, optimise surface brightness sensitivity by removing contaminating sources (i.e. mitigating source confusing) and reimaging at low resolution in order to search for faint large-scale emission such as that expected from cosmic filaments or Galactic structures
(e.g. Oei et al. in prep).

\section{Summary}
\label{sec:summary}

We have presented the second data release from LoTSS which is much
higher fidelity and spans 13 times the area of our first data
release. This data release is the largest yet from our collaboration,
and the largest radio survey in terms of total source numbers so far carried out.
It includes a catalogue of 4,396,228 radio sources and 120-168\,MHz
Stokes I, Q, U and V images that span 27\% of the northern sky.  A
characterisation of our Stokes I images reveals that we: resolve
approximately 8.0\% of the sources; have an astrometric accuracy of
0.2$\arcsec$; have a flux density scale accuracy of 10\%; suffer from severe (15\%
increase in noise) dynamic range limitations in only 2.5\% of the
surveyed area; have an estimated point-source completeness of 90\% at
0.8\,mJy/beam; recover even completely unmodelled diffuse
emission at the $>60\%$ power level; and are able to derive in-band spectral indices, although with very limited precision. Our Stokes Q and U image cubes have a
sensitivity of 2.2\,mJy/beam for each 97.6\,kHz image plane whilst our
Stokes V 120-168\,MHz continuum images reach a sensitivity of
95$\mu$Jy/beam. The Stokes I to Q, U or V leakage is characterised to
be less than 0.2\% but the polarisation images will be further
evaluated in forthcoming publications. The data from this release are publicly available with DOI:10.25606/SURF.LoTSS-DR2 and can be accessed via the collaboration's webpage\footnote{\url{https://lofar-surveys.org/}}, the ASTRON Virtual Observatory\footnote{\url{https://vo.astron.nl}} and the SURF Data Repository\footnote{\url{https://repository.surfsara.nl/}}.

Our aim is now to secure the observing time required to complete LoTSS  whilst ensuring we are able to process the data in a way that maximises the scientific opportunities. To this end we have secured observations that will extend our coverage from 67\% to 85\% of the northern hemisphere by May 2023.

\section{Acknowledgements}

LOFAR is the Low Frequency Array designed and constructed by ASTRON. It has observing, data processing, and data storage facilities in several countries, which are owned by various parties (each with their own funding sources), and which are collectively operated by the ILT foundation under a joint scientific policy. The ILT resources have benefited from the following recent major funding sources: CNRS-INSU, Observatoire de Paris and Université d'Orléans, France; BMBF, MIWF-NRW, MPG, Germany; Science Foundation Ireland (SFI), Department of Business, Enterprise and Innovation (DBEI), Ireland; NWO, The Netherlands; The Science and Technology Facilities Council, UK; Ministry of Science and Higher Education, Poland; The Istituto Nazionale di Astrofisica (INAF), Italy. 

This research made use of the Dutch national e-infrastructure with support of the SURF Cooperative (e-infra 180169) and NWO (grant 2019.056). The Jülich LOFAR Long Term Archive and the German LOFAR network are both coordinated and operated by the Jülich Supercomputing Centre (JSC), and computing resources on the supercomputer JUWELS at JSC were provided by the Gauss Centre for Supercomputing e.V. (grant CHTB00) through the John von Neumann Institute for Computing (NIC).

This research made use of the University of Hertfordshire high-performance computing facility and the LOFAR-UK computing facility located at the University of Hertfordshire and supported by STFC [ST/P000096/1], and of the Italian LOFAR IT computing infrastructure supported and operated by INAF, and by the Physics Department of Turin university (under an agreement with Consorzio Interuniversitario per la Fisica Spaziale) at the C3S Supercomputing Centre, Italy. The data are published via the SURF Data Repository service which is supported by the EU funded DICE project (H2020-INFRAEOSC-2018-2020 under Grant Agreement no. 101017207).

MJH acknowledges support from STFC [ST/R000905/1, ST/V000624/1]. PNB and JS are grateful for support from the UK STFC via grants ST/R000972/1 and ST/V000594/1. WLW  acknowledges support from the CAS-NWO programme for radio astronomy with project number 629.001.024, which is financed by the Netherlands Organisation for Scientific Research (NWO). AB acknowledges support from the VIDI research programme with project number 639.042.729, which is financed by the Netherlands Organisation for Scientific Research (NWO). AD acknowledges support by the BMBF Verbundforschung under the grant 05A20STA. RJvW and RT acknowledge support from the ERC Starting Grant ClusterWeb 804208. M. Br\"uggen and FdG acknowledge support from the Deutsche Forschungsgemeinschaft under Germany's Excellence Strategy - EXC 2121 “Quantum Universe” - 390833306. KJD acknowledges funding from the European Union’s Horizon 2020 research and innovation programme under the Marie Sk\l{}odowska-Curie grant agreement No. 892117 (HIZRAD). CLH acknowledges support from the Leverhulme Trust through an Early Career Research Fellowship. MH acknowledges funding from the European Research Council (ERC) under the European Union's Horizon 2020 research and innovation programme (grant agreement No 772663). LKM is grateful for support from the UKRI Future Leaders Fellowship (grant MR/T042842/1). DJS and RJD acknowledge support by the BMBF Verbundforschung under grant 05A20PB1 and 05A20PC3, respectively. The research of OS is supported by the South African Research Chairs Initiative of the Department of Science and Technology and National Research Foundation. GJW gratefully acknowledges support of an Emeritus Fellowship from the Leverhulme Trust.  LA acknowledges support from the STFC through a ScotDIST Intensive Data Science Scholarship. A. Bonafede acknowledges support from ERC Stg `DRANOEL' n. 714245 and MIUR FARE grant “SMS”.  E. Bonnassieux acknowledges support from the ERC-Stg grant `DRANOEL', n.714245. M. Brienza acknowledges financial support from the ERC Stg “MAGCOW”, no. 714196 and from ERC Stg `DRANOEL' no. 714245. M. Bilicki is supported by the Polish National Science Center through grants no. 2020/38/E/ST9/00395, 2018/30/E/ST9/00698 and 2018/31/G/ST9/03388, and by the Polish Ministry of Science and Higher Education through grant DIR/WK/2018/12. M. Bonato and IP acknowledge support from INAF under the SKA/CTA PRIN “FORECaST” and the PRIN MAIN STREAM “SAuROS” projects and from the Ministero degli Affari Esteri e della Cooperazione Internazionale - Direzione Generale per la Promozione del Sistema Paese Progetto di Grande Rilevanza ZA18GR02. VC acknowledges support from the Alexander von Humboldt Foundation. JHC and BM acknowledge support from the UK Science and Technology Facilities Council (STFC) under grants ST/R00109X/1, ST/R000794/1, and ST/T000295/1. HE acknowledges supported by the DFG under grant number 427771150. KE is a Jansky Fellow of the National Radio Astronomy Observatory, USA. MHaj acknowledges the MSHE for granting funds for Polish contribution to the International LOFAR Telescope (decision no.~DIR/WK/2016/2017/05-1) and for maintenance of the LOFAR PL-612 Baldy (decision no.~59/E-383/SPUB/SP/2019.1), and the NAWA Bekker fellowship (grant No PPN/BEK/2019/1/00431). GDG acknowledges support from the Alexander von Humboldt Foundation. DNH acknowledges support from the ERC through the grant ERC-Stg `DRANOEL' n. 714245. MH acknowledges support by the BMBF Verbundforschung under the grant 05A20STA. VJ acknowledges support by the Croatian Science Foundation for a project IP-2018-01-2889 (LowFreqCRO). RK acknowledges support from the Science and Technology Facilities Council (STFC) through an STFC studentship via grant ST/R504737/1. MKB and AW acknowledge support from the National Science Centre, Poland under grant no. 2017/26/E/ST9/00216. K.M. has been supported by the National Science Centre (UMO-2018/30/E/ST9/00082). FM is supported by the ``Departments of Excellence 2018 - 2022'' Grant awarded by the Italian Ministry of Education, University and Research (MIUR) (L. 232/2016).  SJN is supported by the Polish National Science Center through grant UMO-2018/31/N/ST9/03975. TP is supported by the BMBF Verbundforschung under grant number 50OR1906. KR acknowledges financial support from the ERC Starting Grant “MAGCOW”, no. 714196.  CJR acknowledges financial support from the ERC Starting Grant `DRANOEL', number 714245. MV acknowledges financial support from the Inter-University Institute for Data Intensive Astronomy (IDIA), a partnership of the University of Cape Town, the University of Pretoria, the University of the Western Cape and the South African Radio Astronomy Observatory, and from the South African Department of Science and Innovation's National Research Foundation under the ISARP RADIOSKY2020 Joint Research Scheme (DSI-NRF Grant Number 113121) and the CSUR HIPPO Project (DSI-NRF Grant Number 121291). XZ acknowledges support from China Scholarship Council  The Dunlap Institute is funded through an endowment established by the David Dunlap family and the University of Toronto. J.L.W. acknowledge the support of the Natural Sciences and Engineering Research Council of Canada (NSERC) through grant RGPIN-2015-05948, and of the Canada Research Chairs program.

\label{lastpage}

\end{document}